\documentclass[aps,prd,superscriptaddress,nofootinbib,amsmath,amsfonts, preprintnumbers, groupedaddress, showpacs,
10pt,english]{revtex4-1}
\usepackage{amsmath}
\usepackage{amssymb}
\usepackage{babel}
\usepackage{wrapfig}
\usepackage{cancel}

\usepackage{relsize,exscale}
\makeatletter

\usepackage{array,multirow,graphicx}
\usepackage{dcolumn}
\usepackage{newlfont}
\usepackage{bm}
\usepackage[colorlinks,citecolor=blue,urlcolor=blue,linkcolor=blue]{hyperref}
\usepackage[figtopcap]{subfigure}
\usepackage{color}

\newcommand{\beq}{\begin{equation}}
\newcommand{\eeq}{\end{equation}}
\newcommand{\bea}{\begin{eqnarray}}
\newcommand{\eea}{\end{eqnarray}}

\newcommand{\alp}{\alpha}

\newcommand{\del}{\delta}
\newcommand{\bet}{\beta}

\newcommand{\dd}{\text{d}}
\newcommand{\comm}[1]{}

\newcommand{\Ri}{\mathcal{R}}
\usepackage{scalerel}
\usepackage{tikz}
\usetikzlibrary{svg.path}
\definecolor{orcidlogocol}{HTML}{A6CE39}
\tikzset{
  orcidlogo/.pic={
    \fill[orcidlogocol] svg{M256,128c0,70.7-57.3,128-128,128C57.3,256,0,198.7,0,128C0,57.3,57.3,0,128,0C198.7,0,256,57.3,256,128z};
    \fill[white] svg{M86.3,186.2H70.9V79.1h15.4v48.4V186.2z}
                 svg{M108.9,79.1h41.6c39.6,0,57,28.3,57,53.6c0,27.5-21.5,53.6-56.8,53.6h-41.8V79.1z M124.3,172.4h24.5c34.9,0,42.9-26.5,42.9-39.7c0-21.5-13.7-39.7-43.7-39.7h-23.7V172.4z}
                 svg{M88.7,56.8c0,5.5-4.5,10.1-10.1,10.1c-5.6,0-10.1-4.6-10.1-10.1c0-5.6,4.5-10.1,10.1-10.1C84.2,46.7,88.7,51.3,88.7,56.8z};}}
\newcommand\orcid[1]{\href{https://orcid.org/#1}{\mbox{\scalerel*{
\begin{tikzpicture}[yscale=-1,transform shape]
\pic{orcidlogo};
\end{tikzpicture}
}{|}}}}
\begin{document}
\allowdisplaybreaks[4]
\tolerance=5000

\date{\today}
\title{Confront $f(R,T)=\Ri+\beta T$ modified gravity with the massive pulsar  ${\textit PSR J0740+6620}$}

\author{G.~G.~L.~Nashed~\orcid{0000-0001-5544-1119}}
\email{nashed@bue.edu.eg}
\affiliation {Centre for Theoretical Physics, The British University in Egypt, P.O. Box
43, El Sherouk City, Cairo 11837, Egypt}

\begin{abstract}
Many physically inspired general relativity (GR) modifications predict significant deviations in the properties of spacetime surrounding massive neutron stars. Among these modifications is $f(\Ri, { \mathbb{T}})$, where $\Ri$ is the Ricci scalar, $ { \mathbb{T}}$ represents the trace of the energy-momentum tensor, the gravitational theory that is thought to be a neutral extension of GR. Neutron stars with masses above 1.8 $M_\odot$ expressed as radio pulsars are precious tests of fundamental physics in extreme conditions unique in the observable universe and unavailable to terrestrial experiments. We obtained an exact analytical solution for spherically symmetric anisotropic perfect-fluid objects in equilibrium hydrostatic  using the frame of the form of $f(\Ri, { \mathbb{T}})=\Ri+\beta  { \mathbb{T}}$ where  $\beta$ is a dimensional parameter. We show that the dimensional parameter $\beta$ and the compactness, $C=\frac{ 2GM}{Rc^2}$ can be used to express all physical quantities within the star. We fix the dimensional parameter $\beta$ to be at most\footnote{ Here ${\mathrm \kappa^2}$  is  the  coupling constant of Einstein which is figured as ${\mathrm \kappa^2=\frac{8\pi G}{c^4}}$,  the Newtonian constant of gravitation is denoted as $ {\mathrm F}$ while  ${\mathrm c}$ represents the speed of light.} $\beta_1=\frac{\beta}{\kappa^2}= 0.1$  in positive values through the use of observational data from NICER and X-ray Multi-Mirror telescopes on the pulsar ${\textit PSR J0740+6620}$, which provide information on its mass and radius.  The  radius and mass of the pulsar ${\textit PSR J0740+6620}$ were determined by analyzing data obtained from NICER and X-ray Multi-Mirror telescopes. It is important to mention that no assumptions about equations of state were made in this research. Nevertheless, the model demonstrates a good fit with linear patterns involving bag constants. Generally, when the dimensional parameter $\beta$ is positive, the theory predicts that a star of the same mass will have a slightly larger size than what is predicted by GR.  It has been explained that the hydrodynamic equilibrium equation includes an additional force resulting from the coupling between matter and geometry. This force partially reduces the effect of gravitational force.  As a result, we compute the maximum compactness allowed by the strong energy condition for $f(\Ri, { \mathbb{T}})=\Ri+\beta { \mathbb{T}}$ and for GR, which are $C = 0.757$ and $0.725$, respectively. These values are approximately 3\% higher than the prediction made by GR. Moreover, we calculate the maximum mass to be approximately $4.26 M_\odot$ at a radius of around $15.9$ km for a boundary density of $\rho_{\text{nuc}} = 2.7\times 10^{14}$ g/cm$^3$.
\end{abstract}
\keywords{Massive stars, Non-standard theories of gravity, Stellar structures}

\maketitle
\section{Introduction}\label{Sec:Introduction}

Neutron stars (NSs) provide a special environment for studying matter at high densities, which are several times higher than the nuclear saturation density of $\rho_\text{nuc} \approx 2.7 \times 10^{14}$ g/cm$^3$. Such high-density matter cannot be studied on Earth. Recent observations of X-ray and gravitational wave signals  provide novel opportunities for determining the masses and radii of NSs. The primary objective of the NICER mission revolves around ascertaining the mass and radius of millisecond pulsars (MSPs), and consequently their Equation of State (EoS). This is accomplished by examining the deflection of light due to gravity and studying the patterns of X-ray emission from the rotating heated regions found on the surfaces of millisecond pulsars (MSPs) \citep{Bogdanov:2019ixe,Bogdanov:2019qjb}. NICER has been monitoring some MSPs with masses around 1.5 solar masses, including PSR J0030+0451, which has a mass of $M=1.44^{+0.15}{-0.14} M\odot$ according to \citep{Miller:2019cac}, and $M= 1.34^{+0.15}{-.16} M\odot$ according to \citep{Raaijmakers:2019qny}. Other MSPs such as ${\textit PSR J0437-4715}$, with a mass of $M=1.44 \pm 0.07 M_\odot$ according to \citep{Reardon:2015kba}, are also being examined \citep{2014HEAD...1411607B}. The radius of the isolated pulsar ${\textit PSR J0030+0451}$ has been determined using two separate analysis of data from NICER. The first analysis yielded a radius of $R= 13.02_{-1.06}^{+1.24}$ km according to \citep{Miller:2019cac}, while the second analysis produced a radius of $R= 12.71^{+1.14}_{-1.19}$ km \citep{Raaijmakers:2019qny}. Although different assumptions are made regarding the characteristics of the hot spots in these studies, both approaches yield consistent results when it comes to measuring the radius of the pulsar.

Moreover, a few MSP with mass $\sim 2 M_\odot$, for example${\textit  PSR J1614-2230}$ with mass $M= 1.908 \pm 0.016 M_\odot$ \citep{Demorest:2010bx,Fonseca:2016tux,NANOGRAV:2018hou}, ${\textit PSR J0348+0432}$ with  $M= 2.01 \pm 0.04 M_\odot$ \citep{Antoniadis:2013pzd} and ${\textit PSR J0740+6620}$ with mass $M= 2.08 \pm 0.07 M_\odot$ using the relativistic Shapiro time delay \citep{NANOGrav:2019jur,Fonseca:2021wxt}, are particularly intriguing because the measurements obtained fall within the maximum limit for a NS . The pulsar ${\textit PSR J0740+6020}$,  is in a binary structure that offers individual readings of mass. As a result, the mass can be determined with greater accuracy, though the lower NICER count rate presents a challenge when compared to ${\textit PSR J0348+0432}$.  The radius of  PSR J0740+6020 was determined using data from both the NICER and X-ray Multi-Mirror (XMM) Newton observatories. In a study by Miller et al. (2021), the radius was calculated as $R=13.7_{-1.5}^{+2.6}$ km. Another study by Riley et al. (2021) independently derived a radius of $R=12.39_{-0.98}^{+1.30}$ km at a $68\%$ confidence level. Additionally, combining data from NICER and XMM using a Gaussian process and non-parametric equation of state (EoS) method, Legred et al. (2021) obtained a mass of $M=2.07 \pm 0.11 M_\odot$ and a radius of $R=12.34^{+1.89}_{-1.67}$ km. These findings are consistent with the results presented by Landry et al. (2020) at a $68\%$ confidence level.

Numerous solutions to the field equations for spherically symmetric fluid spheres in general relativity (GR) have been well-documented in the literature \cite{Stephani:2003tm}. Nevertheless, it was identified that only nine (out of 127) fulfill the basic criteria for physical plausibility~\cite{Delgaty:1998uy}. The Buchdahl~\cite{1967ApJ...147..310B}, Nariai IV~\cite{1999GReGr..31..945N}, and the Tolman VII (T-VII)~\cite{Tolman:1939jz},  Krori-Barua \cite{1975JPhA....8..508K} solutions are frequently used. Krori and Barua's (KB) \cite{1975JPhA....8..508K} singularity space-time metric meets the physical requirements of a real star. { The main motivation for considering KB spacetime is to investigate gravitational theories or models that deviate from the standard general relativity. Krori and Barua spacetime, was proposed by the physicists Krori and Barua, represents an alternative metric that can be used to describe the geometry of spacetime in specific scenarios. By studying this spacetime, researchers aim to explore its implications and potential applications in understanding gravity, cosmology, and other related phenomena. The motivation lies in expanding our understanding of gravitational theories beyond the conventional framework and potentially uncovering new insights into the nature of spacetime and the universe.} As a consequence, the KB model has garnered considerable attention in the literature, with numerous intriguing properties highlighted in \cite{Roupas:2020mvs, Nashed:2023pxd, Rahaman:2011cw, MonowarHossein:2012ec, Kalam:2012sh, Rahaman:2010mr}.

To address these unanswered concerns, a variety of modifications of the  gravitational lagrangian of Einstein-Hilbert (EH)  have been studied~\cite{Nojiri:2017ncd}. Notably, Einstein made his attempt to change the Lagrangian of EH when he proposed the cosmological constant\footnote{The impact of the repulsive cosmological constant within the cosmological framework was summarized in~\cite{Carroll:2000fy}, specifically, there have been particular considerations about accretion events in compact objects and dark matter halos in the work of \cite{Stuchlik:2020rls, Boehmer:2004nu, Stuchlik:2016xiq, Novotny:2021zlq}, and the motion of relative  galaxies in~\cite{Stuchlik:2011zz}.} to enable his field equations to have a static universe solution~\cite{Einstein:1917ce,ORaifeartaigh:2017uct}. One can consider a later modification of the gravity sector by treating the EH  term as an elevated linear function of the Ricci scalar $\Ri$. One of the simplest and most effective extensions of  GR is achieved by the replacement of the EH expression  with   arbitrary   function of $\Ri$ in  $f(\Ri)$ theory ~\cite{Buchdahl:1970ynr, Sotiriou:2008rp}. This theory allows for the inclusion of the accelerated phases of the universe~\cite{Copeland:2006wr, DeFelice:2010aj, Nojiri:2010wj}. Additionally, compact object solutions have been found within this theory, as seen in \citep[see][]{DeFelice:2010aj} and the references therein.

The so-called $f(\Ri, { \mathbb{T}})$ gravity~\cite{Harko:2011kv}, which includes the trace of the energy-momentum tensor (EMT) $ { \mathbb{T}}\equiv g^{\mu\nu} { \mathbb{T}}_{\mu\nu}$ in the construction of the modified gravity sector, is another generalization of $f(\Ri)$ gravity that has gained significant interest in recent years.  The generalized curvature-matter coupling theories, which include the $f(\Ri, { \mathbb{T}})$ theory of gravity, produce intriguing phenomenology when used in cosmological contexts~\cite{Harko:2014gwa,Goncalves:2021vci,Goncalves:2022ggq}. Yet, in $f(\Ri, { \mathbb{T}})$ gravity, solutions to relativistic compact stars have also been studied in~\cite{Hansraj:2018jzb,Bhar:2021uqr,Kumar:2021vqa,Feng:2022bvk}. { In the frame of $f(\Ri, { \mathbb{T}})$  an  anticipation of the presence of quark stars   composed of quark matter in the color-flavor-locked  phase of color superconductivity has been studied in \cite{Tangphati:2022arm}. Moreover, the properties of quark stars  have been investigated in the context of $f(\Ri, { \mathbb{T}})$ \cite{Tangphati:2022mur}.}

 Throughout this investigation, our focus is directed towards the KB spacetime~\cite{1975JPhA....8..508K}, a model proven to be pertinent in exploring the interior structure of neutron stars (NSs). We derive an analytical extension within the linear approximation of $f(\Ri, { \mathbb{T}})=\Ri+\beta { \mathbb{T}}$ gravity for this model and examine the influence of the dimensional parameter $\beta$ on the internal structure of the stellar object. Furthermore, we employ numerical methods to solve the modified Tolman-Oppenheimer-Volkoff (TOV) system of equations with a uniform density profile, emphasizing the impact of the extra force arising from this theory.

{ Anisotropic spacetime is relevant in alternative theories of gravity beyond general relativity. Various modified gravity theories propose modifications to the Einstein field equations, allowing for anisotropic contributions to the gravitational field. By studying anisotropic spacetime, researchers can test and refine these alternative theories, exploring their implications and potential observational signatures.}

The arrangement of this article is as follows: In Sec.~\ref{Sec:Overview_of_f(R,T)} we give a summary of the construction of $f(\Ri, { \mathbb{T}})$ gravitational theory, where  we present the generalized form of the field equations. As a result, in Sec.\ref{int}, we adopt KB's approach \citep{1975JPhA....8..508K} to derive an analytic solution for the field equations within the framework of linear $f(\Ri, { \mathbb{T}})$ gravitational theory. Additionally, in Sec. \ref{Sec:Stability}, we utilize the observations of the PSR ${\textit PSR J0740+6620}$ made by the X-ray telescopes NICER and XMM to determine the values of the parameters used in the model.Furthermore, we examine the intrinsic characteristics of the pulsar PSR ${\textit PSR J0740+6620}$ and evaluate its stability by considering the limitations imposed by the existing model. In Sec. \ref{S8}, we compare our model to additional data from pulsars, which is presented in Tables \ref{Table1} and \ref{Table2}.  In Sec. \ref{MR}, we calculate the highest permissible compactness based on physical constraints and subsequently generate curves of mass versus radius  for various surface densities. For each category, the maximum attainable mass for a stable structure serves as its representation. In Sec.~\ref{Sec:Conclusion}, we present the conclusions of this study.

\vspace{0.15cm}\noindent\textbf{Nomenclature and notation:} We adopt the metric signature $(-,+,+,+)$ throughout this work. Unless explicitly mentioned, primes denote derivatives w.r.t, the radial coordinate.


\section{Brief of  $f(\Ri, { \mathbb{T}})$ gravitational theory}\label{Sec:Overview_of_f(R,T)}

In the following section, we will present a brief summary of $f(\Ri, { \mathbb{T}})$ gravity within the framework of metric construction, using the Levi-Civita connection as the basis, as outlined in  ~\cite{Harko:2011kv}.  It's important to note that in the frame of Palatini construction, the connection is considered  as independent variable through the process of  the variation of the Lagrangian \citep[see ][]{Wu:2018idg}. The action for $f(\Ri, { \mathbb{T}})$ theory takes the following form:
\beq
S=\int{\dd^4x\sqrt{-g}}\left[ \frac{f(\Ri, { \mathbb{T}})}{\kappa^2}+\mathcal{L}_m \right]\,,
\label{f(R,T) action}
\eeq
where ${\mathrm \kappa^2}$ represents the Einstein coupling constant, determined as ${\mathrm \kappa^2=\frac{8\pi G}{c^4}}$, with ${\mathrm G}$ denoting the Newtonian gravitational constant and ${\mathrm c}$ representing the speed of light.
If we assume $\mathcal{L}_m$ to depend on the metric, then energy-momentum tensor of matter is defined by:
\beq
 { \mathbb{T}_{\mu\nu}}=\frac{-2}{\sqrt{-g}}\frac{\del \left(\sqrt{-g} \mathcal{L}_m \right)}{\del g^{\mu\nu}}\,.
\eeq
The EMT of an anisotropic fluid takes the following form:
\begin{equation}\label{Tmn-anisotropy}
    { \mathbb{T}{^\mu}{_\nu}=  (p_{t}+\rho c^2)w{^\mu} w{_\nu}+p_{t} \delta ^\mu _\nu + (p_{r}-p_{t}) v{^\mu} v{_\nu}}\,.
\end{equation}
In this particular scenario, the symbol ${\mathrm  \rho=\rho(r)}$ denotes the  density of the fluid. The term  $\mathrm {{ p}_{r}={ p}_{r}(r)}$ refers to the pressure exerted in the radial direction (i.e., in alignment with the time-like four-velocity ${\mathrm w^\alpha}$'s trajectory), while $\mathrm{{ p}{t}={ p}{t}(r)}$ denotes the tangential pressure exerted orthogonal to ${\mathrm w^\alpha}$. Moreover, ${\mathrm {v}^\alpha}$ represents the normalized spatial vector aligned radially.  Consequently, the energy-momentum tensor can be written in a diagonal form as ${ \mathrm {T}{^\alpha}_{\beta}=diag(-\rho c^2,\,p_{r},\,p_{t},\,p_{t})}$. By varying the metric with respect to Eq.~\eqref{f(R,T) action}, the resulting equations correspond to the modified Einstein Eqs.~\cite{Harko:2011kv}:
\beq
f_\Ri \Ri_{\mu\nu}-\frac{1}{2}f\, g_{\mu\nu}+S_{\mu\nu}f_\Ri=\kappa^2 T_{\mu\nu}-f_T\left(T_{\mu\nu}+ \Theta_{\mu\nu} \right)\equiv { \mathbb{T}}_{\mu\nu}\,,
\label{f(R,T) EOM}
\eeq
where $f_\Ri \equiv \frac{\partial f(\Ri, { \mathbb{T}})}{\partial \Ri}$, $f_ { \mathbb{T}} \equiv \frac{\partial f(\Ri, { \mathbb{T}})}{\partial { \mathbb{T}}}$, $S_{\mu\nu}$ stands for  $S_{\mu\nu} \equiv \left(g_{\mu\nu} \Box-\nabla_{\mu}\nabla_{\nu} \right)$ and
\beq
\Theta_{\mu\nu} \equiv g^{\alp\bet}  \frac{\del{ \mathbb{T}}_{\alp\bet}}{\del g^{\mu\nu}}=-2 { \mathbb{T}}_{\mu\nu}+g_{\mu\nu} \mathcal{L}_m-2g^{\alp\bet}\frac{\partial^2 \mathcal{L}_m}{\partial g^{\mu\nu} \partial g^{\alp \bet}}\,.
\eeq
Following the approach given by Harko \emph{et al.}~\cite{Harko:2011kv}, we assume $\mathcal{L}m=p$  (see~\cite{Harko:2011kv}, $\mathcal{L}m=-p$). Following such assumption, we get $\Theta{\mu\nu}=-2 { \mathbb{T}}{\mu\nu}+pg{\mu\nu}$. The trace of Eq.~\eqref{f(R,T) EOM} yields:
\beq
f_\Ri \Ri-2 f+3 \Box f_\Ri=\kappa^2  { \mathbb{T}}\,.
\label{eq:trace_eom}
\eeq
In $f(\Ri,{ \mathbb{T}})$ theory, Equation (\ref{eq:trace_eom}) becomes nonlinear in $\Ri$, resulting in a second-order differential equation, which is not algebraic as in the case of pure General Relativity (GR). Consequently, nonlinear forms of $f(\Ri,{ \mathbb{T}})$ lead to a non-vanishing of $\Ri$. This is because of the non-vanishing value  of the trace of EMT in the action of $f(\Ri, { \mathbb{T}})$  gravity. As a result, the -divergence of  EMT is given by~\cite{BarrientosO:2014mys}
\beq
\nabla^{\mu}{ \mathbb{T}}_{\mu\nu}=\frac{f_{ \mathbb{T}}}{ \kappa^2-f_{ \mathbb{T}}} \left[ \left({ \mathbb{T}}_{\mu\nu}+ \Theta_{\mu\nu} \right) \nabla^{\mu} \ln{f_{ \mathbb{T}}} + \nabla^{\mu}\Theta_{\mu\nu}-\frac{1}{2}\nabla_{\nu}{ \mathbb{T}} \right]\,.
\label{Tmn divergence}
\eeq
The non-conservation of the energy-momentum tensor (EMT) of matter in $f(\Ri,{ \mathbb{T}})$ theory leads to an additional force. This extra force could have effects for gravitational  beyond those predicted by  GR and could be significant at galactic scales and in the solar system~\cite{Harko:2011kv}.

The conservation of the EMT is verified when $f(\Ri, { \mathbb{T}})=\Ri$. In such cases, the form of $f(\Ri,{ \mathbb{T}})$ can be constructed based on this condition~\citep[see also][]{Pretel:2021kgl}. However, the corresponding field equations can only be solved in a numerical way. Since our main focus in this study is on obtaining analytical solutions, we will not delve into discussing these specific models of $f(\Ri,{ \mathbb{T}})$ gravitational theory.

It is appropriate to recast Eq.~\eqref{f(R,T) EOM} into an equivalent  form of  GR, where the left-hand side contains the Einstein tensor $G_{\mu \nu}$, and the right-hand side combines the effective  EMT  taking into account contributions from both matter fields and curvature terms, given by:
\beq
G_{\mu\nu}=\frac{1}{f_\Ri}\left[{ \mathbb{T}}_{\mu\nu} +\frac{f-\Ri\,f_\Ri}{2}g_{\mu\nu}-S_{\mu\nu}f_\Ri \right] \equiv { \mathbb{T}}^{(\text{eff})}_{\mu\nu}\,.
\label{Gmn_Teff}
\eeq
In general, for an arbitrary $f(\Ri,{ \mathbb{T}})$ theory, the components of ${ \mathbb{T}}^{(\text{eff})}_{\mu\nu}$ will be quite complex, making it exceedingly challenging to derive an exact analytical solution in this theory. It should be noted that for the special model
\beq
f(\Ri,T)=\Ri+h(T)\,,
\label{f(R,T) separable}
\eeq
that is the specific case where the $f(\Ri,{ \mathbb{T}})$ theory is linear in the Ricci scalar and does not involve mixing terms between $\Ri$ and ${ \mathbb{T}}$, the contributions of curvature fluids in the effective energy-momentum tensor (EMT) simplify considerably. As a result, the formulation of the field equations becomes more manageable and takes the form:
\beq
G_{\mu\nu} = \kappa^2 { \mathbb{T}}_{\mu\nu}+\frac{h}{2}g_{\mu\nu}+h_{ \mathbb{T}} \left({ \mathbb{T}}_{\mu\nu}-p\,g_{\mu\nu} \right)\,.
\label{GR+T effects EOM}
\eeq
In this study and  in order to explain the minimal extension of Einstein GR in the framework of $f(\Ri,{ \mathbb{T}})$ gravity, and  to avoid overly complicated field equations that could be prevent the search of an exact analytic solution, we limit our discussion to the case of linear $f(\Ri,{ \mathbb{T}})$ gravity in which $f(\Ri,{ \mathbb{T}})$ takes the form:
\beq
f(\Ri,{ \mathbb{T}})=\Ri+\beta { \mathbb{T}}\,.
\label{eq:linear_f(R,T)}
\eeq
In this linear case, the field equations take a more straightforward form, with $\beta$ representing a dimensional constant parameter of the theory. For this specific linear scenario, an analytic solution can be easily derived for the field equations (\citep[see also][]{Hansraj:2018jzb, Bhar:2021uqr, Pretel:2020oae, Pretel:2021kgl}).

However, it is important to note that for the more complicated non-linear forms of $f(\Ri,{ \mathbb{T}})$ gravity, the field equations generally become more complex. One way to understand the complexity is by exploring the dynamical equivalence of scalar-tensor representations of the theory and employing the Palatini construction~\citep[see also][]{Rosa:2021teg, Rosa:2022cen}.
\section{Interior spherically symmetric solution}\label{int}
In this section we are going to derive a spherically symmetric interior solution in the frame of $f(\Ri,{ \mathbb{T}})=\Ri+\beta\,{ \mathbb{T}}$. For this aim we use the  general  spherically symmetric geometry and static which is by the following line element \cite{Nashed:2004pn,Nashed:2006yw,Nashed:2020kjh}:
\beq
ds^2=-e^{\alpha(r)}dt^2+e^{\alpha_1(r)}dr^2+r^2d\theta^2+r^2\sin^2\theta d\phi^2\,,
\label{line element}
\eeq
where $\alpha(r)$ and $\alpha_1(r)$ are functions of the radial coordinate.

As Eq.\eqref{eq:linear_f(R,T)} represents a particular form of the theories described by Eq.\eqref{f(R,T) separable}, when we substitute Eqs.\eqref{Tmn-anisotropy} and \eqref{line element} into Eq.~\eqref{GR+T effects EOM}, we obtain the following system of differential equations:
\begin{eqnarray}\label{sys}
\mathrm {\kappa^2 \rho c^2}&=&\mathrm{\frac{1}{1+3\,\beta_1
 }\left[\frac {\alpha'_1 r+{
e^{\alpha_1}-1}}{{r}^{2}{e^{\alpha_1}}}+\frac{5\kappa \beta_1}{3} \left(p_r  +2\,p_t  \right)\right]}\,,\nonumber\\
\mathrm{\kappa^2 p_r}&=&\mathrm{\frac{3}{3+11\,\beta_1
 }\left[\frac {\alpha'r -{
e^{\alpha_1}+1}}{{r}^{2}{e^{\alpha_1}}}+\frac{\kappa \beta_1}{3} \left(3\rho c^2  -10p_t  \right)\right]}\,,\nonumber\\
\mathrm{\kappa^2 p_t}&=&\mathrm{\frac{3}{3+16\,\beta_1
 }\left[\frac {2\alpha'' r+\alpha'[\alpha'-\alpha'_1]+2}{4{r}{e^{\alpha_1}}}+\kappa \beta_1\left(\rho c^2-\frac{5}{3} p_r  \right)\right]} \,.
\label{eq:Feqs}
\end{eqnarray}
Here, the prime symbol ($'$) indicates differentiation with respect to the radial coordinate, i.e., $r$ and $\beta_1=\frac{\beta}{\kappa^2}$.  By utilizing the aforementioned equations, we can derive the anisotropy parameter, denoted as $\mathrm{\Delta(r)}$, which is defined as the difference between the tangential pressure $\mathrm{p_t}$ and the radial pressure $\mathrm{p_r}$ as:
\begin{equation}\label{eq:Delta1v}
\mathrm{\Delta(r) =\frac {2\alpha'' {r}^{2}+\alpha'^2{r}^{2}- \left[{\alpha'_1}{r}-2 \right] \,r\alpha' +4 e^{\alpha_1}-2{\alpha_1}r-4}{4\kappa\left(2\beta_1+1 \right){r}^{2}e^{\alpha_1}}}.
\end{equation}
Surprisingly, the connection between matter and geometry that results from  trace component  does not effect the anisotropy as described by Eq. \eqref{eq:Delta1v}, unless we suppose that the spacetime configuration is spherically symmetric \citep{Nashed:2022zyi}. Nonetheless, if $\beta_1 \neq 0$, we can anticipate a minor alteration in anisotropy. Hence, different anisotropic effects are unable to conceal discrepancies from GR that are produced by the connection between matter and geometry.  If the dimensionless parameter $\beta_1$ is equal to zero, the differential equations \eqref{eq:Feqs} will be the same as the field equations of  GR theory for a spherically symmetric interior spacetime \citep[c.f.,][]{Roupas:2020mvs}.

 The system of differential equations (\ref{eq:Feqs}) has three independent non-linear differential equations and  five unknowns $\mathrm{\alpha}$, $\mathrm{\alpha_1}$, $\mathrm{\rho}$, $\mathrm {p_r}$ and $\mathrm{p_t}$. Therefore, we required two extra conditions  to limit the  system. Therefore, we will  introduce the Krori-Barua (KB)  ansatz as \cite{1975JPhA....8..508K}:
\begin{equation}\label{eq:KB}
    \alpha(r)=\frac{\mu\, r^2}{R^2}+\nu \equiv \mu \,x^2+\nu,\,  \qquad \qquad \qquad {\alpha_1}(r)=\frac{\lambda\, r^2}{R^2}\equiv \lambda\, x^2,
\end{equation}
where the dimensionless  radius $x$ is defined as \[0 \leq \frac{r}{R}=x \leq 1\,.\] In this context, $R$ corresponds to the stellar radius. Furthermore, we establish the values of the dimensionless coefficients {$\mu, \nu, \lambda$} by enforcing matching criteria at the stellar's surface. Moreover, we present dimensionless quantities to simplify the analysis.
\begin{equation}
    \widetilde{\rho}(r)=\frac{\rho(r)}{\rho_{\ast}}\,, \qquad \widetilde{p}_r(r)=\frac{p_r(r)}{\rho_{\ast} c^2}\,, \qquad \widetilde{p}_t(r)=\frac{p_t(r)}{\rho_{\ast} c^2}\,, \qquad \widetilde{\Delta}(r)=\frac{\Delta(r)}{\rho_{\ast} c^2}\,,
\end{equation}
where $\rho_{\ast}$ is the characteristic density defined as:
\begin{equation}
    \rho_{\ast}=\frac{1}{\kappa^2 c^2 R^2(1+10\beta_1+16\beta_1{}^2)}\,.
\end{equation}
Therefore, the equation of motions \eqref{eq:Feqs} can be expressed as follows:
\begin{eqnarray}
\widetilde{\rho}&=& \frac{e^{-\lambda x^2}}{x^2}(e^{\lambda x^2}-1+2\lambda x^2) \nonumber\\
&+&\frac{2\beta_1}{x^2}\left[\left(5\mu(\mu-\lambda)x^4 +(16\lambda+15\mu)x^2 -8\right)e^{-\lambda x^2} +8\right],\nonumber\\[8pt]
\widetilde{p}_r&=&\frac{e^{-\lambda x^2}}{x^2}(1-e^{\lambda x^2}+2\mu x^2) \nonumber\\
&-&\frac{2\beta_1}{x^2}\left[\left(5\mu (\mu-\lambda) x^4 - (9 \lambda-8 \mu) x^2 -8\right)e^{-\lambda x^2} +8\right],\nonumber\\[8pt]
\widetilde{p}_t&=& e^{-\lambda x^2}(2 a_0-\lambda +\mu (\mu - \lambda) x^2) \nonumber\\
&-&\frac{2\beta_1}{x^2}\left[\left(7\mu (\mu-\lambda) x^4 - (9 \lambda-4 \mu) x^2 +8\right)e^{-\lambda x^2} - 8\right],\nonumber \\
\label{eq:Feqs2}
\end{eqnarray}
and anisotropy parameter \eqref{eq:Delta1v} takes the following form:
\begin{equation}\label{eq:Delta2}
   \widetilde{\Delta}=\frac{e^{-\lambda x^2}}{x^2}(1+8\beta_1)\left[e^{\lambda x^2}-1+\mu(\mu-\lambda)x^4 -\lambda x^2\right].
\end{equation}
The parameter characterizing the cumulative mass contained inside a given radius $r$ is referred to as the  function of mass, and its definition is as follows:
\begin{equation}\label{11}
    \mathbb{M}(r)=4 \pi \int_0^r \rho(\widetilde{r}) \, \widetilde{r}{^2} d\widetilde r.
\end{equation}
Substitute the density given by Eq.~\eqref{eq:Feqs2} in Eq.~ \eqref{11} we get:
\begin{equation}\label{eq:Mass}
    \mathbb{M}(x)=\frac{M}{C(1+10\beta_1+16\beta_1{}^2)}e^{-\lambda x^2}\left[x(e^{\lambda x^2}-1)+\beta_1 \, \eta(x)\right].
\end{equation}
Here, $C$ denotes the compactness and   $\eta(x)$ is defined as:
\begin{eqnarray}\label{COMP}
C&=&\frac{2GM}{c^2 R}, \nonumber\\
\eta(x)&=& \frac{\lambda^{-\frac{9}{2}}}{3}\left\{\left(16 x \lambda^\frac{9}{2}+\frac{15}{4} \mu \sqrt{\pi} \lambda^2 (\lambda+\mu) \text{erf}(\sqrt{\lambda} x) \right) e^{\lambda x^2}\right.\nonumber\\
&-&\left. x \left[\left(\frac{15}{2} \mu \lambda^\frac{5}{2}+5\lambda^\frac{7}{2} (\mu x^2 +\frac{3}{2})\right)\mu\lambda^\frac{5}{2} -(5\mu x^2-16) \lambda^\frac{9}{2}\right]\right\}\,,\nonumber\\
\end{eqnarray}
where $erf$ is the error function defined as
\[erf(y)=\frac{2}{\sqrt{\pi}}\int_{t=0}^{t=y} e^{-t^2} dt\,.\]
Equations \eqref{eq:Mass} reduces to the model of GR  when $\beta_1=0$ \citep{Roupas:2020mvs}.
\subsection{Matching conditions}\label{Sec:Match}

The fact that both Einstein's General Relativity and the modified theory $f(\Ri,{ \mathbb{T}})=\Ri+\beta_1 { \mathbb{T}}$ share identical vacuum solutions is widely recognized. In the case of the exterior solution, it is described by Schwarzschild's solution, which can be expressed as follows:
\begin{equation}
    ds^2=-\left(1-\frac{2GM}{c^2r}\right) c^2 dt^2+\frac{dr^2}{\left(1-\frac{2GM}{c^2 r}\right)}+r^2 d\theta^2+r^2\,\sin^2 \theta d\phi^2.
\end{equation}
Remembering the interior spacetime given by Eq.~\eqref{eq:Feqs2}, therefore, we assume
\begin{equation}\label{eq:bo}
    \alpha(x=1)=\ln(1-{C}),\, \alpha_1(x=1)=-\ln(1-{C}).
\end{equation}
Furthermore, we assume that  $p_r$  becomes zero at the surface of the star,i.e.,
\begin{equation}
    \bar{p}_r(x=1)=0.
\end{equation}
Using the KB assumption \eqref{eq:KB}, the radial pressure \eqref{eq:Feqs2}, and the boundary conditions mentioned above, we obtain:
\begin{eqnarray}
\mu&=&\frac {  9-5\ln \mathfrak{C}}{ 10}+\frac{1}{10\mathfrak{C}^{1/2}{\beta_1}} \left( 5\left\{2{\beta_1}^{2} \mathfrak{C}  \left( \ln  \mathfrak{C} \right)-10\,\beta_1\, \mathfrak{C} \left( {\beta_1}+{\frac {3}{25}}\ln  \mathfrak{C}  \right)-{\frac {81}{25}} \left(3C-1 \right) {\beta_1}^{2}-{\frac {6}{25}}\left(  16C-9 \right) \beta_1+{ \frac {9}{25}}\mathfrak{C} \right\}^{1/2}\right.\nonumber \\
 &&\left.-3\mathfrak{C}^{1/2} \right)  \,, \nonumber \\
\nu&=&-\frac{1}{10{ \sqrt{\mathfrak{C}} \beta_1}}\left\{\beta_1\sqrt{\mathfrak{C}}[9-15\,\beta_1\,\ln  \mathfrak{C}]+\left[\mathfrak{C}[9-30\,\beta_1\,\ln  \mathfrak{C} +50{\beta_1}^{2}\ln  \mathfrak{C} ]-84 \,\beta_1\,C-250\mathfrak{C}\,{\beta_1}^{2}\ln  \mathfrak{C}-241\,{\beta_1}^{2}C+54\,\beta_1\right.\right.\nonumber\\
&&\left.\left.+81\,{\beta_1}^{2} \right\}^{1/2}+3\mathfrak{C}^{1/2}\right]\,, \qquad
\lambda=-\ln\mathfrak{C}\,, \qquad\mbox{where} \qquad\mathfrak{C}= (1-C)\,.\label{eq:const}
\end{eqnarray}
As the parameter $\beta_1$ approaches zero, the set of constants \eqref{eq:const} becomes identical to the corresponding values in GR \citep{Roupas:2020mvs}
\begin{equation}
     \mu=\frac{C}{2\mathfrak{C}}\,, \qquad \nu=\ln\mathfrak{C}-\frac{C}{2\mathfrak{C}}\,, \qquad \lambda=-\ln\mathfrak{C}\,.
\end{equation}
It is fascinating that all the measurable properties of the KB spacetime in the star, where $0\leq x\leq 1$, can be rephrased  using the dimensionless equations that involve $\beta_1$ and the compactness parameters ${C}$. These non-dimensional expressions include $\widetilde{\rho}(\beta_1,C)$, $\widetilde{p}_r(\beta_1,C)$, and $\widetilde{p}_t(\beta_1,C)$. Generally, the  data of observation fixes a particular numerical value for the compactness parameter, which indirectly limits the non-minimal coupling proposed by $\beta_1$. Furthermore, this presents a chance to set maximum limit on the allowable value of $C$  for  NS.  Consequently, the upper numerical value  mass can achieved.  A detailed examination of this subject will be presented in Section \ref{MR}.

\subsection{Radial derivatives}
In this subsection, we compute the radial changes in the density and pressures of the fluid, denoted by:
\begin{align}\label{eq:dens_grad}
&\widetilde{\rho}'=\frac{2e^{-a_2 x^2}}{x^3}\left\{1- e^{\lambda x^2}-2 \lambda^2 x^4+ \lambda x^2-\frac{2\beta_1}{3}\left[8e^{\lambda x^2}-8-5\mu a_2(\lambda-\mu)x^6+[5(4\lambda-\mu)\mu-16a_2^2] x^4-8\lambda x^2\right]\right\}\,,\nonumber\\
\end{align}
\begin{align}\label{eq:pr_grad}
  & \widetilde{p}'_r=\frac{-2e^{-\lambda x^2}}{x^3}\left\{1- e^{\lambda x^2}+2 \mu \lambda x^4+ \lambda x^2-\frac{2\beta_1}{3}\left[8-e^{\lambda x^2}-5\mu \lambda(\lambda-\mu)x^6-[4(2\lambda-\mu)\mu-5\mu^2] x^4-8\lambda x^2\right]\right\}\,,
\end{align}
\begin{align}\label{eq:pt_grad}
&\widetilde{p}'_t=\frac{2e^{-\lambda x^2}}{x^3}\left\{\mu \lambda(\lambda-\mu) x^6+ [\lambda(\lambda-3 \mu)+\mu^2] x^4+\frac{2\beta_1}{3}\left[4-4e^{\lambda x^2}+7\mu \lambda(\lambda-\mu)x^6-[(16\lambda-7\mu)\mu-4\mu^2] x^4\right.\right.\nonumber\\
&\left.\left.+4\lambda x^2\right]\right\}\,,\nonumber\\
\end{align}
where $'\equiv \frac{d}{dx}$. Equations~\eqref{eq:dens_grad}, \eqref{eq:pr_grad} and \eqref{eq:pt_grad} describe the gradients of the components of the EMT. These equations play a crucial role in demonstrating the physical stability of the compact object.
\section{Observational limits and the equilibrium  conditions}\label{Sec:Stability}
In the present analysis, we apply the pulsar ${\textit PSR J0740+6620}$, the radius and mass determined from NICER+XMM based on a non-parametric equation of state method. The stellar system under consideration has a radius of $R=12.34^{+1.89}{-1.67}$ km and a mass of $M=2.07 \pm 0.11 M\odot$ according to a study by Legred \cite{Legred:2021hdx}, with the solar mass denoted as $M_\odot=1.9891\times 10^{30}$ kg. To be considered physically legitimate, a star model must comply with a group of requirements called (\textbf{I}) to (\textbf{XIII}):
\subsection{Matter sector}\label{Sec:matt}
\begin{figure*}
\centering
\subfigure[~The energy-density]{\label{fig:density}\includegraphics[scale=0.25]{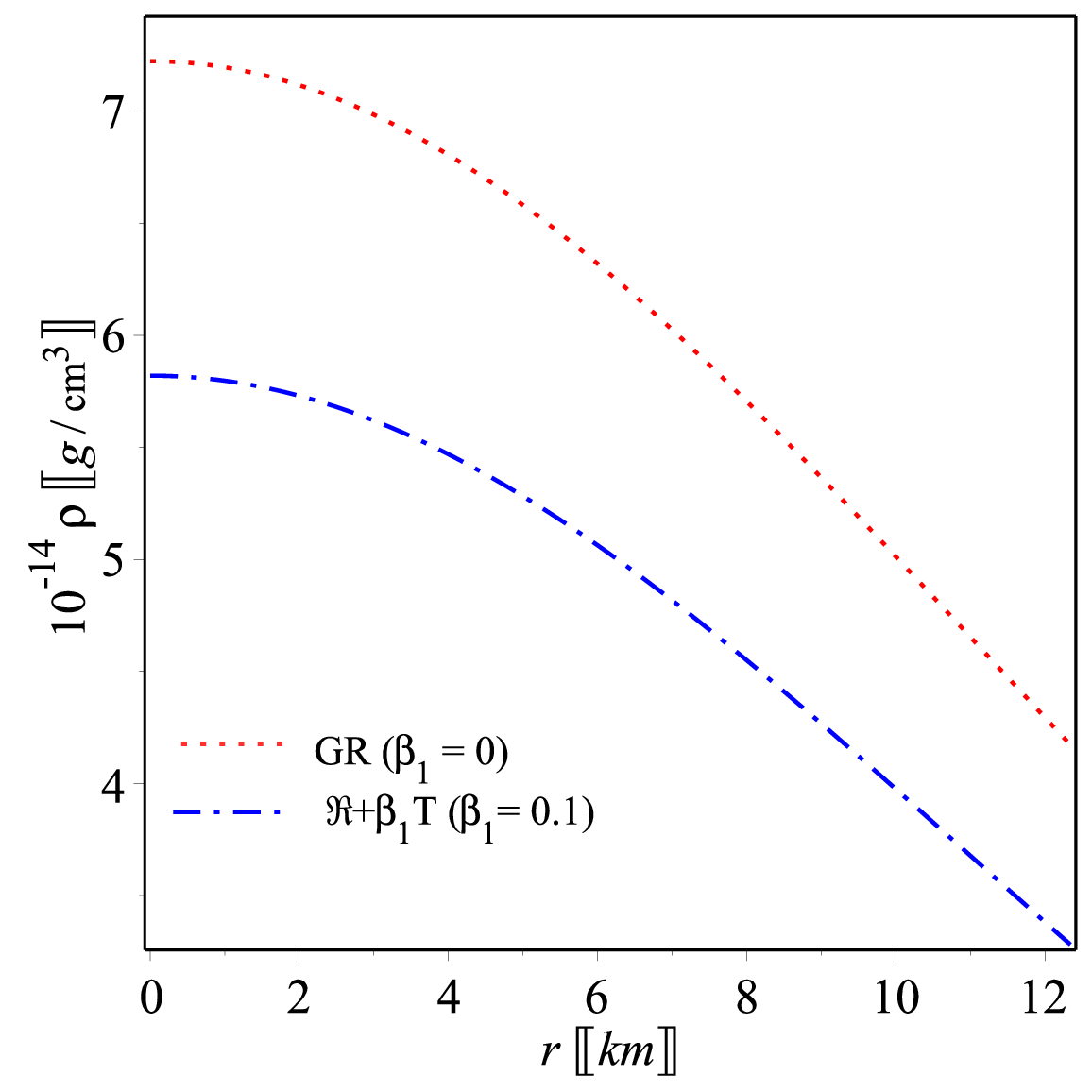}}
\subfigure[~The radial pressure]{\label{fig:radpressure}\includegraphics[scale=0.25]{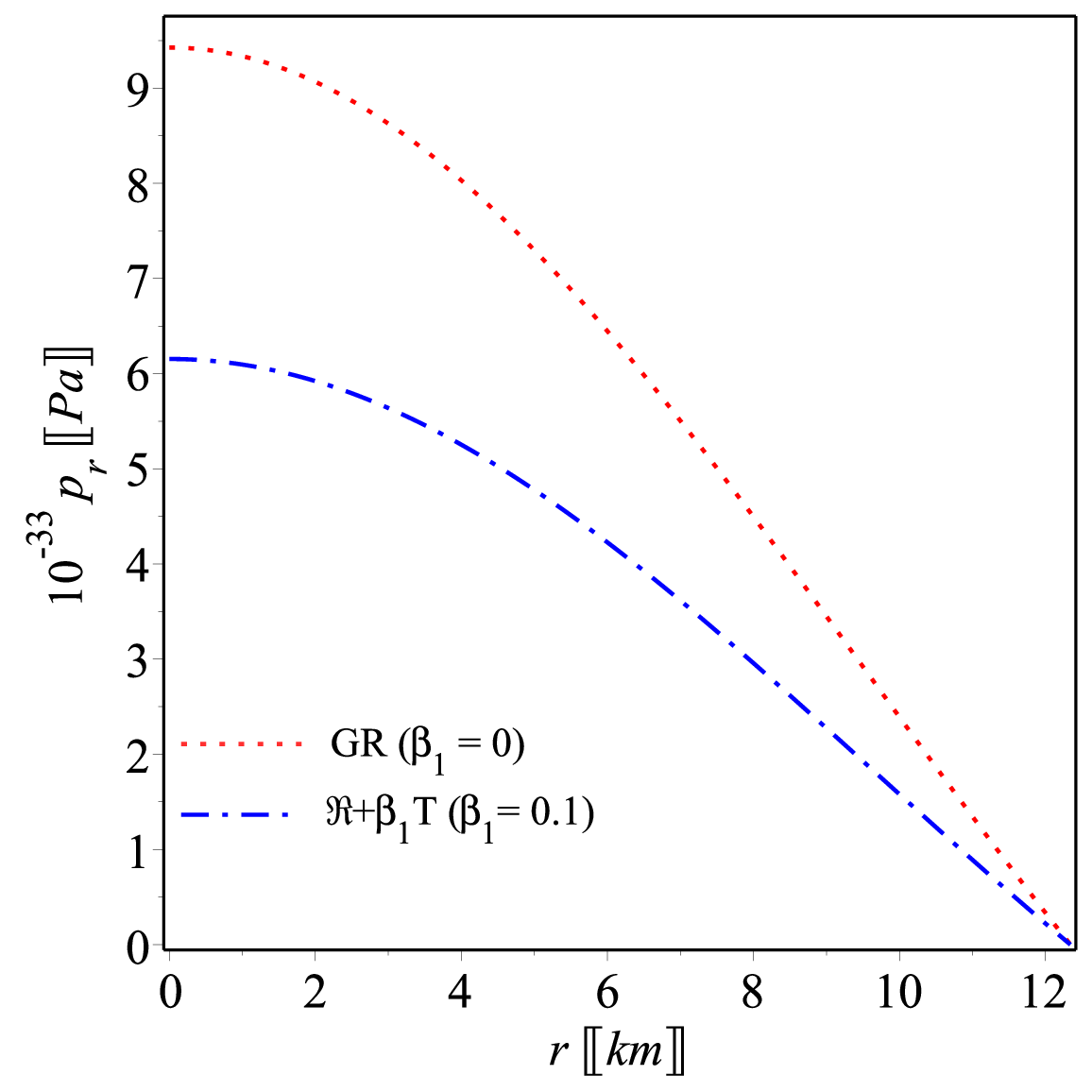}}
\subfigure[~The tangential pressure]{\label{fig:tangpressure}\includegraphics[scale=0.25]{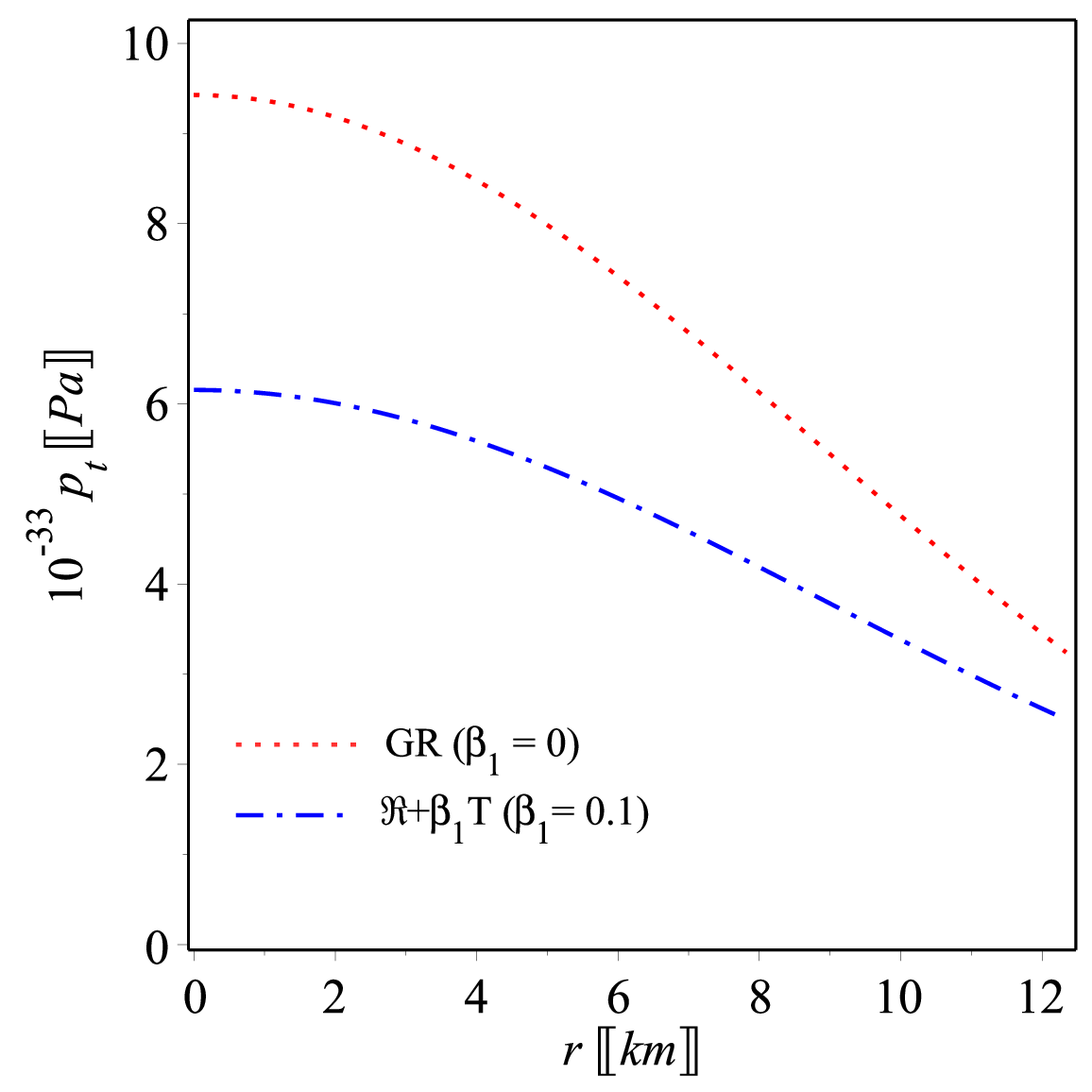}}\\
\subfigure[~The gradients (GR)]{\label{fig:GRgrad}\includegraphics[scale=0.25]{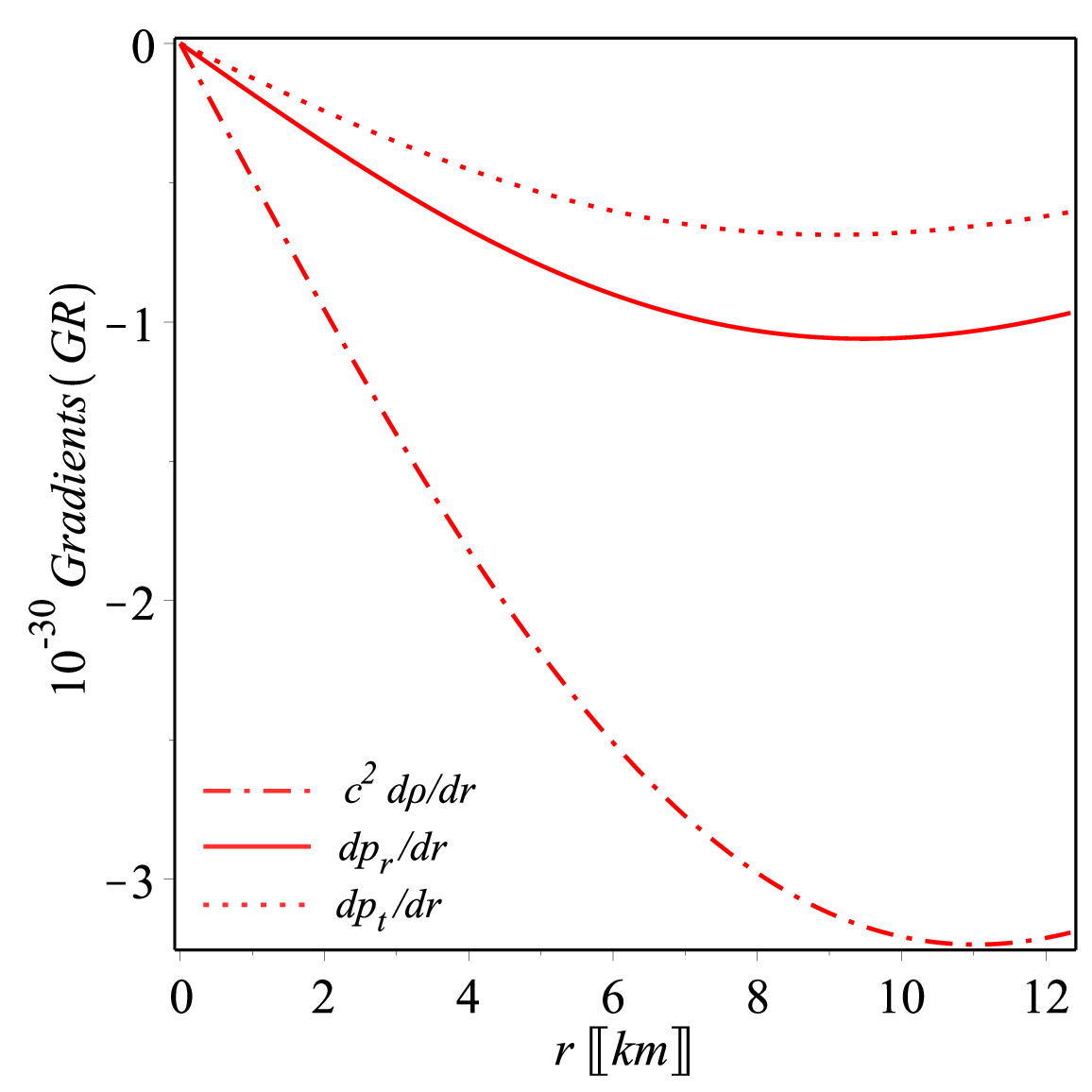}}
\subfigure[~The gradients f(R,T)]{\label{fig:RTgrad}\includegraphics[scale=0.25]{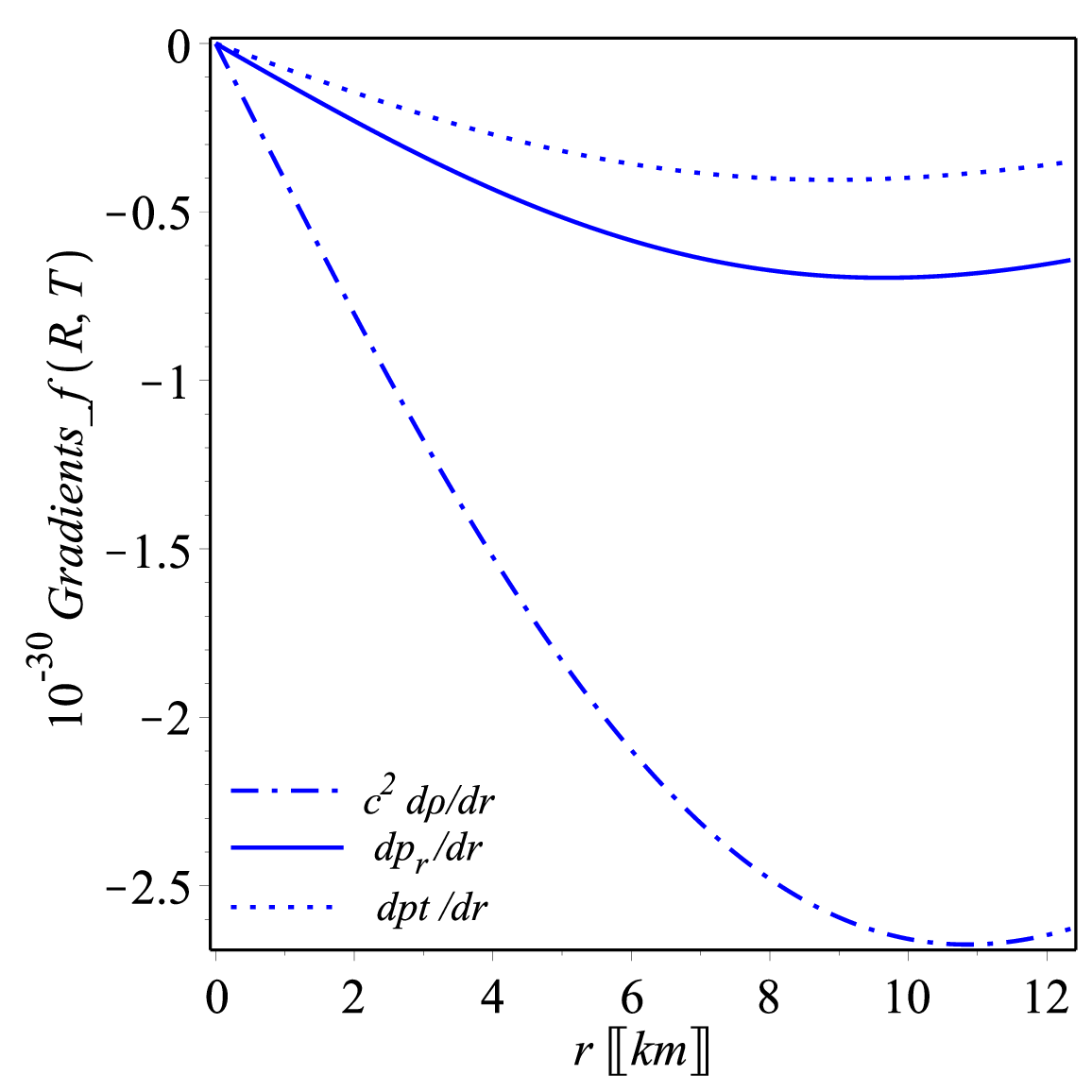}}
\subfigure[~The anisotropy (GR)]{\label{fig:anisotg}\includegraphics[scale=0.25]{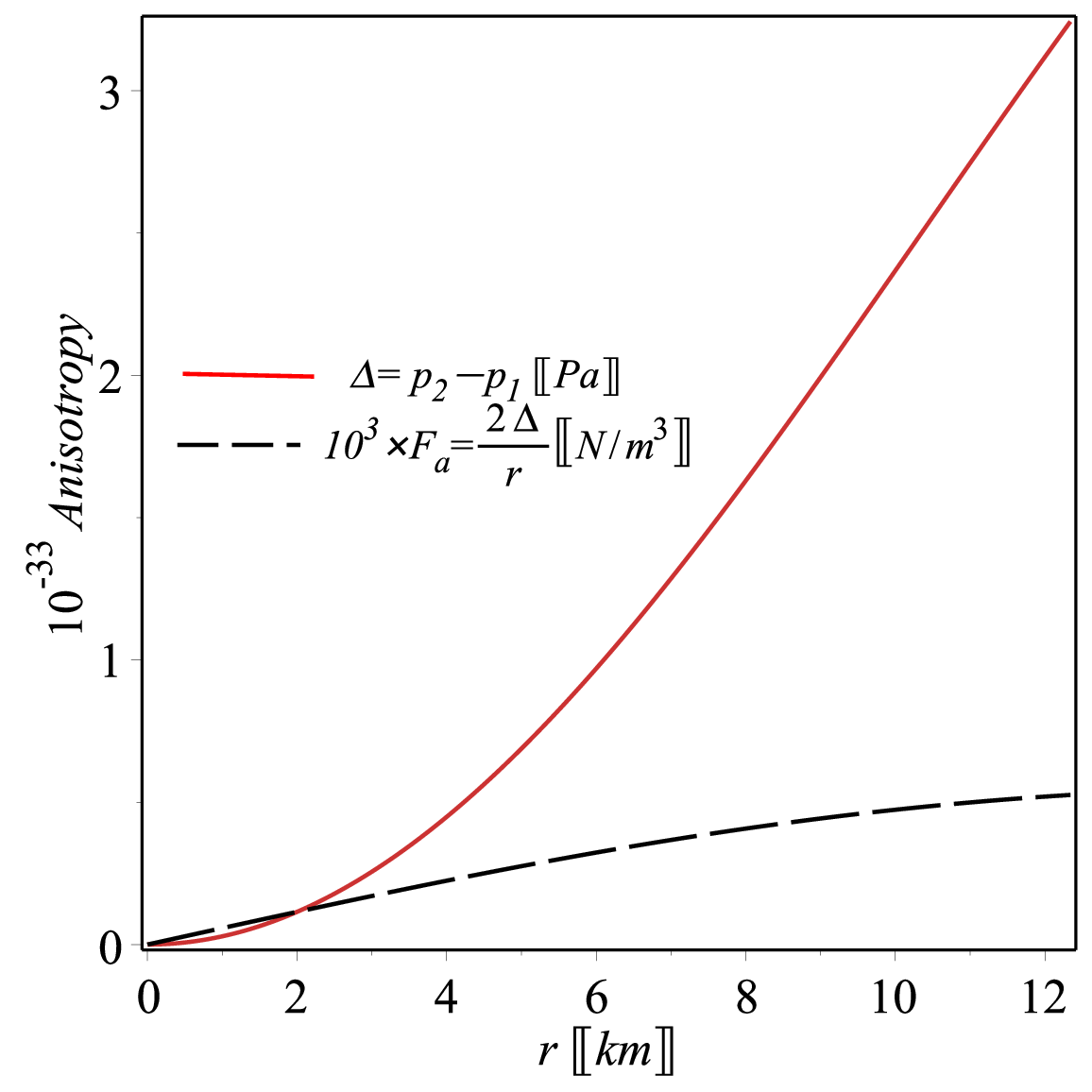}}
\subfigure[~The anisotropy f(R,T)]{\label{fig:anisotf}\includegraphics[scale=0.25]{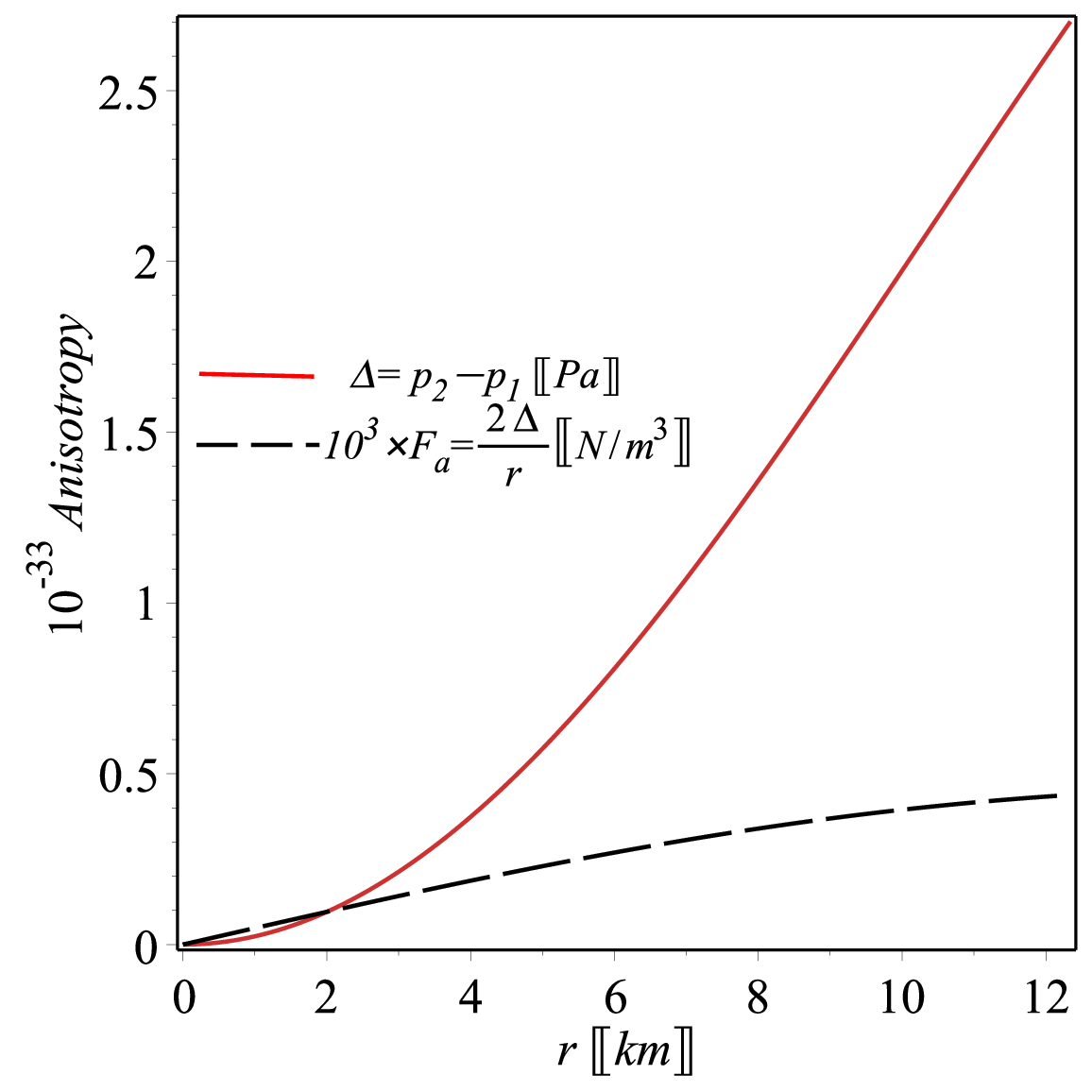}}
\caption{The patterns of the components of energy-momentum tensor as described by Eqs. \eqref{eq:Feqs2}  of the pulsar ${\textit PSR J0740+6620}$ are plotted in  Figs. \subref{fig:density}--\subref{fig:tangpressure}. Figures \subref{fig:GRgrad} and \subref{fig:RTgrad} illustrate the radial profiles of the gradients of these quantities in both the General Relativity (GR) and $f(\Ri,{ \mathbb{T}})$ scenarios.  Figures \subref{fig:anisotg} and \subref{fig:anisotf} display the variation of   $\Delta$ for both the   $f(\Ri,{ \mathbb{T}})$ and GR scenarios, as described by Eq. \eqref{eq:Delta2}.  All the graphs presented in Fig. \ref{Fig:dens_press} verify the fulfillment of the constraints indicated from (\textbf{I}) to (\textbf{IV}).}
\label{Fig:dens_press}
\end{figure*}
\noindent Constrain (\textbf{I}): The components of the EMT, energy content, radial and tangential stress, within the stellar fluid  ($0 < r/R < 1$), should maintain a positive outlook, specifically  ${\mathrm \rho(0 < r/R < 1)>0}$, ${\mathrm  p_r(0 < r/R < 1)>}0$ and ${\mathrm  p_t(0 < r/R <1)>0}$.

\noindent Constrain (\textbf{II}):  Interior solutions must have a regular, non-singular pattern.Hence, it is required that the density, radial stress, and tangential stress  investigated in this study remain regular  at the core. Furthermore, these quantities must exhibit regular behavior throughout the star's interior, i.e.,
\begin{itemize}

  \item[A-] ${\mathrm \rho(r=0)>0}$, \qquad ${\mathrm\rho'=0}$, \qquad ${\mathrm \rho''<0}$,\qquad and \qquad ${\mathrm \rho'(0< x=r/R \leq1)< 0}$\,,
  \item[B-] ${\mathrm p_r(r=0)>0}$, \qquad ${\mathrm p'_r(r=0)=0}$, \qquad ${\mathrm p''_r(r=0)<0}$, \qquad and \qquad ${\mathrm p'_r(0< x=r/R  \leq 1)< 0}$\,,
  \item[C-] ${\mathrm p_t(r=0)>0}$, \qquad ${\mathrm p'_t(r=0)=0}$, \qquad ${\mathrm p''_t(r=0)<0}$, \qquad and \qquad ${\mathrm p'_t(0<x =r/R  \leq 1)< 0}$\,.
\end{itemize}

\noindent Constrain (\textbf{III}): At the star's boundary, the  radial pressure must be zero, denoted as ${\mathrm p_r=0}$.  However, it is not necessary for $p_t$ to be zero at the boundary.

 The constraints (\textbf{I})--(\textbf{III}) are confirmed for ${\textit PSR J0740+6620}$, as depicted in Fig. \ref{Fig:dens_press}. As displayed in Fig. \ref{Fig:dens_press} \subref{fig:density},  the present study approximates a NS core density ${\mathrm \rho_\text{core}\approx 5.84\times 10^{14}}$ g/cm$^{3} \approx 3.26\rho_\text{nuc}$ for the pulsar ${\textit PSR J0740+6620}$. Thus, the present model does not rule out the potential existence of neutrons in the core of the pulsar. Moreover, the value of the energy density at the  center  reinforces the hypothesis that the fluid adopts an anisotropic configuration.

\noindent Constrain (\textbf{IV}): The anisotropy parameter, $\Delta$, must reach a value of zero at the core of the pulsar, meaning that ${\mathrm p_r=p_t}$.  Moreover,  $\Delta$  must have an increasing value at the boundary of the pulsar, i.e. $\Delta'(0 \leq r/R\leq 1)>0$. Thus, the anisotropic force ${\mathrm F_a=2\frac{\Delta}{r}}$ should be vanishing at the core. As shown in Eq.~\eqref{eq:Delta2}, taking the limit  $r\to 0$  gives $\Delta\to 0$.

 The constraint (\textbf{IV}) has been verified for  {\textit PSR J0740+6620}, as demonstrated in Figs. \ref{Fig:dens_press}\subref{fig:anisotg} and \ref{Fig:dens_press}\subref{fig:anisotf}.

We can also demonstrate that  ${\mathrm \Delta(r>0) >} 0$  ensuring    $\mathrm {p_t>p_r}$.  This condition is crucial  to make the  anisotropy  force    repulsive allowing  larger NS sizes than for an isotropic fluid. Nevertheless, when considering $\beta_1>0$, it indicates that $\Delta$ in $f(\Ri,{ \mathbb{T}})=\Ri+ \beta_1 { \mathbb{T}}$ is marginally lower than in  GR.
\subsection{The condition of Zeldovich}
\noindent Constraint (\textbf{V}): Based on the study by \cite{1971reas.book.....Z}, the radial pressure at the core should be:
\begin{equation}\label{eq:Zel}
  \mathrm  {\frac{\widetilde{p}_r(0)}{\widetilde{\rho}(0)}\leq 1.}
\end{equation}
At the center of the pulsar, we derive equations for the density, the pressure in the radial direction, and the pressure in the tangential direction as:
\newpage
\begin{align}
 &\mathrm {{\widetilde\rho}(x=0) =3\lambda+2\beta_1(8\lambda+5\mu)}\,, \nonumber \\
 & \mathrm {\widetilde p_r(x=0) = {\widetilde p}_t(x=0)={2\mu(1+3\beta_1)-\lambda}}\,.\end{align}
We calculate the compactness parameter of the pulsar $\textit{PSR J0740+6620}$ to be $C =0.491+\pm 0.0547087977$. Thus, we can use the Zeldovich condition \eqref{eq:Zel} to determine a valid range of values for the dimensionless parameter $\beta_1$, which is $0\leq \beta_1 \leq 0.1$. In this interval, the value of $\beta_1$ is expected to be consistent with that of GR when $\beta_1=0$.
\subsection{The radius and mass  observational limits of  $\textit{PSR J0740+6620}$ }
\begin{figure}
\centering
\subfigure[~The Mass function]{\label{Fig:Mass}\includegraphics[scale=0.25]{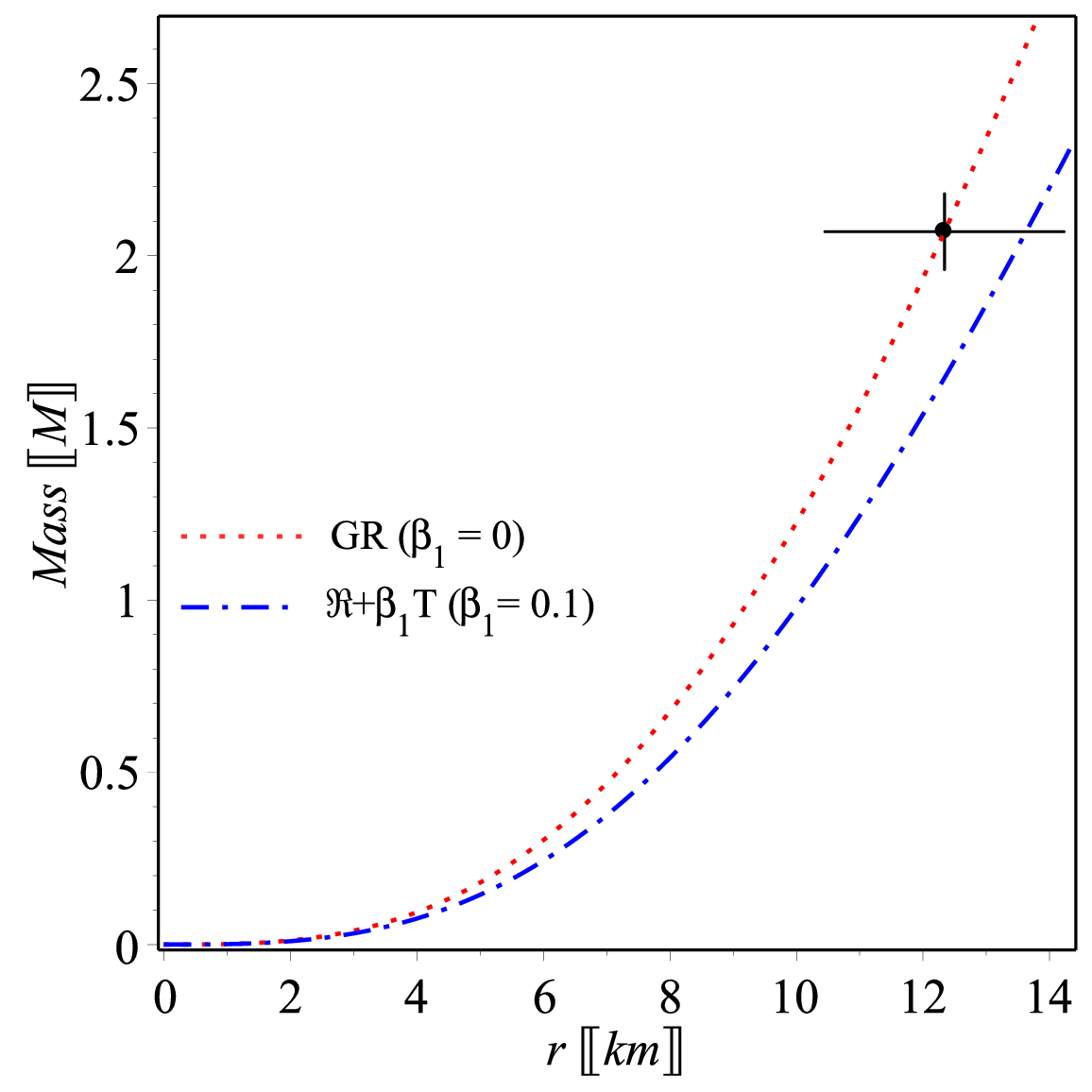}}
\subfigure[~The Compactness parameter]{\label{Fig:Comp}\includegraphics[scale=0.25]{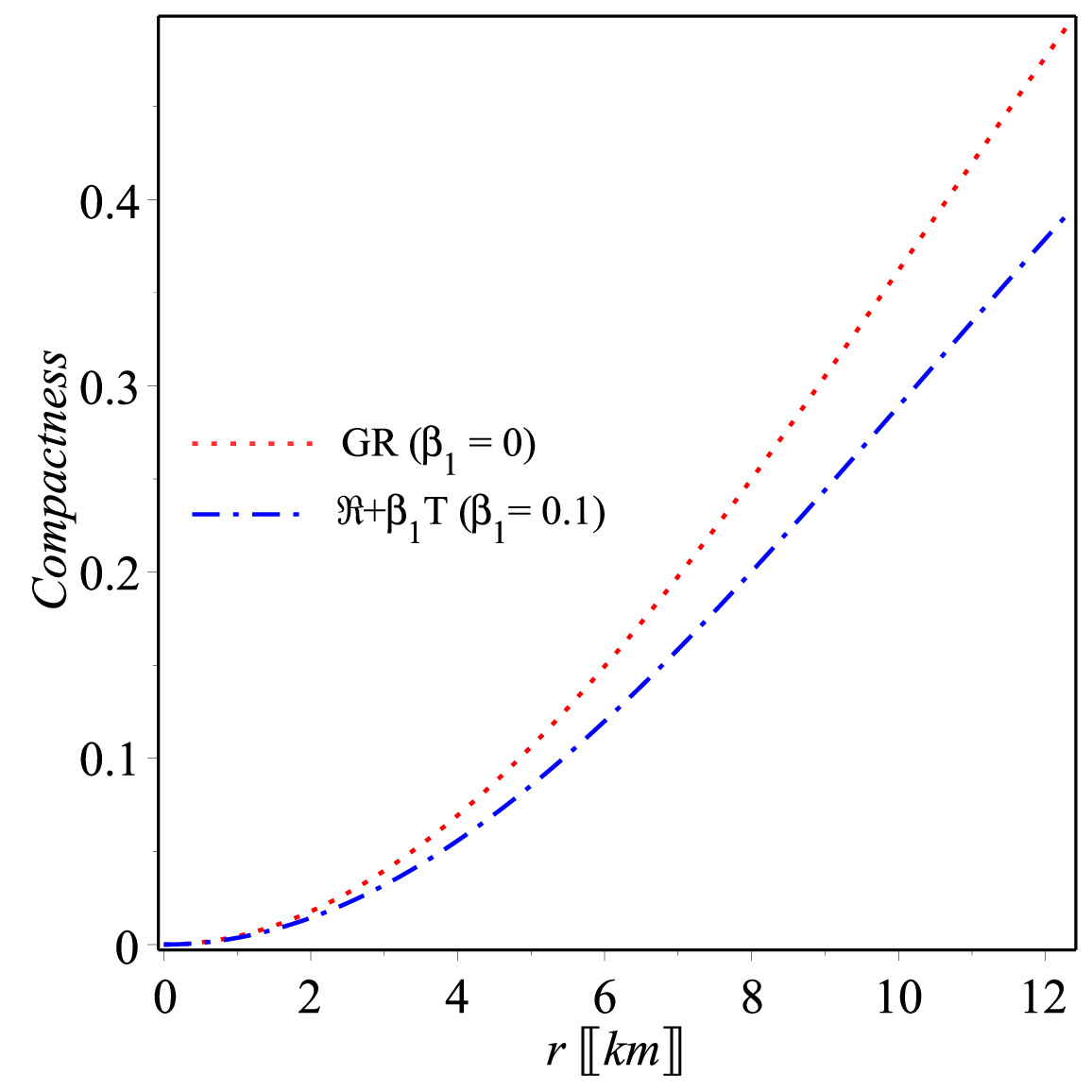}}
\caption{We have generated plots for the mass and the compactness, represented by Eqs. \eqref{eq:Mass} and \eqref{COMP}, respectively, pertaining to  $\textit{PSR J0740+6620}$.  The resulting curves are in excellent harmony with the data of ($M=2.07\pm 0.11 M_\odot$ and $R=12.34\pm1.89$ km) \cite{Legred:2021hdx}. For $f(\Ri,{ \mathbb{T}})=\Ri+\beta_1,{ \mathbb{T}}$ case, we assign the following values to the parameters: {$\beta_1=0.1$, $\kappa=2.302\times 10^{-43},N^{-1}$, $\mu =0.4248802917$, $\nu =-1.109209$, $\lambda =0.6843287084$}. In the case of GR, we set $\beta_1=0$, and the corresponding values for the parameters are { $\mu =0.4912202964$, $\nu =-1.175549005$}.}
\label{Fig:Mass1}
\end{figure}
Taking into account the restrictions imposed by the mass and radius of ${\textit PSR J0740+6620}$, we find that the mass function given by Eq. \eqref{eq:Mass} predicts a total mass of $M=2.12 M_\odot$ at  $R=13.65$ km,  corresponding to ${C}=0.3214945055$, when $\beta_1$ is set to $0.1$. This is consistent with the  values of ($M=2.07\pm0.11 M_\odot$ and $R=12.34\pm1.89$ km) \cite{Legred:2021hdx}. This determines the values of the set of constants \{ $\mu =0.4248802917$, $\nu =-1.109209$, $\lambda =0.6843287084$, $\kappa=2.302\times 10^{-43}\,N^{-1}$\}. These   values  unequivocally confirm the fulfillment of Eq. \eqref{eq:Zel}.
In Figs. \ref{Fig:Mass1} \subref{Fig:Mass} and \subref{Fig:Comp}, we present the trends of the mass function and compactness, respectively, to demonstrate the model's agreement with the observed mass and radius of the pulsar.  In comparison to  GR, the model $f(\Ri,{ \mathbb{T}})=\Ri+\beta_1,{ \mathbb{T}}$ with a positive $\beta_1$ predicts the same mass but at a larger radius. Consequently, the  $f(\Ri,{ \mathbb{T}})=\Ri+\beta_1,{ \mathbb{T}}$ anticipates a higher compactness for a given mass. This demonstrates that $f(\Ri,{ \mathbb{T}})=\Ri+\beta_1,{ \mathbb{T}}$ has the capacity to accommodate higher masses or greater compactness values while still adhering to stability conditions.
\subsection{Geometric sector}\label{Sec:geom}
\noindent  Constrain (\textbf{VI}): The metric potentials ${\textit g_{rr}}$ and ${\textit g_{tt}}$  should have no physical or coordinate  singularities within the stellar interior   $0\leq x\leq 1$. The metric \eqref{eq:KB} satisfies  this constraint because:  At the core, ${\mathrm g_{tt}(x=0)=e^\nu }$ and ${\mathrm g_{rr}(x=0)=1}$, i.e, they remain  finite.

\noindent Constrain (\textbf{VII}): The solutions for the metric potentials in both the exterior and interior regions must match smoothly at the star's boundary.
\begin{figure}
\centering
{\includegraphics[scale=0.3]{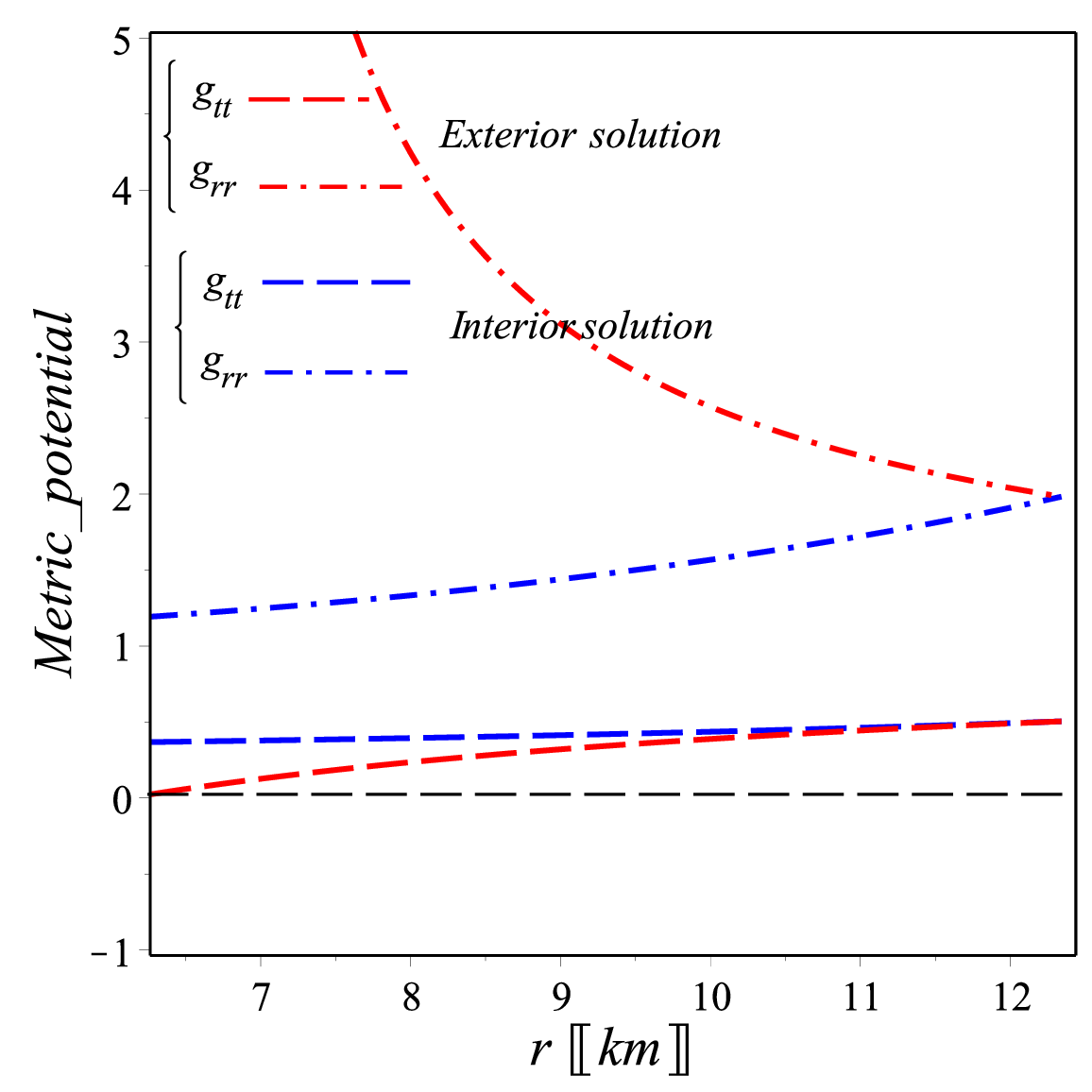}}
\caption{The metric components, $g_{rr}$ and $g_{00}$,  of the star ${ \textit PSR J0740+6620}$ have finite values within its interior of the pulsar and seamlessly match with the Schwarzschild vacuum solution outside. The above plots provide  confirmation of these conditions.}
\label{Fig:Matching}
\end{figure}

It is evident from Fig. that conditions (\textbf{VI}) and (\textbf{VII}) are fulfilled using  $\textit{PSR J0740+6620}$.

\noindent Constrain (\textbf{VIII}): The red-shift of the metric  \eqref{eq:KB} is figured as:
\begin{equation}\label{eq:redshift}
    Z=\frac{1}{\sqrt{-g_{tt}}}-1=\frac{1}{\sqrt{e^{\mu\,x+\nu}}}-1.
\end{equation}
The gravitational redshift must possess the following characteristics: it should be positive ($Z > 0$), have a finite value inside the stellar body, and decrease as one moves towards the stellar surface ($Z' < 0$).
\begin{figure}
\centering
{\includegraphics[scale=0.3]{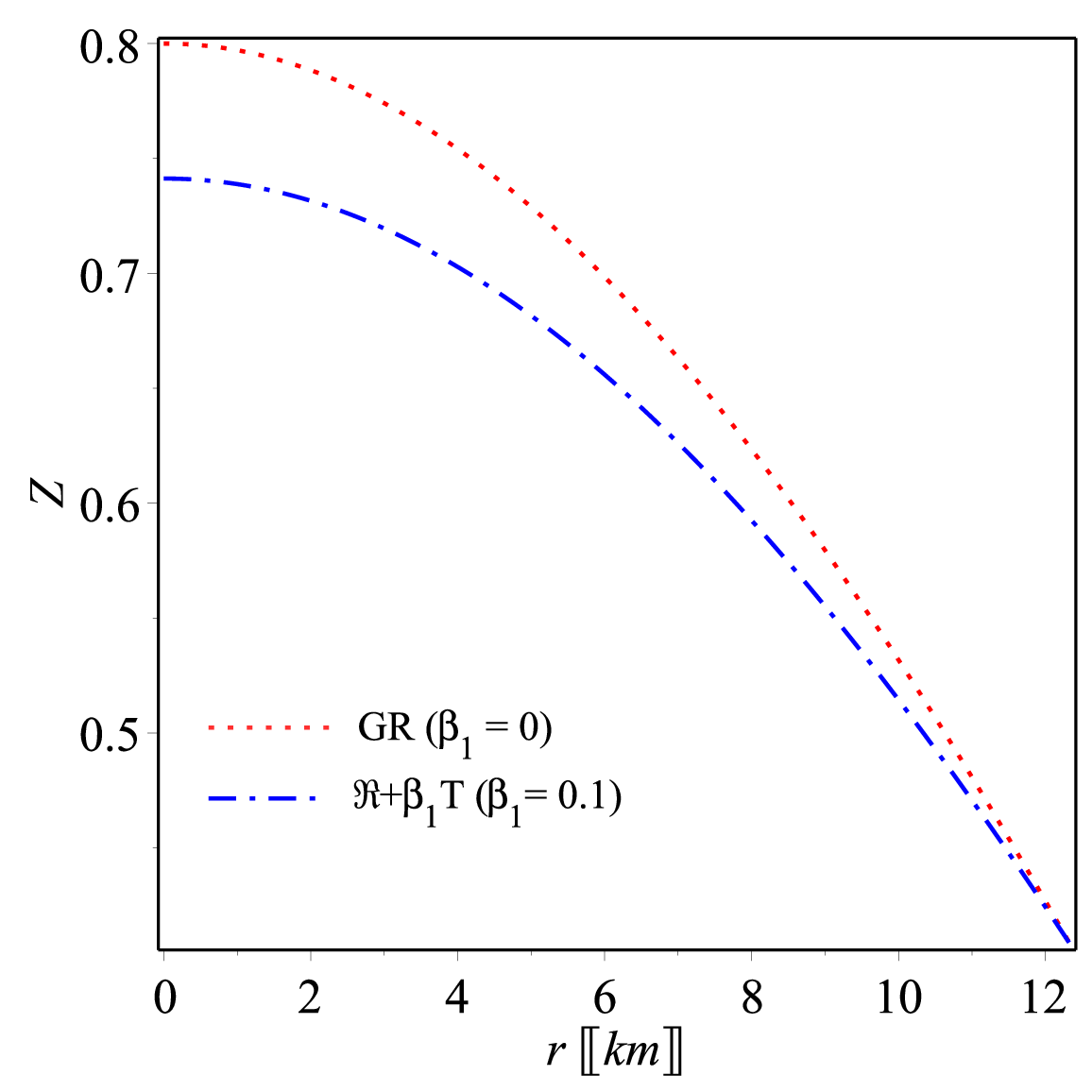}}
\caption{The red-shift of Eq.~\eqref{eq:redshift} of  {\textit PSRJ0740} is illustrated in Fig. \ref{Fig:Redshift}, confirming the satisfaction of condition (\textbf{VIII}).}
\label{Fig:Redshift}
\end{figure}

It is evident from Fig. \ref{Fig:Redshift} that condition (\textbf{VIII}) is fulfilled using  $\textit{PSR J0740+6620}$.

 At   center: Gravitational red-shift factor is $Z(0)\approx 0.7412521838$ (the value of $Z(0)$ is below the corresponding value in GR, which is approximately $0.7999781079$.). This central value is the maximum red-shift within the star. At the stellar  surface:  The red-shift factor is $Z_R\approx 0.4079916880$ (around the same value as in GR). This surface value is below the upper bound of $Z_R=2$ \cite{Buchdahl:1959zz}, which has been explored in other studies, such as \citep{Ivanov:2002xf, Barraco:2003jq} for anisotropic analyses and \cite{Boehmer:2006ye} in the context of a cosmological constant.  The maximum red-shift limit does not provide a tight constraint and cannot properly bound compactness alone, as discussed in other works. Less restrictive constraints are needed \citep[c.f.,][]{Ivanov:2002xf,Barraco:2003jq,Boehmer:2006ye}. Applying energy conditions to the matter sector yields considerably more stringent constraints, as demonstrated in \citep[see,][]{Roupas:2020mvs}.

\subsection{Energy conditions}\label{Sec:Energy-conditions}
In   GR,  focusing theorems relate the positivity of the tidal tensor trace to physical phenomena,   ${\mathrm{R}_{\alpha\beta} w^{\alpha} w^{\beta} \geq 0}$ and ${\mathrm {R}_{\alpha\beta} \ell^{\alpha} \ell^{\beta} \geq 0}$, where, ${\mathrm w^{\alpha}}$ refers to a vector with a timelike component that can take any value, while ${\mathrm \ell^{\alpha}}$ represents any null vector that is directed towards the future. These theorems impose four restrictions on the energy-momentum tensor (EMT)  ${ \mathbb{T}}^{\alpha\beta}$. These constraints are commonly referred to as the energy conditions. The energy conditions can be extended or generalized to encompass modified gravitational theories beyond GR.
They provide limits on the matter sector based on geometric properties. Specifically, for the theory $f(\Ri,{ \mathbb{T}})=\Ri+\beta_1 { \mathbb{T}}$: The energy conditions can be expressed using the (EMT) as:  ${{ \mathbb{T}}{^\alpha}{_\beta}}=\mathrm{diag(-\rho c^2,p_r, p_t, p_t)}$.\\
\begin{figure*}[t!]
\centering
\subfigure[~WEC \& NEC (radial)]{\label{fig:Cond1}\includegraphics[scale=0.27]{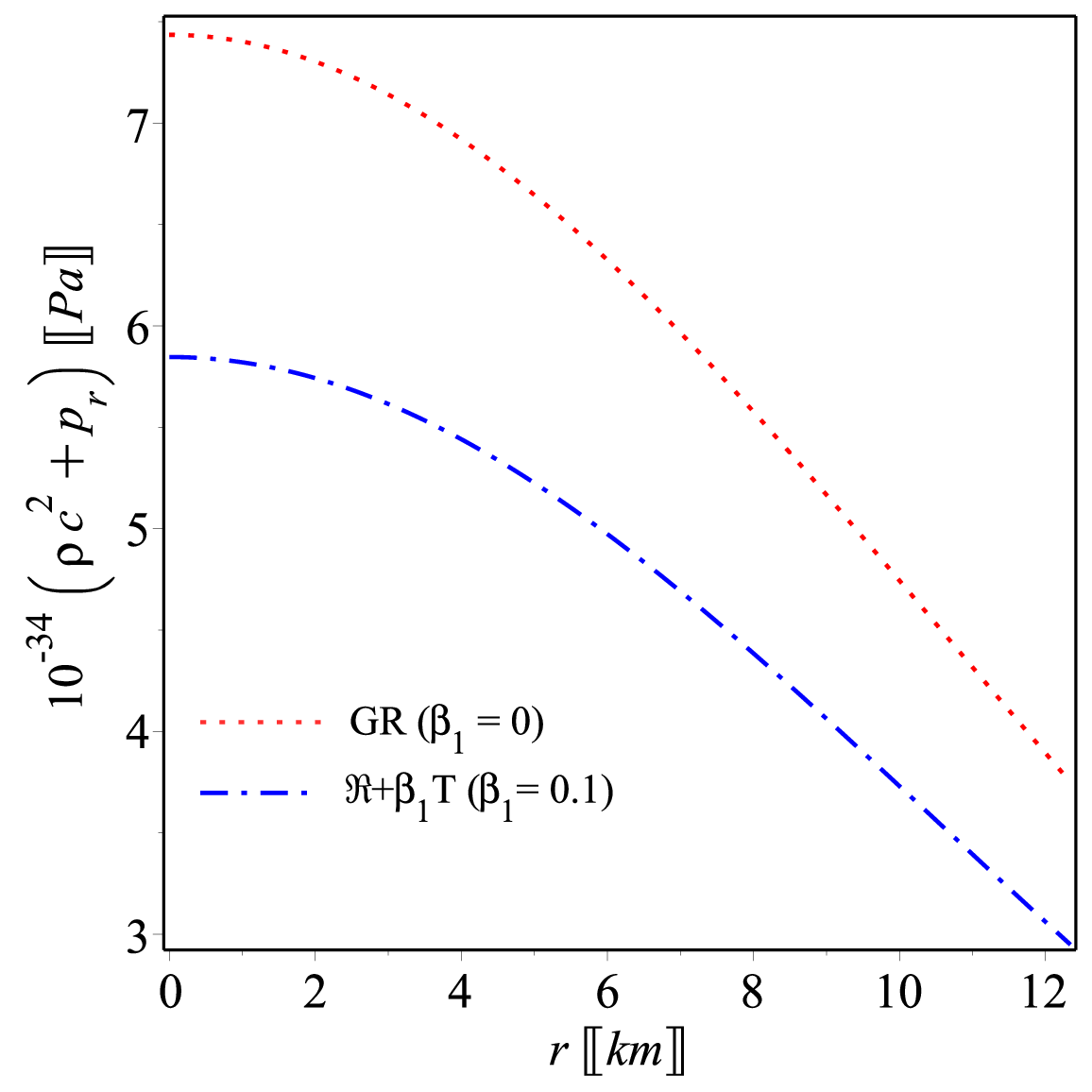}}\hspace{0.1cm}
\subfigure[~WEC \&NEC (tangential)]{\label{fig:Cond2}\includegraphics[scale=0.27]{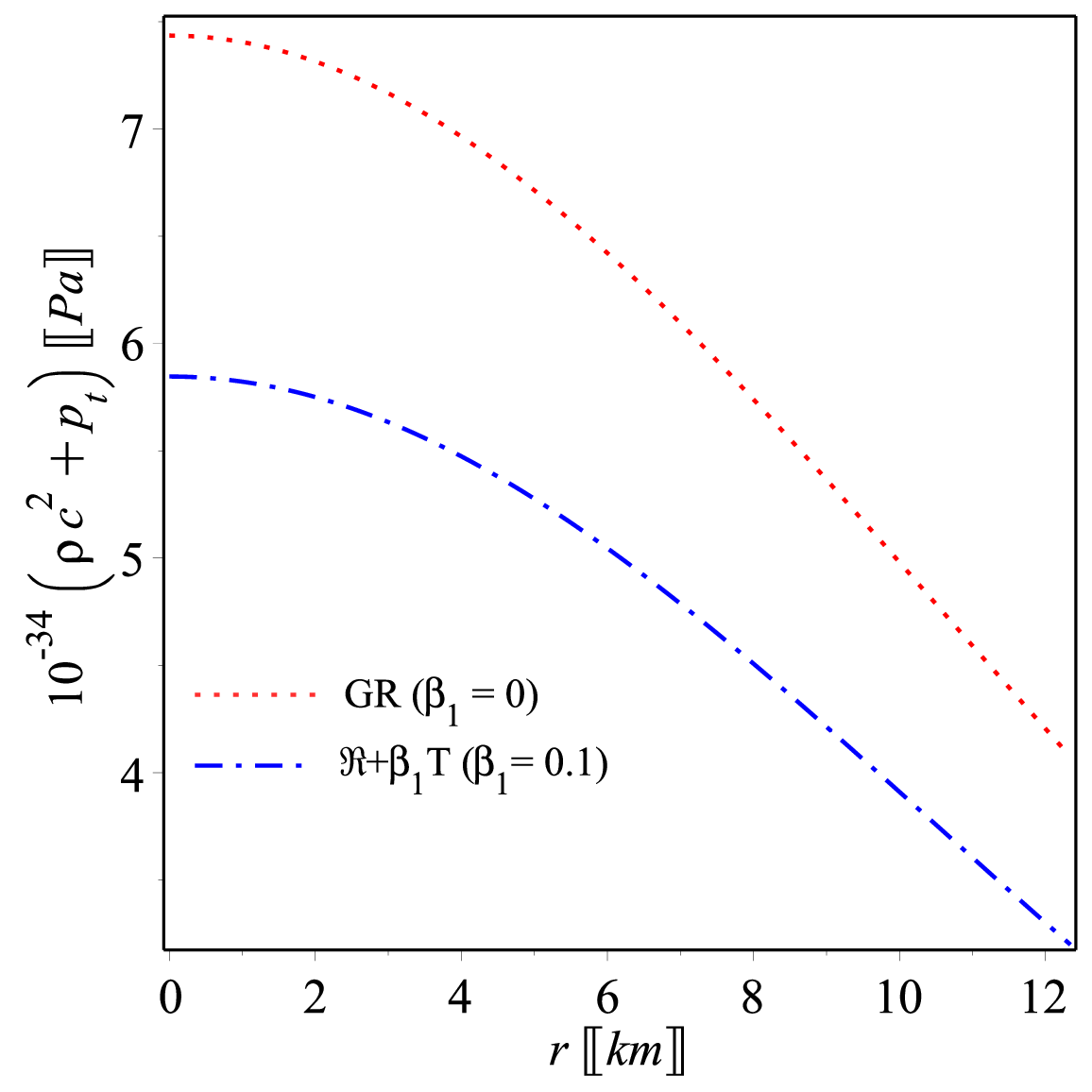}}\hspace{0.1cm}
\subfigure[~SEC]{\label{fig:Cond3}\includegraphics[scale=.27]{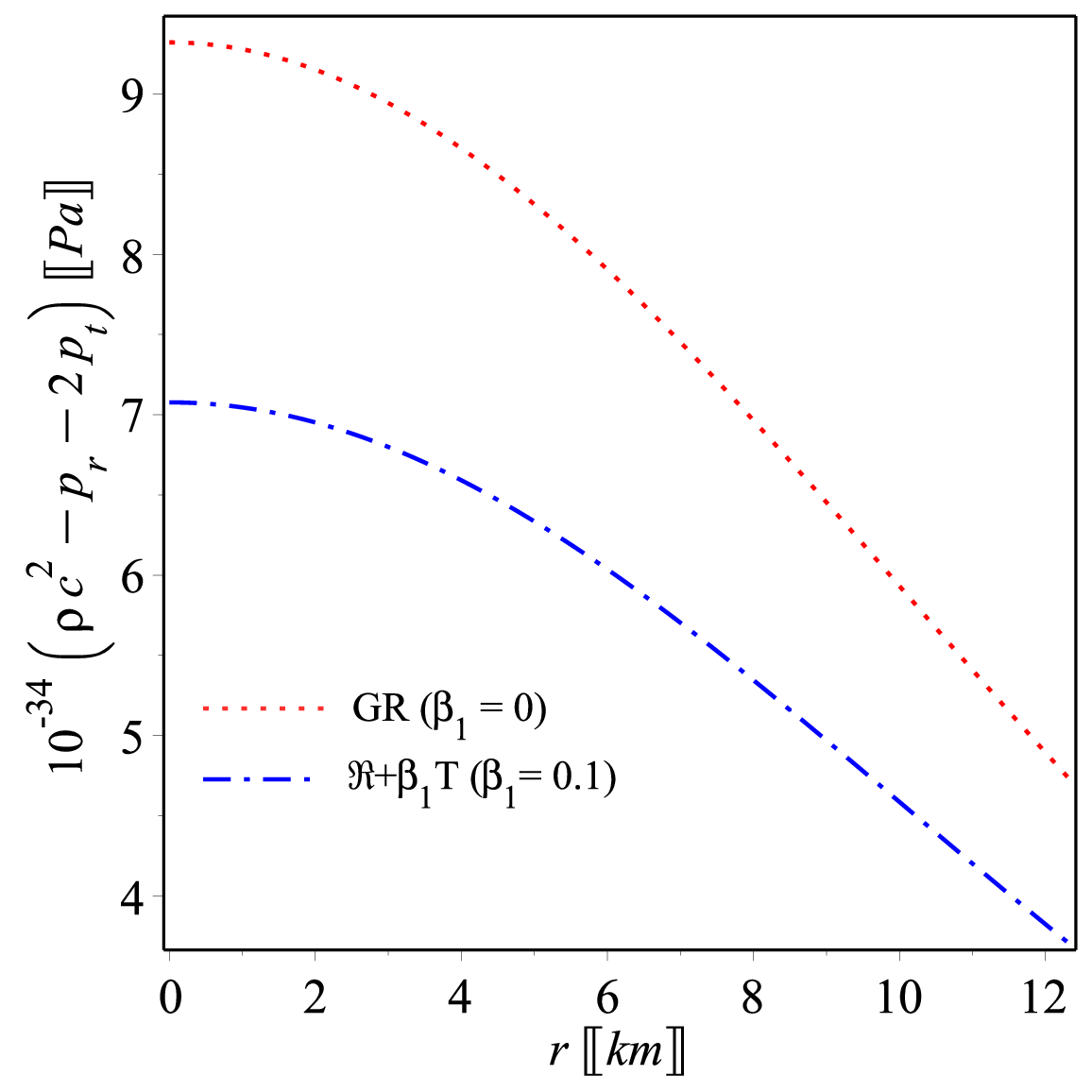}}
\subfigure[~DEC (radial)]{\label{fig:DEC}\includegraphics[scale=.27]{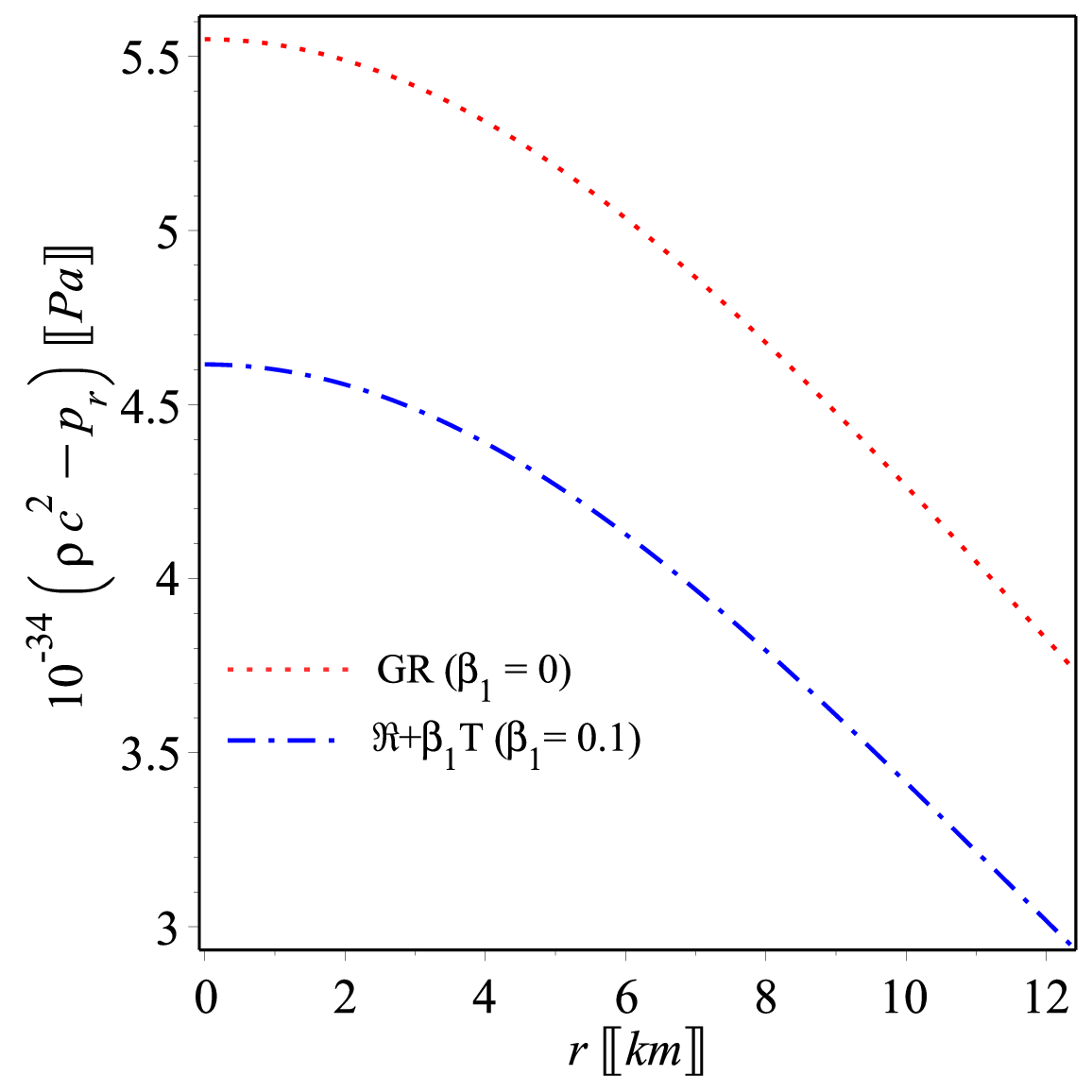}}\hspace{2cm}
\subfigure[~DEC (tangential)]{\label{fig:DEC2}\includegraphics[scale=.27]{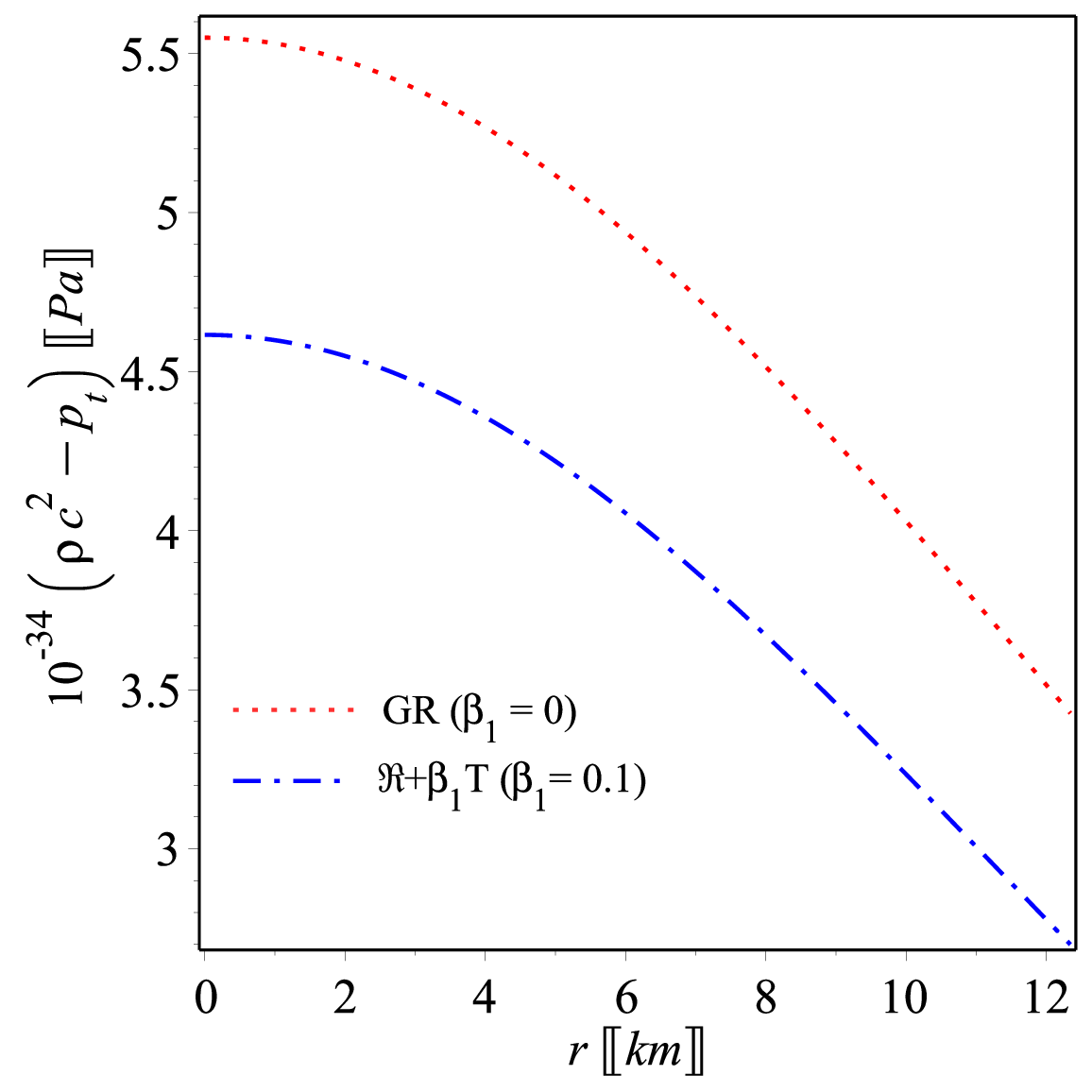}}
\caption{he figures presented in Fig. \ref{Fig:EC} for  $\textit{PSR J0740+6620}$ provide evidence that the energy conditions (\textbf{IX}) are satisfied.}
\label{Fig:EC}
\end{figure*}
\noindent  Constrain (\textbf{IX}): Summary of the necessary energy conditions for a physically acceptable stellar model:
\begin{itemize}
 \item[{\textrm a-}] The fulfillment of the weak energy condition (WEC) requires the inequalities to hold true, i.e.,  ${\mathrm \rho\geq 0}$, \,${\mathrm  p_r+\rho c^2 > 0}$, \,${\mathrm\rho c^2+p_t > 0}$\,,
\item[{\textrm b-}] The null energy condition (NEC) needs to adhere to the following inequalities: ${\mathrm \rho c^2+ p_r \geq 0}$, \, ${\mathrm  p_t+\rho c^2 \geq 0}$\,,
 \item[{\textrm c-}] The strong energy condition (SEC) must adhere to the following inequalities: ${\mathrm \rho c^2-p_r-2p_t \geq 0}$, \,${\mathrm p_r+\rho c^2 \geq 0}$, \,${\mathrm\rho c^2+p_t \geq 0}$\,,
 \item[{\textrm d}-] The dominant energy conditions (DEC) must meet the following inequalities: ${\mathrm \rho \geq 0}$, \,${\mathrm\rho c^2-p_r \geq 0}$, and ${\mathrm\rho c^2-p_t \geq 0}$\,.\\
\end{itemize}

Now let us reformulate the field equations \eqref{eq:Feqs} as:
\begin{eqnarray}\label{eq:Eff-dens-press}
\nonumber {\mathrm \rho c^2} &=&{\mathrm  \rho_{GR} c^2 +\frac{5\beta_1}{3(1+3\,\beta_1)
 }\left( p_r  +2\,p_t \right)},\\
\nonumber {\mathrm p_r} &=& {\mathrm {p_r}_{GR} +\frac{\beta_1}{3+11\beta_1} \left(3\rho c^2  -10p_t  \right)},\\
{\mathrm p_t} &=&{\mathrm { p_t}_{GR} + \frac{3\beta_1}{3+16\,\beta_1
 }\left(\rho c^2-\frac{5}{3} p_r  \right)}\,.
\end{eqnarray}
When the dimensionless parameter $\beta_1=0$ we can recover  GR. It is time to analysis constrain  (\textbf{IX}) in analytic way.

By the  use of constrains  (\textbf{I}) and  (\textbf{II}), we can demonstrate that:  density and radial/tangential stresses have  positive values within the interior of the  star. This confirms the NEC is verified. Some additional implications and proofs  can shown:
\begin{align}\label{eq:Ras_EC}
\nonumber& {\mathrm \rho c^2}+{\mathrm p_r} = ({\mathrm \rho_{GR} c^2} +{\mathrm {p_r}_{GR}}) + {\mathrm \beta_1}\frac { \left( 27\,{\beta_1}+9\, \right) {\mathrm \rho c^2}+ \left( 55\,{\beta_1}+15\, \right){\mathrm p_r} +20\,{\mathrm {\beta_1}}{\mathrm p_t}  }{9+99\,{\beta_1}^
{2}+60\,\beta_1},\\
&\nonumber  {\mathrm \rho c^2}+{\mathrm p_t} = ({\mathrm \rho_{GR} c^2} +{\mathrm {p_t}_{GR}})- \beta_1\,\frac{ \left(10{\mathrm p_t}[3
-16\beta_1]- 125\,{\mathrm p_r} \beta_1+9\,
{\mathrm \rho} c^2[1+3 \beta_1] \right) }{3
 \left( 1+3\,\beta_1 \right)  \left(3-16\,\beta_1\right) }
,\\
\nonumber& {\mathrm\rho} c^2-{\mathrm p_r}-2{\mathrm p_t} = ({\mathrm\rho_{GR}} c^2-{\mathrm{p_r}_{GR}}-2 {\mathrm {p_t}_{GR}}) \\
&+\beta_1{\frac { \left( 162\,{\beta_1}^{2}+297{\beta_1}+81
 \right) {\mathrm \rho c^2}- \left( 1870{\beta_1}^{2}+675{
\beta_1}+45 \right){\mathrm p_r} +20{\mathrm p_t} \left( 3-16\beta_1 \right) }{27-1584
\,{\beta_1}^{3}-663\,{\beta_1}^{2}+36\,\beta_1}},\\
\nonumber &{\mathrm {\rho} c^2}- {\mathrm {p}_r} = ({\mathrm \rho_{GR} c^2} - {\mathrm {p_r}_{GR}}) +\beta_1\,\frac { \left( 15\,{\mathrm p_r} +55{\mathrm p_r}\beta_1+60\,p_t +200\,{\mathrm p_t}  \beta_1-9\,{\mathrm \rho} c^2[1+3 \beta_1] \right) }{3 \left( 1+3\,\beta_1 \right)  \left( 3+11\,
\beta_1 \right) }
,\\
\nonumber &{\mathrm {\rho} c^2}- {\mathrm {p}_t} = ({\mathrm \rho_{GR} c^2} - {\mathrm {p_t}_{GR}}) +\beta_1{\frac { \left( {\mathrm p_r}[30 -35
  \beta_1]+{\mathrm p_t}[30  -160{\mathrm p_t}
  \beta_1]-9{\mathrm \rho c^2}[1+\beta_1] \right) }{9-144{\beta_1}^{2}-21\,\beta_1}}
,\\
& {\mathrm \rho c^2}-{\mathrm p_r}-2{\mathrm p_t} = ({\mathrm \rho_{GR} c^2} -{\mathrm {p_r}_{GR}}-2 {\mathrm {p_t}_{GR}}) \\
&+\frac {\beta_1 \left({\mathrm p_r}[ 135  +525  \beta_1+110 {\beta_1
}^{2}]+{\mathrm p_t}[180  -360
\beta_1-3200\,{\beta_1}^{2}] -3\,{\mathrm \rho c^2}[27+99\beta_1+54 {\beta_1}^{2}] \right) }{3\left( 1+3\,\beta_1 \right)  \left( 3+
11\,\beta_1 \right)  \left( 3-16\,\beta_1 \right) }
.\qquad
\end{align}
Given that $\beta_1>0$ (with a value of $\beta_1=0.1$ in this study), and considering that ${\mathrm \rho\geq 0}$, ${\mathrm p_r\geq 0}$, and ${\mathrm p_t\geq 0}$, the only thing left to demonstrate is that  \[ {\mathrm p_r}[30 -35
  \beta_1]+{\mathrm p_t}[30  -160p_t
  \beta_1]>9\,{\mathrm \rho c^2}[1+\beta_1]>0\,,\] \[\left( 162\,{\beta_1}^{2}+297{\beta_1}+81
 \right) {\mathrm \rho c^2} +20{\mathrm p_2} {\beta_1} \left( 3-16\beta_1 \right)>\left( 1870{\beta_1}^{2}+675{
\beta_1}+45 \right)p_r>0\,,\] and \[10{\mathrm p_t}[3
-16\beta_1]+9{\mathrm \rho c^2}[1+3 \beta_1]>125\,{\mathrm p_r} \beta_1>0\,,\]  to ensure that the energy conditions ({\textrm I}--{\textrm IV}) are satisfied.

The information presented in Fig. \ref{Fig:EC}\subref{fig:Cond1}--\subref{fig:DEC2} indicates that   conditions (\textbf{IX}) are satisfied for   $\textit{PSR J0740+6620}$.

It should be noted that the condition of energy density being greater than the sum of $p_r$ and $p_t$ pressures is guaranteed in the theory $f(\Ri,{ \mathbb{T}})=\Ri+\beta_1 { \mathbb{T}}$ just as   GR, since \[{\mathrm {\rho} c^2} - \mathrm{p_r}- 2\mathrm{p_t} =\frac {\beta_1 \left({\mathrm p_r}[ 135  +525  \beta_1+110 {\beta_1
}^{2}]+p_r[180  -360
\beta_1-3200\,{\beta_1}^{2}] -3\,{\mathrm \rho c^2}[27+99\beta_1+54 {\beta_1}^{2}] \right) }{3\left( 1+3\,\beta_1 \right)  \left( 3+
11\,\beta_1 \right)  \left( 3-16\,\beta_1 \right) }.\]  Some authors refer to the requirement that energy density exceeds total pressure as the SEC  \citep[c.f.][]{1988CQGra...5.1329K,Ivanov:2017kyr,2019EPJC...79..853D,Roupas:2020mvs}.

From the above discussion,  we understand that the energy conditions primarily relate to and constrain the matter sector in $f(\Ri,{ \mathbb{T}})$   gravitational theories, though they do not necessarily require the same formulations across all cases. However, certain key conditions are necessary to ensure the others can also be met where density and pressures remain positive: The dominant energy condition (DEC) stipulating ${\mathrm \rho c^2} -{\mathrm p_r} - 2{\mathrm p_t}\geq 0$ is especially crucial. This condition must be verified for the plausibility of the others.
\subsection{ Equilibrium   and causality conditions}
\begin{figure*}
\centering
\subfigure[~Radial speed of sound]{\label{fig:vr}\includegraphics[scale=0.28]{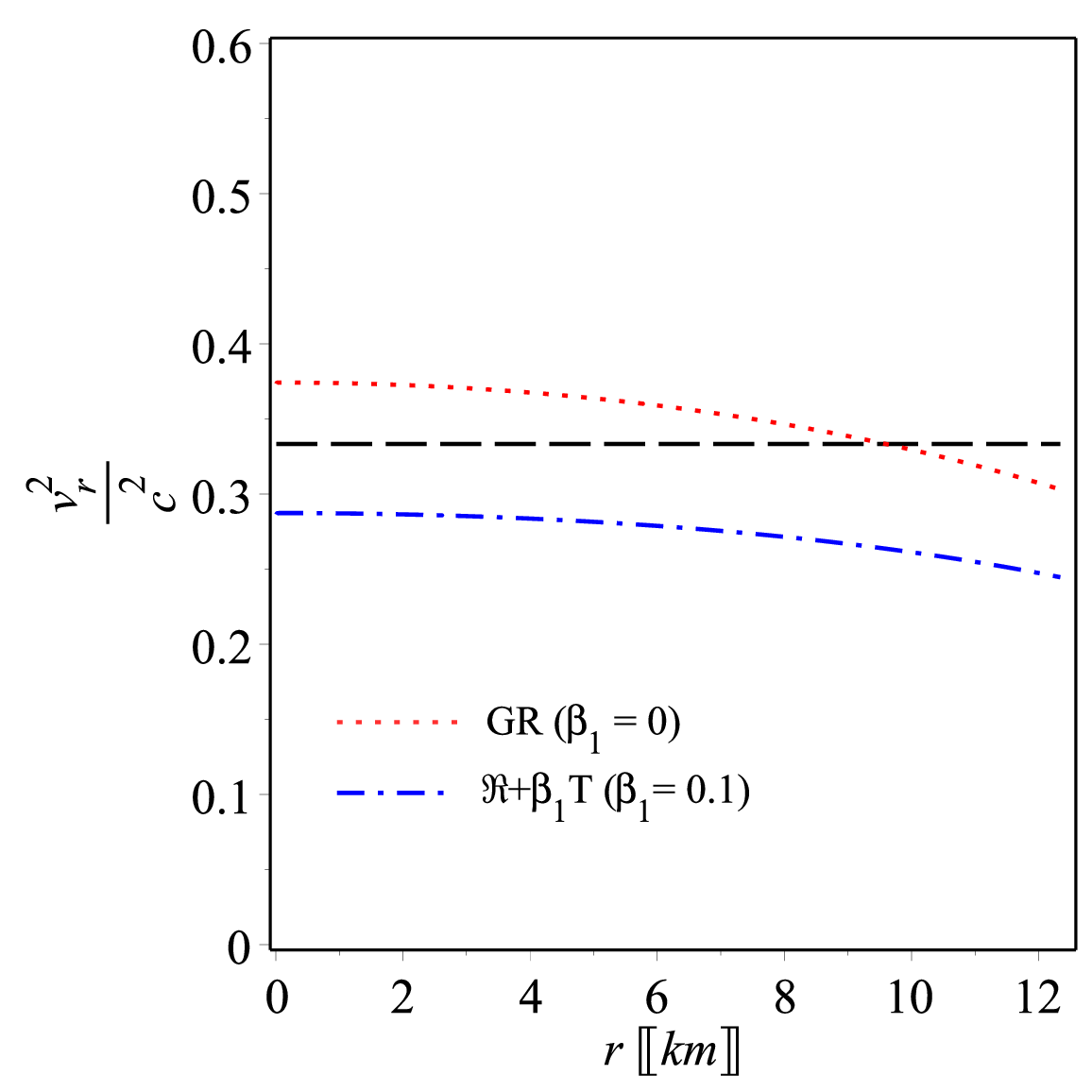}}
\subfigure[~Tangential speed of sound]{\label{fig:vt}\includegraphics[scale=.28]{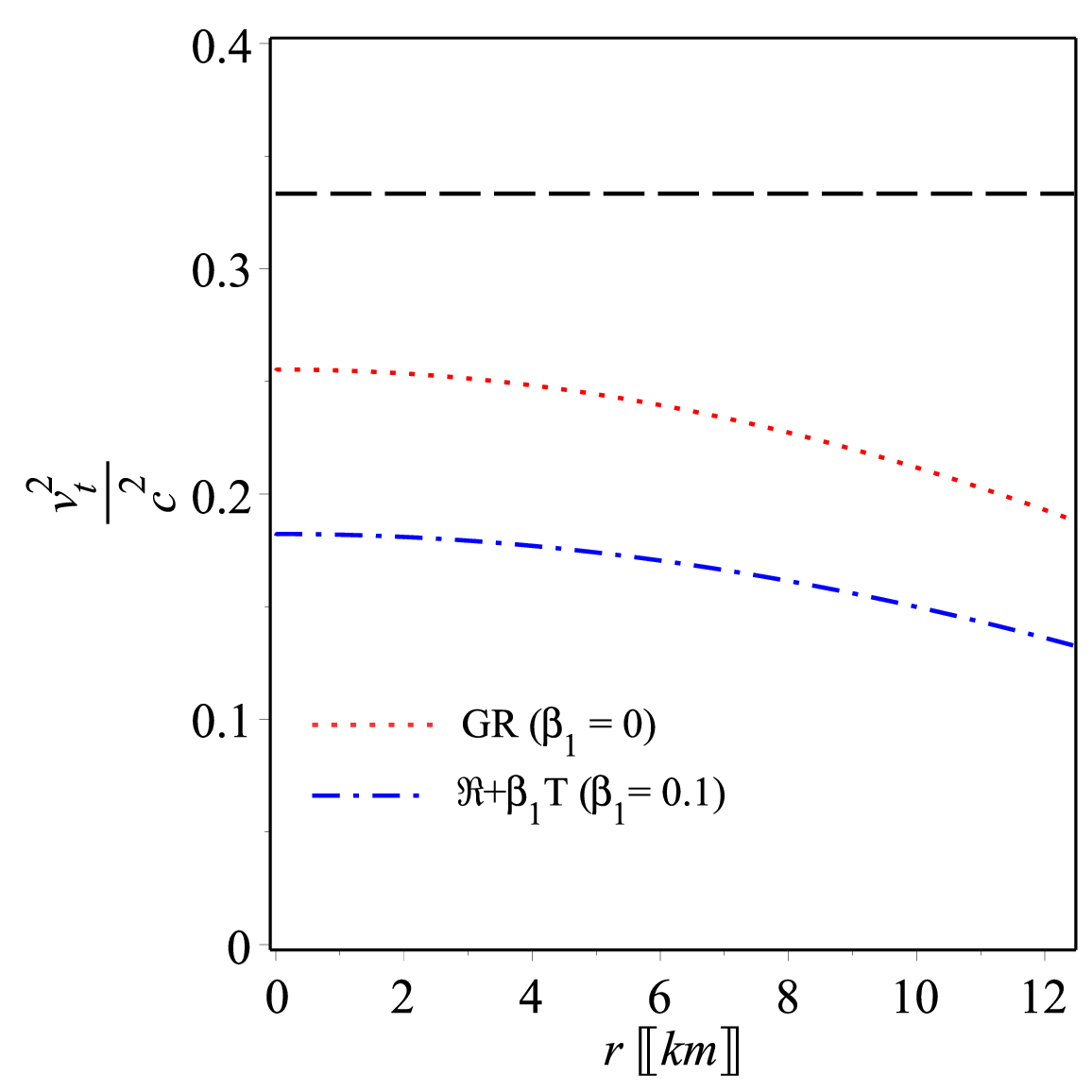}}
\subfigure[~difference between radial and tangential  speed of sounds]{\label{fig:vt-vr}\includegraphics[scale=.28]{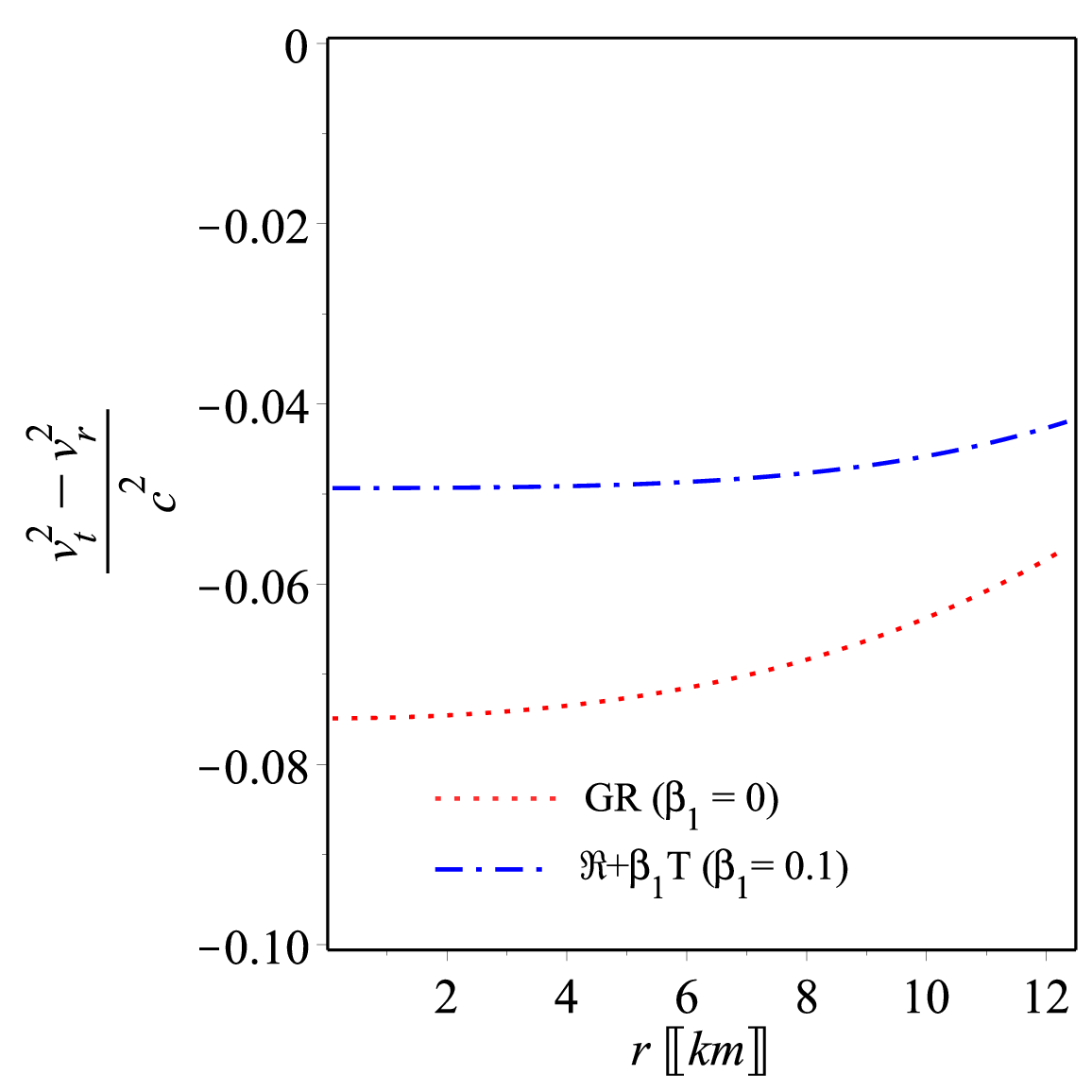}}
\caption{The speeds of sound in the radial and tangential directions given by Eq.~\eqref{eq:sound_speed}     $\textit{PSR J0740+6620}$. These figures assure that the conditions of causality and stability referred to  as (\textbf{XII}) and (\textbf{XIII}) have been  verified.}
\label{Fig:Stability}
\end{figure*}
The tangential and radial sound speeds are figured as:
\begin{equation}\label{eq:sound_speed}
    {\mathrm  v_t^2} = \frac{d{\mathrm  p}_t}{d{\mathrm   \rho}}= \frac{{\mathrm  p'}_t}{{\mathrm \rho'}}\,, \qquad {\mathrm v_r^2} =  \frac{ d{\mathrm p}_r}{d {\mathrm \rho}}=  \frac{{\mathrm  p'}_r}{{\mathrm \rho'}}\,.
\end{equation}
Here  density and  derivative of  radial and tangential components of pressure  are given in Eqs. \eqref{eq:dens_grad}--\eqref{eq:pt_grad}.\\

\noindent Constrain (\textbf{X}):The star's structure must be causal, meaning that it must satisfy the causality condition, which requires that the speeds of sound be positive and less than the speed of light in the stellar region ($0\leq {\mathrm v_t/c} \leq 1$, $0\leq {\mathrm v_r}/c\leq 1$), furthermore, it can be observed that they exhibit a decline as they approach the boundary  (${\mathrm v'_t}{^2}<0$,${\mathrm v'_r}{^2}<0$).

Although the speed of sound near the core of  $\textit{PSR J0740+6620}$, satisfies the conditions of stability and causality, the speed of sound predicted by GR   goes beyond the suggested upper bound for the speed of sound, which is $v_r^2=c_s^2\leq c^2/3$.  Moreover, different models have exhibited inconsistency with the conformal upper threshold for the speed of sound, often through assumptions regarding the EoS  \citep{Cherman:2009tw, Landry:2020vaw} or by using a non-parametric EoS approximation based on Gaussian processes in conjunction with X-ray NICER+XMM observations of  $\textit{PSR J0740+6620}$ \citep{Legred:2021hdx}. In contrast,as depicted in Figure \ref{Fig:Stability}\subref{fig:vr}, the confining threshold for the speed of sound, according to the indicated limit   $\textit{PSR J0740+6620}$ is not contravened  in   $f(\Ri, { \mathbb{T}})=\Ri+\beta_1, { \mathbb{T}}$ model from the center of the star to its boundary.

\noindent Constrain (\textbf{XI }): The  behavior  of stellar must have a stable behavior thus it should verify the equilibrium  limit, i.e.,  $-1< ({\mathrm  v_t^2}-{\mathrm v_r^2})/c^2 < 0$  in the stellar \cite{Herrera:1992lwz}.

\textit{It is clear that the stability and causality conditions} (\textbf{X}) and (\textbf{XI}) \textit{are verified for   $\textit{PSR J0740+6620}$ as  shown in Fig. \ref{Fig:Stability}}.

{\subsection{The  equilibrium of hydrodynamic }
Now we will employ  another extra  exam to test the equilibrium  of the  model in $f(\Ri, { \mathbb{T}})=\Ri+\alpha_1  { \mathbb{T}}$.  We will examine  the validity  of TOV equation \citep{Hansraj:2018jzb,Oppenheimer:1939ne} assuming that the sphere  is  in an equilibrium where the forces acting in the opposite directions on a star are in equilibrium with each other. The TOV equation can be formalized as follows to reflect the modified form caused by the newly introduced force, $F_T$:
\begin{equation}\label{eq:RS_TOV}
{\mathrm F_a}+{\mathrm F_g}+{\mathrm F_h}+{\mathrm F_T=0}\,,
\end{equation}
where  the gravitational force is denoted by ${\mathrm F_g}$ and the hydrostatic force is denoted by ${\mathrm F_h}$.  In this study we figured the various forces as follow:
\begin{eqnarray}\label{eq:Forces}
  {\mathrm F_a} &=&\frac{ 2{\mathrm  \Delta}}{\mathrm r} ,\qquad
  {\mathrm F_g} = -\frac{{\mathrm  M_g}}{r}({\mathrm  \rho c^2}+{\mathrm p_r})e^{{\mathrm  \gamma/2}} ,\qquad
  {\mathrm  F_h} =-{\mathrm  p'_r} ,\qquad
  {\mathrm F_T} = \frac{\beta_1}{3(1-\beta_1)}({\mathrm  c^2 \rho}'-{\mathrm p}'_r-2{\mathrm  p}'_t)\,.
\end{eqnarray}
In this context, we define ${\mathrm \gamma}=\alpha-\alpha_1$, and ${\mathrm M_g}$ is the mass of the isolated  systems, ${\mathrm V}$ (where $t=$ con.) is determined using the mass of Tolman that has the  form \citep{1930PhRv...35..896T}
\begin{eqnarray}\label{eq:grav_mass}
{\mathrm M_g(r)}&=&{\int_{\mathrm V}}\Big(\mathbb{{\mathrm T}}{^r}{_r}+\mathbb{{\mathrm T}}{^\theta}{_\theta}+\mathrm{T}{^\phi}{_\phi}-\mathrm{T}{^t}{_t}\Big)\sqrt{-g}\,dV\nonumber\\
&=&\frac{(e^{\alpha/2})'}{e^{\alpha}} e^{\alpha_1/2} r =\frac{\alpha'}{2} r e^{-\gamma/2}\,,
\end{eqnarray}
 with ${\mathrm F_g}=-\frac{\mu r}{R^2}({\mathrm \rho c^2}+{\mathrm p_r})$ being the force of gravity. Now we ready to put the stability condition the amended TOV equation.
%
%
%
\begin{figure}
\centering
\subfigure[~GR]{\label{fig:GRTOV}\includegraphics[scale=0.3]{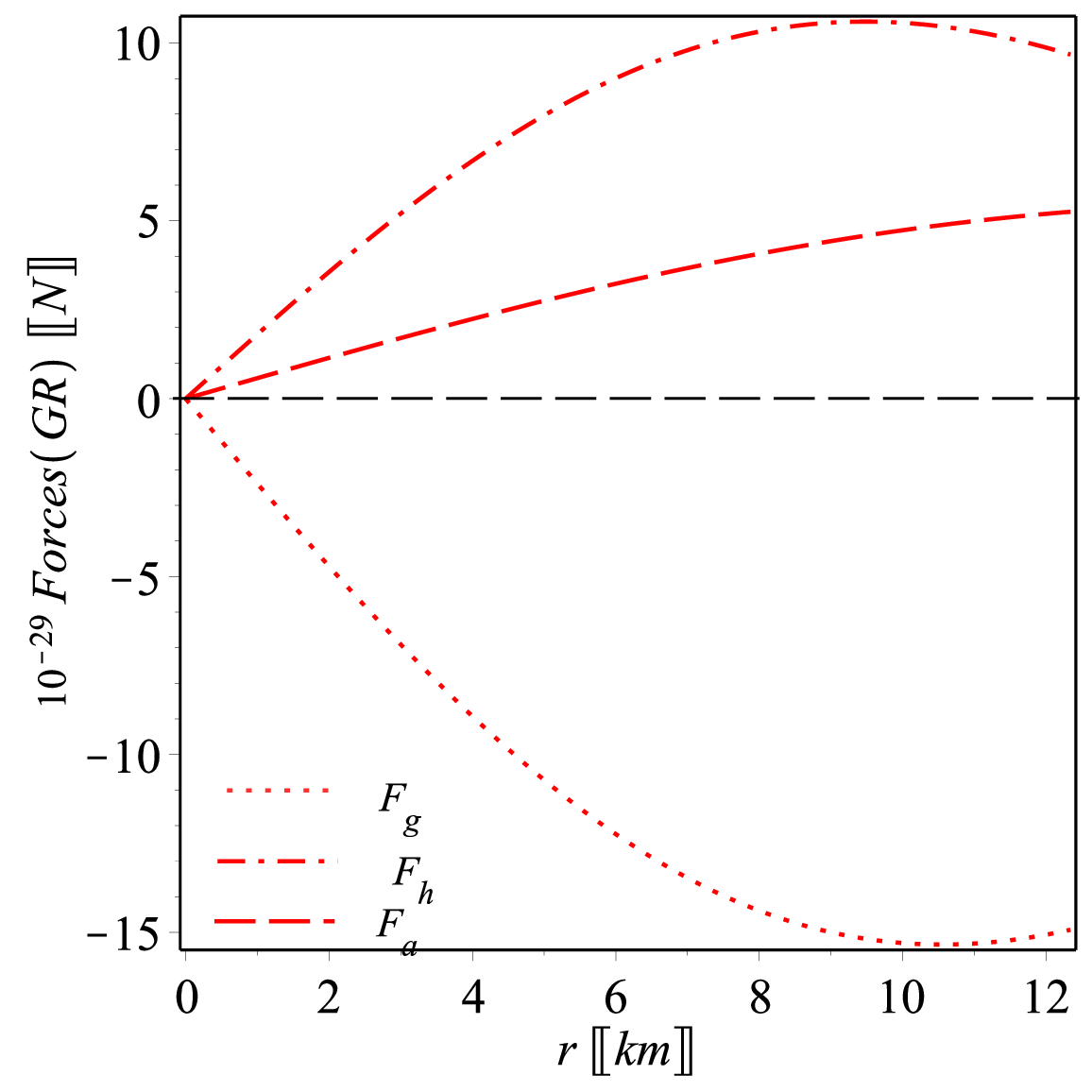}}
\subfigure[~$f(\Ri,{ \mathbb{T}})=\Ri+\beta_1\,{ \mathbb{T}}$]{\label{fig:RTTOV}\includegraphics[scale=0.3]{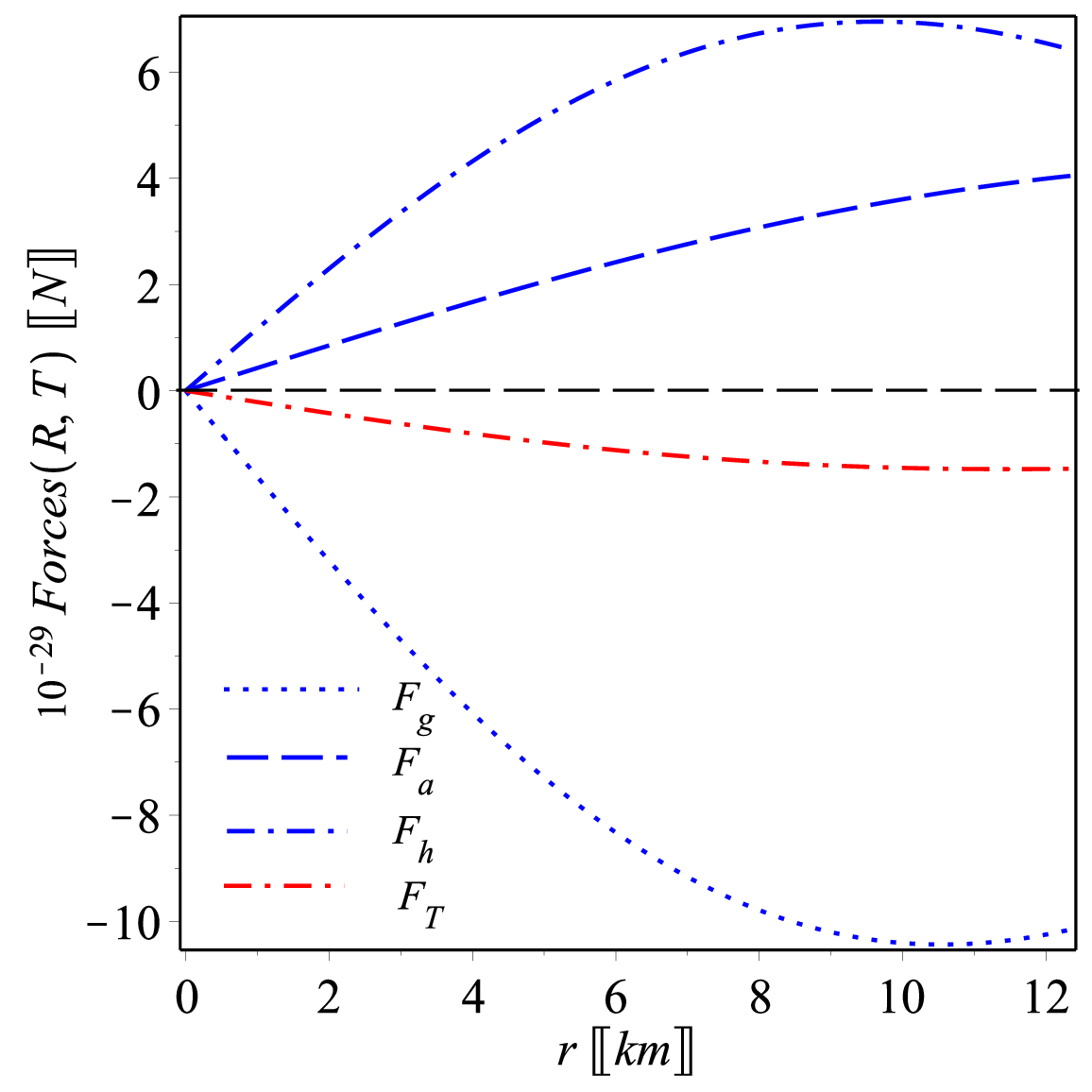}}
\caption{The various forces \eqref{eq:Forces} of the amended TOV equation given by Eq.~\eqref{eq:RS_TOV} of   $\textit{PSR J0740+6620}$.  The figures assure the hydrodynamic stability constrain (\textbf{XIII}) which is hold  true for   $\textit{PSR J0740+6620}$.}
\label{Fig:TOV}
\end{figure}
%



\noindent Constrain (\textbf{XII}):The anisotropic star achieves the equilibrium  when the total forces verified the amended TOV  \eqref{eq:RS_TOV}. We employ Eqs. \eqref{eq:Feqs2} and \eqref{eq:dens_grad}--\eqref{eq:pt_grad} to compute  the total forces \eqref{eq:Forces}   are plotted in Fig. \ref{Fig:TOV} for   the $f(\Ri,{ \mathbb{T}})$ and GR theories. The figures indicate that the   gravitational force (which has a negative direction) compensates the other forces to have a hydrodynamics equilibrium pattern .

\textit{It is obvious that the  fulfillment of condition of hydrodynamic equilibrium (\textbf{XII}) for $\textit{PSR J0740+6620}$}   \textit{is verified for  as shown in Fig. \ref{Fig:TOV}}.
\subsection{EoS}

By the use of the   data of    $\textit{PSR J0740+6620}$, through our derivation, we derived a satisfactory result for $\beta_1=0.1$, which connects the relationship of matter and geometry in the  spacetime. Therefore, we evaluate the central energy-density to be $\rho_R \approx 5.82\times 10^{14}$ g/cm$^{3}$ but at the surface it decreases up to $\rho_\text{core} \approx 3.28\times 10^{14}$ g/cm$^3$.
 The value of the central  density of   $\textit{PSR J0740+6620}$ assumes that  core consists of neutrons.
\begin{figure}[th!]
\centering
\subfigure[~Radial EoS]{\label{fig:REoS}\includegraphics[scale=0.3]{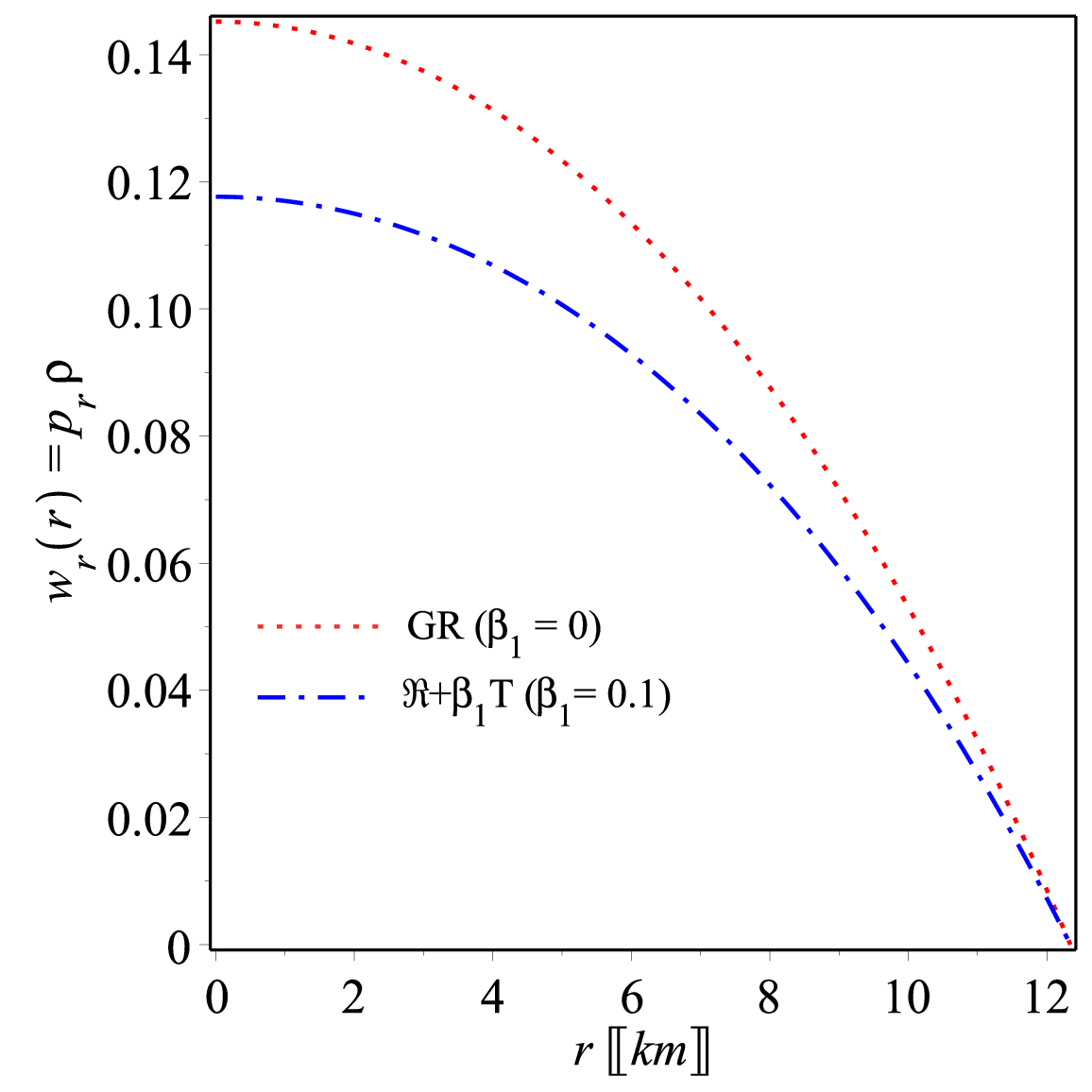}}
\subfigure[~Linear radial EoS]{\label{fig:OREoS}\includegraphics[scale=0.3]{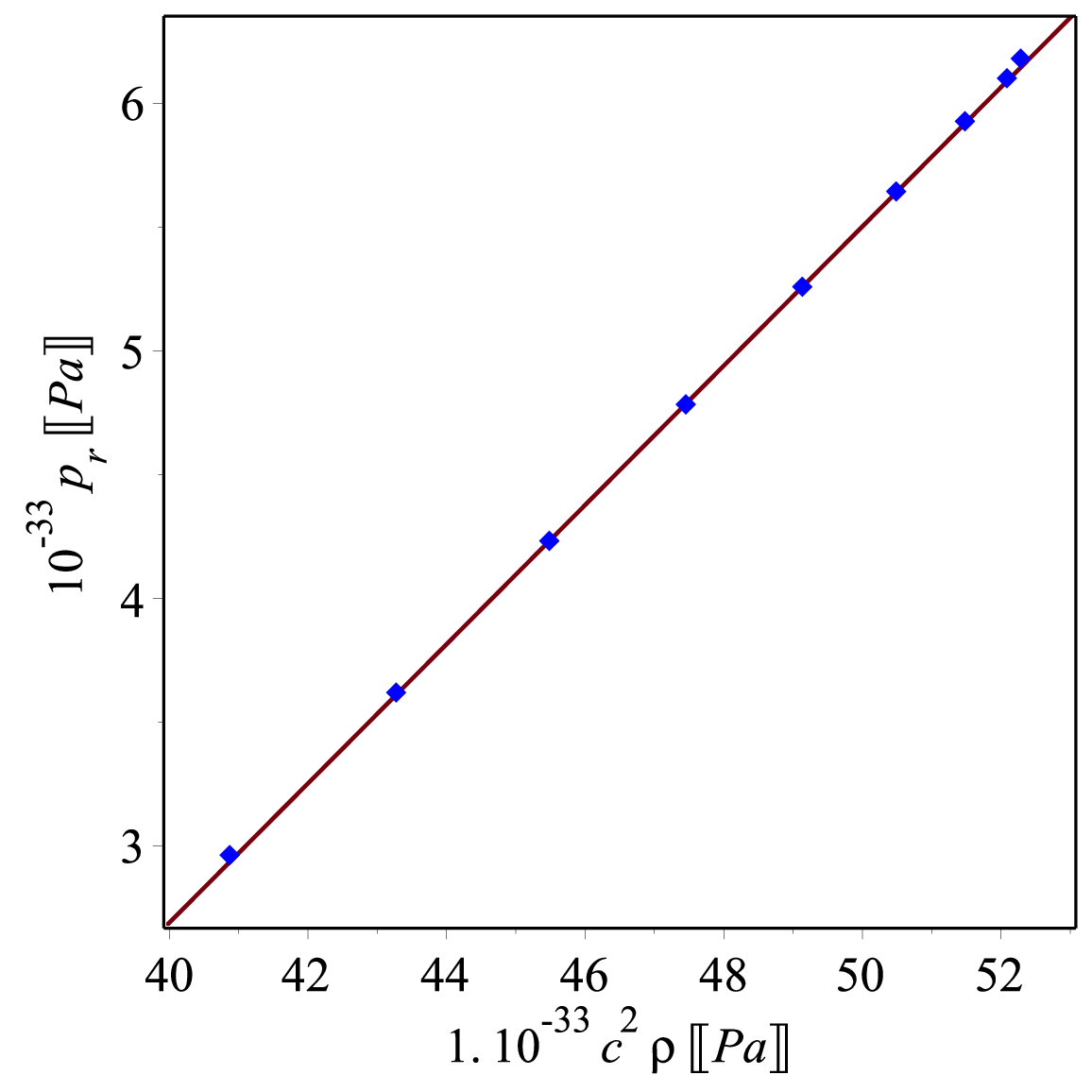}}
\subfigure[~Tangential EoS]{\label{fig:TEoS}\includegraphics[scale=.3]{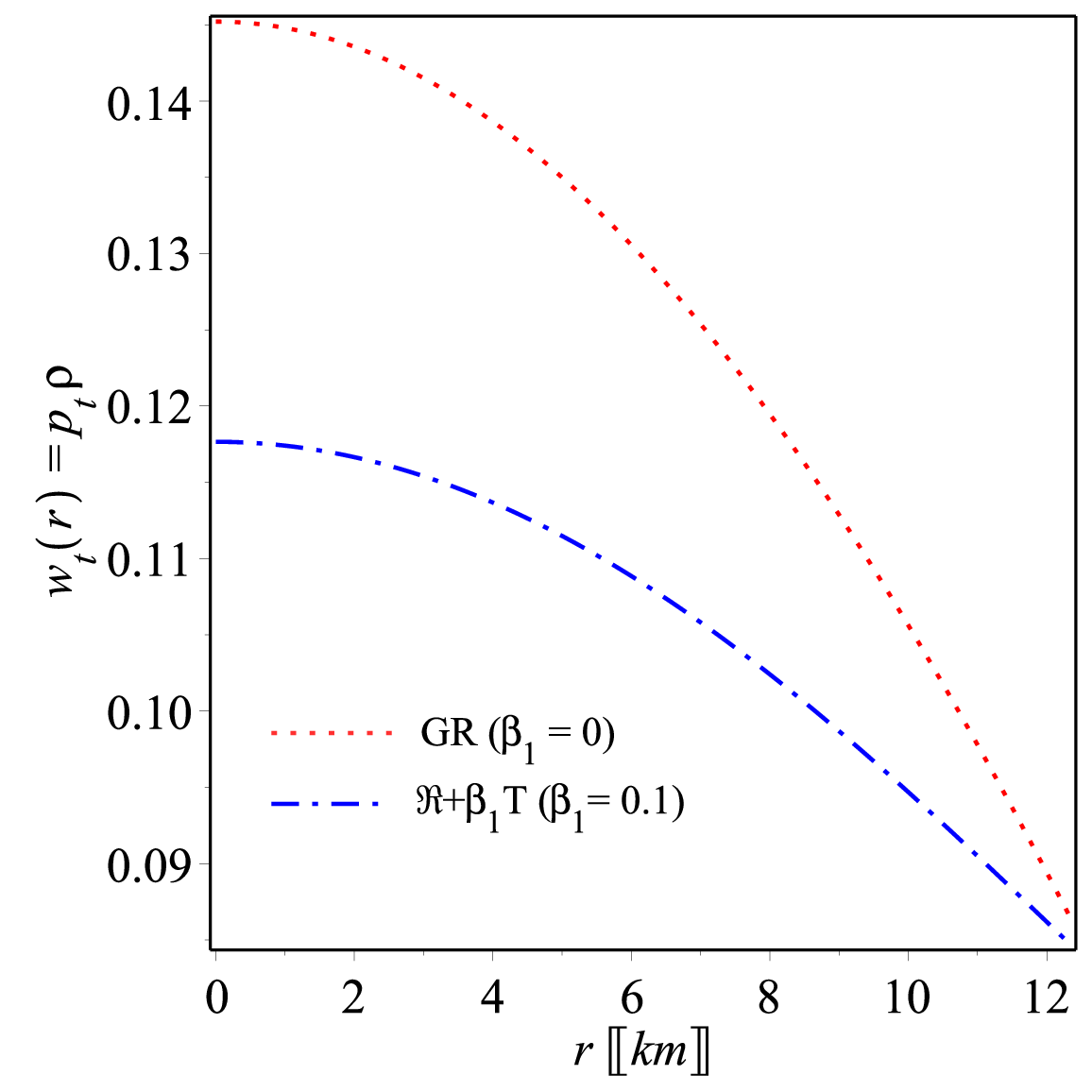}}
\subfigure[~Linear tangential EoS]{\label{fig:OTEoS}\includegraphics[scale=0.3]{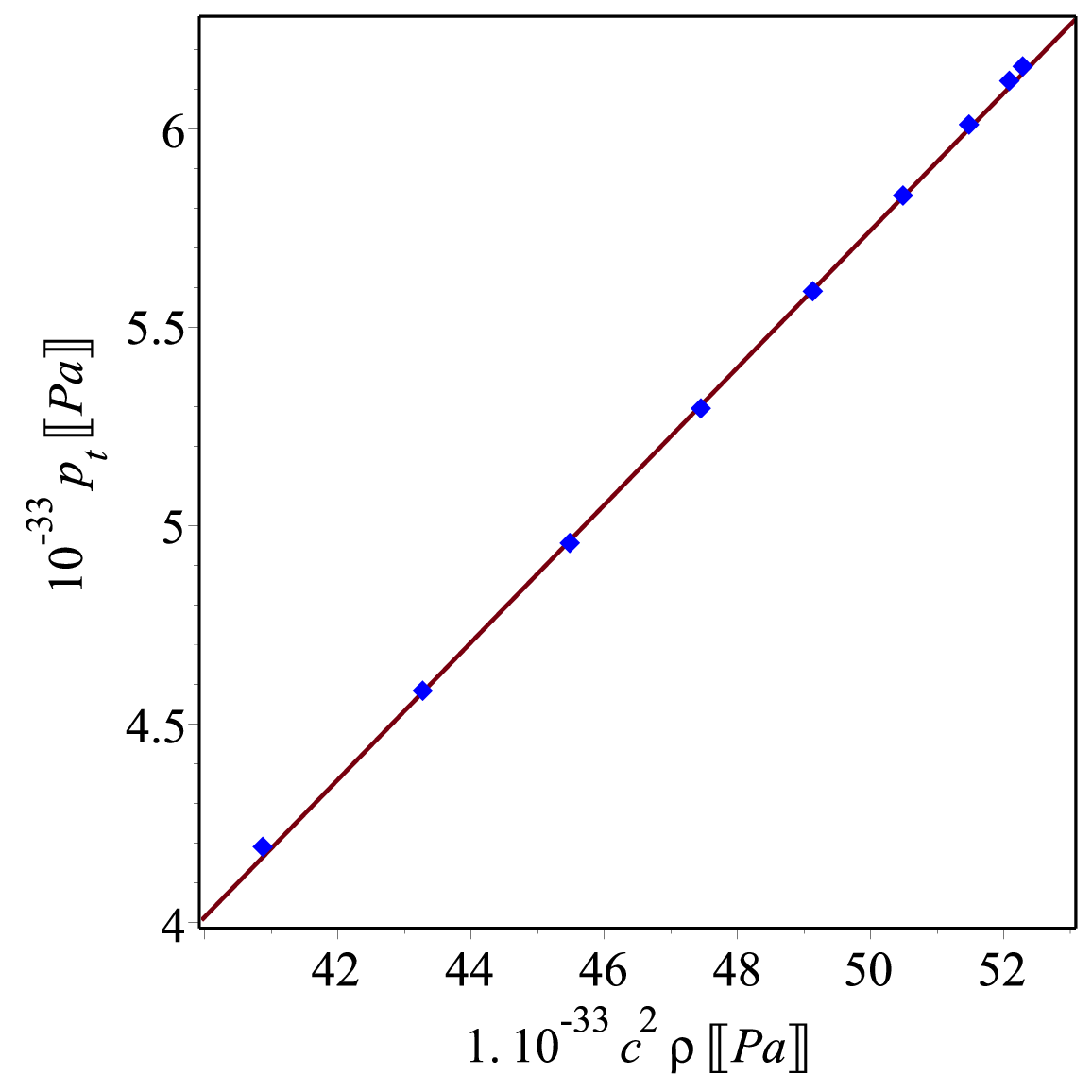}}
\caption{The pressures-density relationship of   {\textit PSRJ0740} is in excellent coherent  with the EoS in its linear form  with a bag constant. Additionally, $dp_1/d\rho=0.281c^2$ and $dp_2/d\rho=0.173c^2$ clearly validate that the nuclear matter within the neutron star adheres to the proposed limit on the sound speed based on the conjecture, i.e., $c_s^2\leq c^2/3$ throughout the star's interior.}
\label{Fig:EoS}
\end{figure}
In Figs. \ref{Fig:EoS}\subref{fig:REoS}, and \subref{fig:TEoS}, when $\beta_1=0$ and $\beta_1\neq 0$, we plot the evolution of the tangential and radial EoSs, i.e.,  $\mathrm {w_r(r)=p_r/\rho}$ and $\mathrm{w_t(r)=p_t/\rho}$ via the radial distance from the center. The figures show that the EoSs slightly change from the maximum at the center with a monotonically decreasing behavior to the surface of the pulsar, where $w_r$ approaches to zero at the surface  as constrain (\textbf{III}) stated. Furthermore, we notice that the behavior of  the EoSs  is more pronounced in $\Ri+\beta_1{ \mathbb{T}}$ gravity compared to GR. This occurrence can be ascribed to the matter-geometry interaction, which is responsible for the observed effect. Moreover, we should note that we did not suppose further  forms of the EoS in the model under construction, \ref{Fig:EoS}\subref{fig:REoS} and \subref{fig:TEoS}, nevertheless, we note that the EoS $p(\rho)$ in the radial and tangential directions are fit well with linear models, where the best fit are given by $p_r\approx 0.281\,\rho-8.56$ and $p_t\approx 0.173\, \rho-2.9$ for   $\textit{PSR J0740+6620}$. Surprisingly, the tangents  of  $dp_r/d\rho=0.281=v_r^2$ and $dp_t/d\rho=0.173=v_t^2$ are fitted lines.  Also, the current anisotropic NS pattern generates an EoS that  not extremely stiff but is following the EoS predicted from gravitational wave, where there isn't any obvious indication of a tidal deformation.
\section{Extra Observational limits}\label{S8}
Next, we will assess the model's applicability to a range of astrophysical observations by comparing it to data from other pulsars. Furthermore, we present the relation between  mass and radius   for various boundary density choices to demonstrate the model's reliability in predicting masses. Additionally, we compare the model's predicted compactness parameter to the Buchdahl compactness limit.
\subsection{Pulsars' data}
Furthermore, besides   $\textit{PSR J0740+6620}$, the same discussion  is applied to more 21 stars, with masses  from $0.8 M_\odot$  to  $2.01 M_\odot$  for heavy pulsars. Assuming, the form $\Ri+\beta_1 { \mathbb{T}}$ with $\beta_1=0.1$, in Table \ref{Table1}, we present   masses and radii, along with their   model parameters $\mu$, $\nu$, and $\lambda$. We guarantee that the pulsar masses predicted by this study l are consistent with the observed values. We present the determined values of  relevant  quantities in Table  \ref{Table2}. 
\begin{table*}
\caption{The measured mass-radius values of twenty pulsars along with their respective model parameters ($\beta=0.1$).}
\label{Table1}
\begin{tabular*}{\textwidth}{@{\extracolsep{\fill}}llcccccc@{}}
\hline\hline
\multicolumn{1}{c}{Star} &{Ref} &  mass & radius & evaluated mass & \multicolumn{1}{c}{$\mu$}
& \multicolumn{1}{c}{$\nu$}& \multicolumn{1}{c}{$\lambda$}\\
   & &  ($M_{\odot}$) &   [{km}] &  ($M_{\odot}$) &      &      &    \\
\hline
\multicolumn{8}{c}{X-ray binaries with high masses}\\
\hline
Her X-1         &\cite{Abubekerov:2008inw}        &  $0.85\pm 0.15$    &  $8.1\pm 0.41$   & $0.664$&  $0.184$    & $-0.555$    & $0.371$     \\
4U 1538-52      &\cite{Gangopadhyay:2013gha}   &  $0.87\pm 0.07$    &  $7.866\pm 0.21$ & $0.68$&  $0.2$    & $-0.596$    & $0.396$     \\
LMC X-4         &\cite{Rawls:2011jw}           &  $1.04\pm 0.09$    &  $8.301\pm 0.2$  & $0.816$&  $0.245$    & $-0.707$    & $0.462$     \\
Cen X-3         &\cite{Naik:2011qc}            &  $1.49\pm 0.49$    &  $9.178\pm 0.13$ & $1.182$&  $0.397$    & $-1.05$    & $0.653$     \\
Vela X-1      &\cite{Rawls:2011jw} &  $1.77\pm 0.08$    &  $9.56\pm 0.08$   & $1.42$&  $0.532$    & $-1.32$    & $0.792$     \\
\hline
\multicolumn{8}{c}{X-ray binaries with low masses exhibiting quiescence and thermonuclear bursts)}\\
\hline
EXO 1785-248    &\cite{Ozel:2008kb}            &  $1.3\pm 0.2$      &  $8.849\pm 0.4$  & $1.026$&  $0.325$    & $-0.895$    & $0.569$     \\
M13             &\cite{Webb:2007tc}            &  $1.38\pm 0.2$     &  $9.95\pm 0.27$  & $1.087$&  $0.292$    & $-0.820$    & $0.527$     \\
X7              &\cite{Bogdanov:2016nle}       &  $1.1\pm 0.35$     &  $12$            & $0.857$&  $0.151$    & $-0.467$    & $0.316$     \\
4U 1820-30      &\cite{Ozel:2015fia}           &  $1.46\pm 0.2$     &  $11.1\pm 1.8$   & $1.147$&  $0.266$    & $-0.758$    & $0.492$     \\
4U 1608-52      &\cite{1996IAUC.6331....1M}    &  $1.57\pm 0.3$     &  $9.8\pm 1.8$    & $1.244$&  $0.386$    & $-1.027$    & $0.641$     \\
KS 1731-260     &\cite{Ozel:2008kb}            &  $1.61\pm 0.37$    &  $10\pm 2.2$     & $1.277$&  $0.39$    & $-1.035$    & $0.646$     \\
EXO 1745-268    &\cite{Ozel:2008kb}            &  $1.65\pm 0.25$    &  $10.5\pm 1.8$   & $1.31$&  $0.371$    & $-0.995$    & $0.624$     \\
4U 1724-207     &\cite{Ozel:2008kb}            &  $1.81\pm 0.27$    &  $12.2\pm 1.4$   & $1.43$&  $0.331$    & $-0.908$    & $0.577$     \\
SAX J1748.9-2021&\cite{Ozel:2008kb}            &  $1.81\pm 0.3$     &  $11.7\pm 1.7$   & $1.432$&  $0.36$    & $-0.97$    & $0.611$     \\
\hline
\multicolumn{8}{c}{Millisecond Pulsars}\\
\hline
PSR J0030+0451  &\cite{Raaijmakers:2019qny}    &  $1.34\pm 0.16$    &  $12.71\pm 1.19$ & $1.0467$&  $0.186$    & $-0.559$    & $0.373$     \\
PSR J0030+0451  &\cite{Miller:2019cac}         &  $1.44\pm 0.16$    &  $13.02\pm 1.24$ & $1.126$&  $0.2$    & $-0.596$    & $0.396$     \\
PSR J0437-4715  &\cite{Reardon:2015kba} \& \cite{Gonzalez-Caniulef:2019wzi}       &  $1.44\pm 0.07$    &  $13.6\pm 0.9$   & $1.125$&  $0.187$    & $-0.562$    & $0.375$     \\
PSR J1614-2230  &\cite{NANOGrav:2017wvv}       &  $1.908\pm 0.016$  &  $13\pm 2$       & $1.51$&  $0.325$    & $-0.893$     & $0.569$     \\
PSR J0348+0432  &\cite{Antoniadis:2013pzd}     &  $2.01\pm 0.04$    &  $13\pm 2$       & $1.59$&  $0.359$    & $-0.969$    & $0.61$     \\
\hline
\multicolumn{8}{c}{Gravitational-wave Signals}\\
\hline
LIGO-Virgo      &\cite{LIGOScientific:2020zkf} &  $1.4$             &  $12.9\pm 0.8$   & $1.094$&  $0.194$    & $-0.58$    & $0.387$     \\
GW170817-1      &\cite{LIGOScientific:2018cki} &  $1.45\pm 0.09$    &  $11.9\pm 1.4$   & $ 1.137$&  $0.234$    & $-0.68$    & $0.446$     \\
GW170817-2      &\cite{LIGOScientific:2018cki} &  $1.27\pm 0.09$    &  $11.9\pm 1.4$   & $0.99$&  $0.19$    & $-.568$    & $0.379$     \\
\hline
\end{tabular*}
\end{table*}

In accordance with the criteria listed in the preceding section, we further assess the model's stability. Table \ref{Table2} presents the essential physical quantities of interest. The results showcase the model's ability to predict stable compact star structures that align with observations.  It is intriguing to observe that the speeds of sound  obtained from this study for all categories align with the conformal limit $c^2 s \leq c^2/3$. It is worth noting that the version of this model predicted $c^2_s=0.34$ at the core above the conformal limit, as derived by \cite{Roupas:2020mvs}, particularly concerning low-mass of $\textit{4U 1608-52}$ and $\textit{KS 1731-260}$. The findings presented in Table \ref{Table2} confirm that the present study ensures avoidance of sound speed violations, aligning with our previous discourse regarding the pulsar's sound speed $\textit{PSR J0740+6620}$. This underscores the possible influence of the interaction between geometry and matter in maintaining this upper limit across the entirety of the compact object.
\begin{table*}
\caption{The computed essential physical parameters}
\label{Table2}
\begin{tabular*}{\textwidth}{@{\extracolsep{\fill}}l|cc|cc|cc|cc|cc@{\extracolsep{\fill}}}
\hline\hline
\multicolumn{1}{c|}{Star}                              &\multicolumn{2}{c|}{$\rho$ [g/cm$^3$]} &   \multicolumn{2}{c|}{$v_r^2/c^2$}  &   \multicolumn{2}{c|}{$v_t^2/c^2$}  & \multicolumn{2}{c|}{DEC [$dyne/cm^2$]} & \multicolumn{1}{c}{$Z_R$}\\ \cline{2-10}
                                        &\multicolumn{1}{c}{core}            &        boundary   &   core        &    boundary      &  core          &  boundary       &  core     & boundary       & {}\\
\hline
\multicolumn{10}{c}{X-ray binaries with high-mass companions}\\
\hline
Her X-1             &7.15$\times10^{14}$     &5.22$\times10^{14}$  &  0.221   &   0.201     &  0.119 & 0.09 & 7.53$\times10^{34}$ & 5.24$\times10^{34}$ & 0.204  \\
4U 1538-52          &8.1$\times10^{14}$     &5.79$\times10^{14}$  &  0.226   &   0.204     &  0.124 & 0.094 & 8.62$\times10^{34}$ & 5.86$\times10^{34}$ & 0.219  \\
LMC X-4             &8.54$\times10^{14}$     &5.77$\times10^{14}$  &  0.239   &   0.213     &  0.137 & 0.1&9.37$\times10^{34}$ & 5.98$\times10^{34}$ & 0.26  \\
Cen X-3             &10$\times10^{14}$     &5.79$\times10^{14}$  &  0.28   &   0.24     &  0.176 & 0.129 & 1.2$\times10^{35}$ & 6.44$\times10^{34}$ & 0.386  \\
Vela X-1       &11.3$\times10^{14}$     &5.86$\times10^{14}$  &  0.314   &   0.262     &  0.205 & 0.152 & 14.5$\times10^{34}$ & 6.92$\times10^{34}$ & 0.486  \\
\hline
\multicolumn{10}{c}{Binaries with low-mass companions, exhibiting quiescence and thermonuclear bursts}\\
\hline
EXO 1785-248        &9.32$\times10^{14}$     &5.77$\times10^{14}$  &  0.261   &   0.227     &  0.159 & 0.116 & 10.7$\times10^{34}$ & 6.22$\times10^{34}$ & 0.329  \\
M13                 &6.81$\times10^{14}$     &4.36$\times10^{14}$  &  0.252   &   0.222     &  0.15 & 0.111 & 7.69$\times10^{34}$ & 4.63$\times10^{34}$ & 0.302  \\
X7                  &2.76$\times10^{14}$     &2.11$\times10^{14}$  &  0.211   &   0.194     &  0.108 & 0.084 & 2.84$\times10^{34}$ & 2.08$\times10^{34}$ & 0.171  \\
4U 1820-30          &5.09$\times10^{14}$     &3.36$\times10^{14}$  &  0.245   &   0.217     &  0.143 & 0.106 & 5.66$\times10^{34}$ & 3.52$\times10^{34}$ & 0.279  \\
4U 1608-52          &8.61$\times10^{14}$     &5.03$\times10^{14}$  &  0.277   &   0.238     &  0.173 & 0.127 & 10.3$\times10^{34}$ & 5.57$\times10^{34}$ & 0.378  \\
KS 1731-260         &8.33$\times10^{14}$     &4.85$\times10^{14}$  &  0.278   &   0.239     &  0.174 & 0.128 & 9.95$\times10^{34}$ & 5.38$\times10^{35}$ & 0.381  \\
EXO 1745-268        &7.29$\times10^{14}$     &4.32$\times10^{14}$  &  0.273   &   0.235     &  0.17 & 0.124 & 8.62$\times10^{34}$ & 4.75$\times10^{34}$ & 0.366  \\
4U 1724-207         &4.97$\times10^{14}$     &3.06$\times10^{14}$  &  0.263   &   0.229     &  0.16 & 0.117 & 5.75$\times10^{34}$ & 3.31$\times10^{34}$ & 0.334  \\
SAX J1748.9-2021    &5.74$\times10^{14}$     &3.44$\times10^{14}$  &  0.27   &   0.233     &  0.167 & 0.122 & 6.74$\times10^{34}$ & 3.76$\times10^{34}$ & 0.357  \\
\hline
\multicolumn{10}{c}{Millisecond Pulsars}\\
\hline
PSR J0030+0451      &2.92$\times10^{14}$     &2.13$\times10^{14}$  &  0.221   &   0.201     &  0.119 & 0.091 & 3.08$\times10^{34}$ & 2.14$\times10^{34}$ & 0.205  \\
PSR J0030+0451      &2.96$\times10^{14}$     &2.11$\times10^{14}$  &  0.262   &   0.204     &  0.124 & 0.094 & 3.15$\times10^{34}$ & 2.14$\times10^{34}$ & 0.219  \\
PSR J0437-4715      &2.56$\times10^{14}$     &1.86$\times10^{14}$  &  0.222   &   0.202     &  0.120 & 0.091 & 2.71$\times10^{34}$ & 1.88$\times10^{34}$ & 0.206  \\
PSR J1614-2230      &4.31$\times10^{14}$     &2.67$\times10^{14}$  &  0.261   &   0.227     &  0.158 & 0.116 & 4.97$\times10^{34}$ & 2.88$\times10^{34}$ & 0.329  \\
PSR J0348+0432      &4.65$\times10^{14}$     &2.78$\times10^{14}$  &  0.27   &   0.233     &  0.167 & 0.122 & 5.46$\times10^{34}$ & 3.04$\times10^{34}$ & 0.357 \\
\hline
\multicolumn{10}{c}{Gravitational-wave Signals}\\
\hline
LIGO-Virgo          &2.94$\times10^{14}$     &2.12$\times10^{14}$  &  0.224   &   0.203     &  0.122 & 0.093 & 3.12$\times10^{34}$ & 2.14$\times10^{34}$ & 0.213  \\
GW170817-1          &4$\times10^{14}$     &2.74$\times10^{14}$  &  0.235   &   0.211     &  0.134 & 0.1 & 4.36$\times10^{34}$ & 2.83$\times10^{34}$ & 0.25
  \\
GW170817-2          &3.38$\times10^{14}$     &2.45$\times10^{14}$  &  0.222   &   0.202     &  0.121 & 0.092 & 3.58$\times10^{34}$ & 2.47$\times10^{34}$ & 0.208\\
\hline
\end{tabular*}
\caption{The speed of sound the center of 4U 1608-52 and KS 1731-260 aligns with the proposed limit on the conformal sound speed, $c_s^2\leq c^2/3$, in contradistinction to the version in GR \citep{Roupas:2020mvs}. This observation validates the significance of the coupling of matter-geometry in verified the upper limit for the conformal sound speed.}
\end{table*}

\section{Mass-Radius relation}\label{MR}
We derive  boundary densities $1.86\times 10^{14} \lesssim \rho_{R} \lesssim 5.86\times 10^{14}$ g/cm$^{3}$ as shown in Table \ref{Table2}; and therefore, we select  three boundary densities  as initial  conditions: $\rho_R=\rho_\text{nuc}=2\times 10^{14}$ g/cm$^3$ at nuclear saturation density, $\rho_R=3.5\times 10^{14}$ g/cm$^3$ and $\rho_R=5.5\times 10^{14}$ g/cm$^3$ to cover the density of nuclear solidification. Therefore, with $\beta_1=0.1$, we establish a relationship between ss $C$ and   $R$ for any surface constraint on the boundary density, utilizing Eq.~\eqref{eq:Feqs2}, where $\rho(r=R)=\rho_R$.  In Figs. \ref{Fig:CompMRGR}  and \ref{Fig:CompMR}, we draw the relation between compactness and radius curves using  certain surface density for the GR case and $f(\Ri,{ \mathbb{T}})=\Ri+\beta_1{ \mathbb{T}}$ and get $C=1$ and $0.868$ respectively.
%
\begin{figure*}
\centering
\subfigure[~Figure of Compactness and Radius relation of the GR case]{\label{fig:CompGR}\includegraphics[scale=0.28]{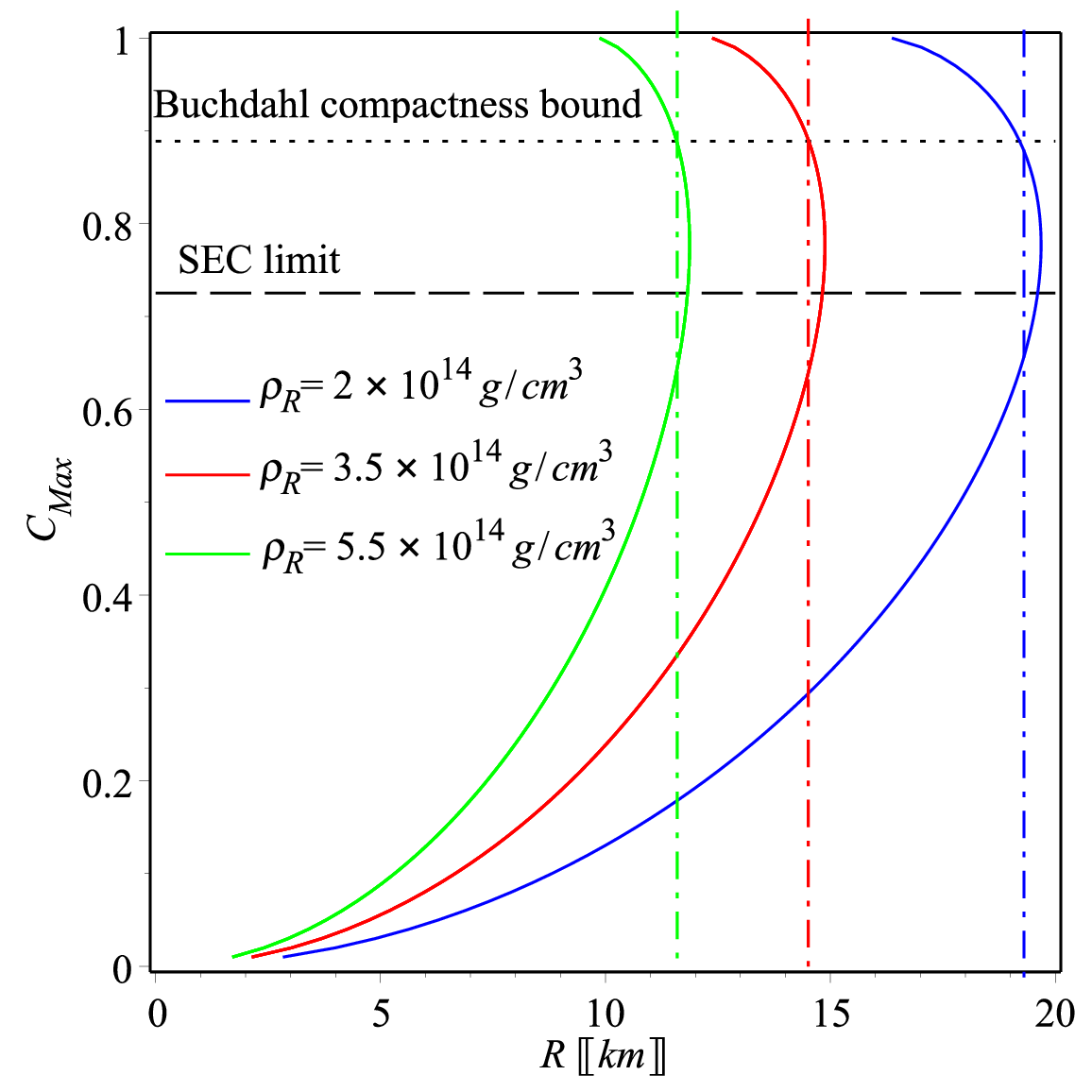}}
\subfigure[~Figure of Mass and Radius relation of the GR case]{\label{fig:MRGR}\includegraphics[scale=.28]{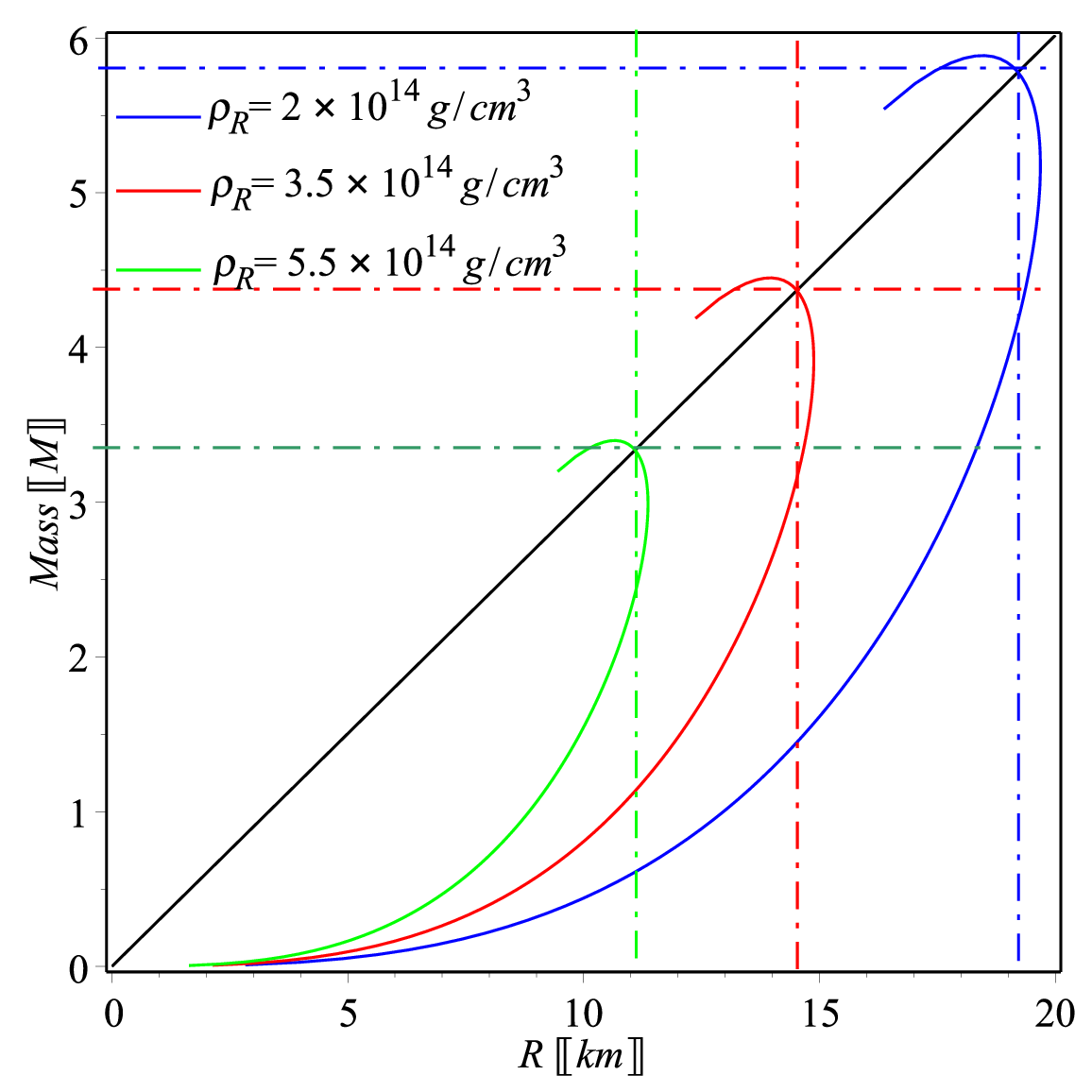}}
\subfigure[~LIGO-Virgo and NICER  restrictions of the GR case]{\label{fig:NICERGR}\includegraphics[scale=.28]{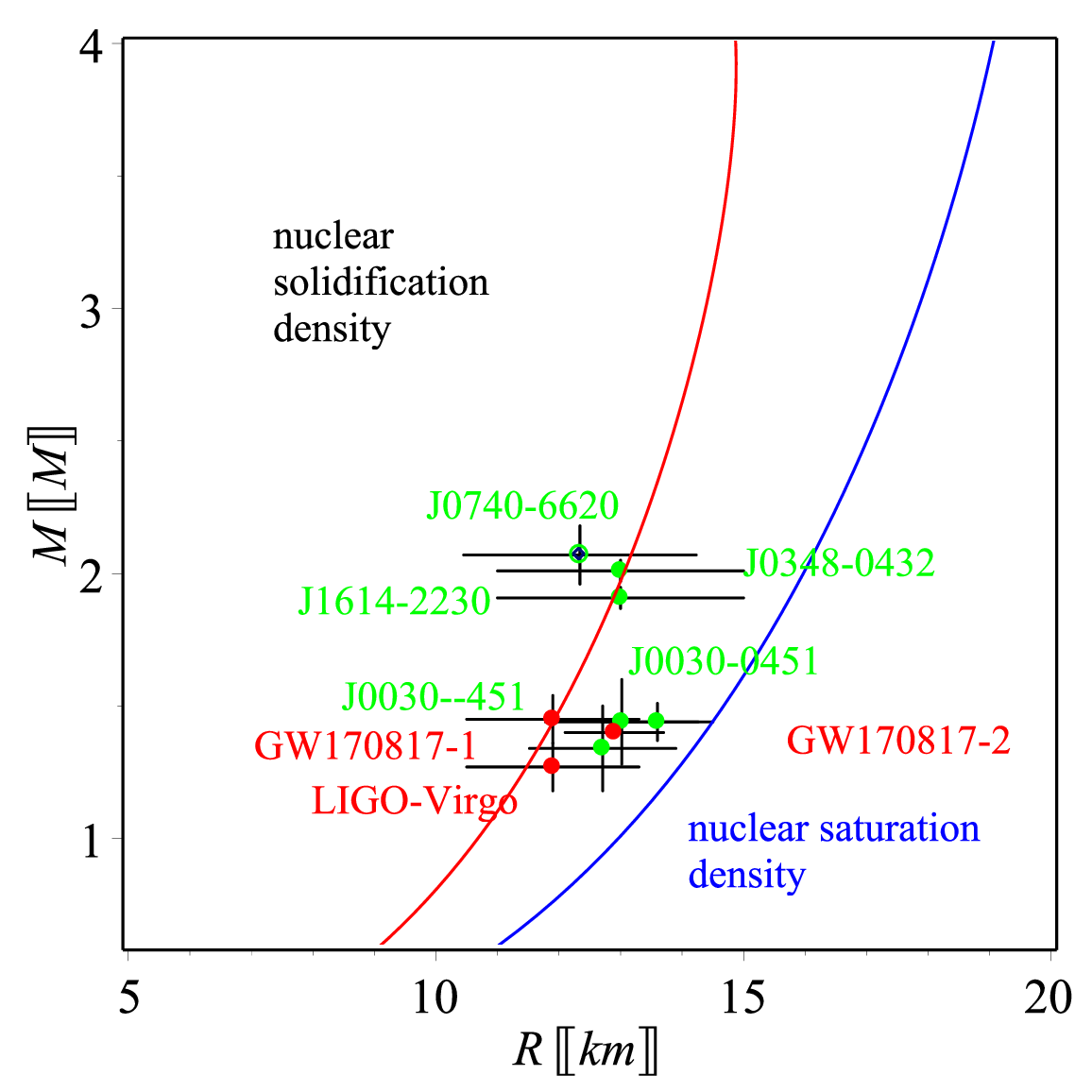}}
\caption{\subref{fig:Comp} The compactness-radius curves of the GR case.
\subref{fig:MR} The mass-radius curves of the GR case.
Green circles represent MSP,  solid red circles represent GW signals, and blue circles represent  X-ray binaries as given from Table \ref{Table1}.
\subref{fig:NICER} A closer look at some  of the most interesting pulsars. }
\label{Fig:CompMRGR}
\end{figure*}
\begin{figure*}
\centering
\subfigure[~Figure of Compactness and Radius relation]{\label{fig:Comp}\includegraphics[scale=0.28]{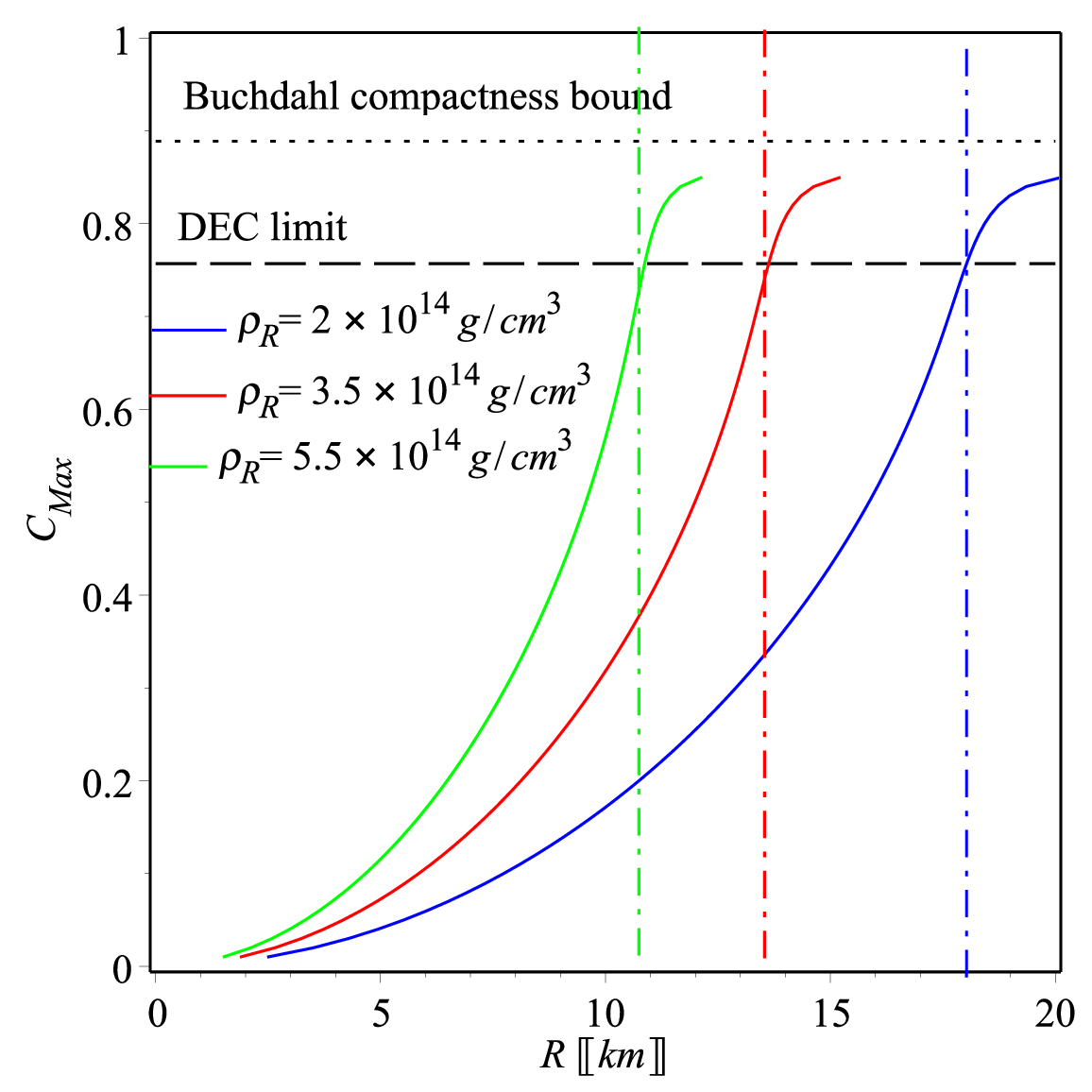}}
\subfigure[~Figure of Mass and Radius relation]{\label{fig:MR}\includegraphics[scale=.28]{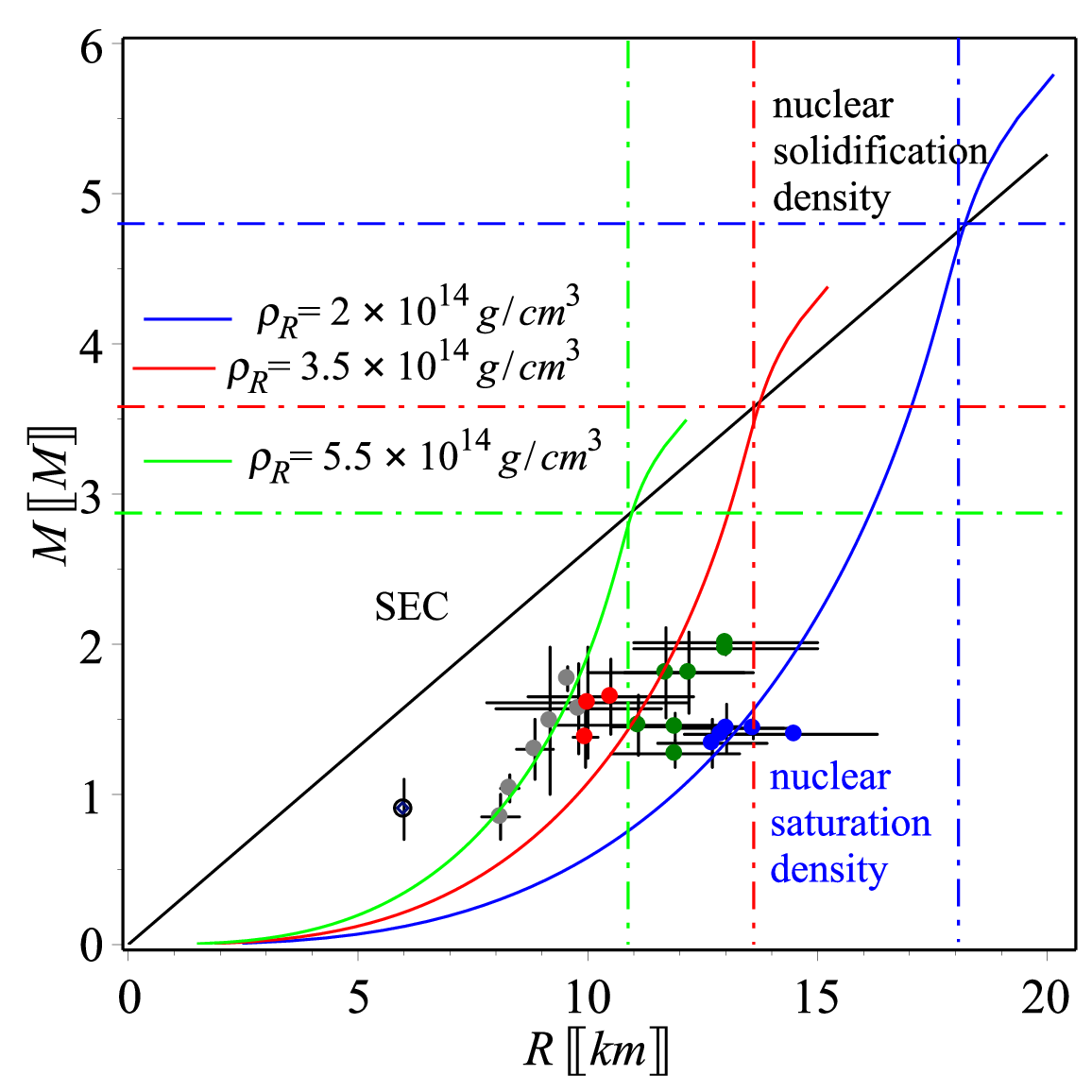}}
\subfigure[~LIGO-Virgo and NICER  restrictions]{\label{fig:NICER}\includegraphics[scale=.28]{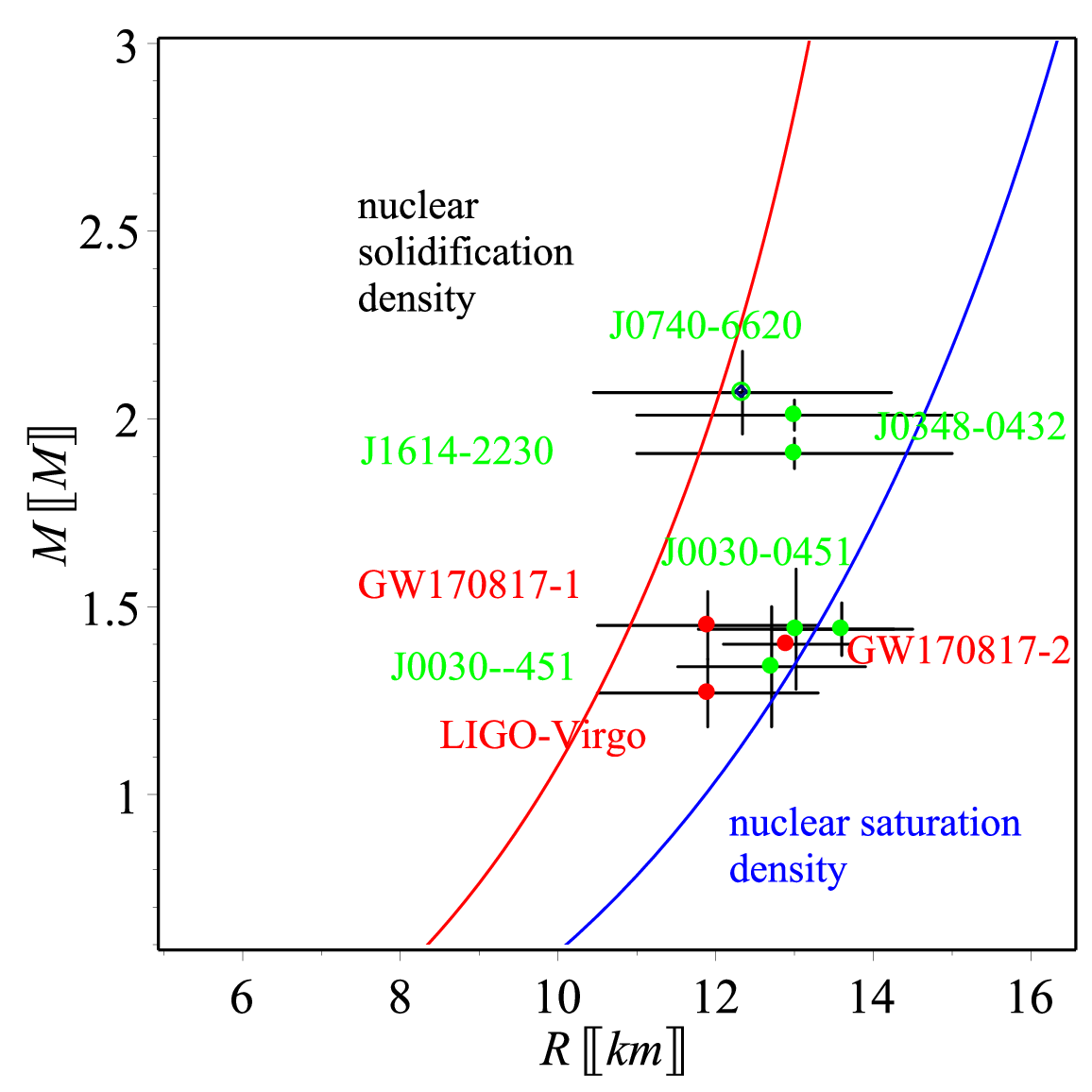}}
\caption{\subref{fig:Comp} The relation between compactness and radius. Fig. \ref{fig:MR}, the relation between  mass and radius curves. In Fig. \ref{fig:NICER}, we focus on some of the most intriguing pulsars in the context of GR.}
\label{Fig:CompMR}
\end{figure*}
%

Next, the mass-radius curves for every  class are then shown,  with its the relevant observational data using Table \ref{Table1} as fixed by Figs. \ref{Fig:CompMRGR} \subref{fig:MRGR} and \ref{Fig:CompMR} \subref{fig:MR}   for the GR situation, and for $f(\Ri,{ \mathbb{T}})=\Ri+\beta_1{ \mathbb{T}}$ respectively. With the help of the DEC condition (shown by  line), we determine the highest  permitted masses  $M_\text{max}= 5.9  M_\odot$  and  $M_\text{max}=4.86  M_\odot$ at  maximum radii of  $R_\text{max}=19.6$ km and $R_\text{max}=18.3$ km, respectively, for the surface density  $\rho_\text{nuc}=2\times 10^{14}$ g/cm$^3$ for the GR case and $f(\Ri,{ \mathbb{T}})=\Ri+\beta_1{ \mathbb{T}}$ respectively \citep{Roupas:2020mvs}. With a positive parameter value of $\beta_1=0.1$,  $f(\Ri,{ \mathbb{T}})=\Ri+\beta_1 { \mathbb{T}}$   predicts roughly the same mass inside a $\sim 1.3$ km smaller size than GR predictions. Similar to this, for the surface densities  $\rho_R=5.5\times 10^{14}$ g/cm$^3$ and $\rho_R=5.5\times 10^{14}$ g/cm$^3$, respectively, we ascertain the utmost masses and radii in the subsequent manner: ($M_\text{max}=3.57 M \odot, R_\text{max}=13.7$ km) and ($M_\text{max}=2.83 M_\odot, R_\text{max}=10.85$ km). These values align with the nuclear solidification density, demonstrating compatibility.
It is noteworthy that the present study can yield neutron stars (NS) in the mass range of $2.83$ to $4.86 M_\odot$. In this particular scenario, the model predicts that the NS possesses a boundary density consistent at the density of nuclear saturation and follows a linear equation of state (EoS).

In Fig. \ref{Fig:CompMR} \subref{fig:NICERGR}, we focus on the investigation of the most intriguing pulsars. The stares $\textit{PSR J0740+6620}$ and $\textit{PSR J0030+0451}$ are found to fix the $\rho_\text{nuc}$-curve in a precise manner.

\section{Conclusion}\label{Sec:Conclusion}

According to \cite{Harko:2011kv}, we looked into the effects of the non-minimal connection between matter and geometry, for $f(\Ri,{ \mathbb{T}})=\Ri+\beta_1{ \mathbb{T}}$ gravitational theory, on the relationship between mass and radius for compact objects. In  presence of curved spacetime, the theory pre supposes a local violation of the law of conservation of energy; otherwise, it degenerates into GR. Considering the substantial curvature of spacetime, it becomes crucial to thoroughly investigate this effect on the structures of compact objects such as neutron stars.The accurate measurements of the radius and mass of $\textit{PSR J0740+6620}$, obtained through NICER observations, have been instrumental in determining a robust estimate for the parameter $\beta_1$.

Our demonstration illustrated that $\Delta$ in $ \Ri+\beta_1{ \mathbb{T}}$ gravity is equivalent to that in  GR . The equivalence between the physical quantities simplifies the detection of deviations from general relativity (GR) with regards to matter-geometry coupling. To represent all the relevant physical quantities, we utilized the KB ansatz and incorporated the parameters $\beta_1$  $C$. Due to the accurate  observational constraints on radius and mass obtained for $\textit{PSR J0740+6620}$, we determined the parameter $\beta_1$ with a fixed positive value of $\beta=0.1$. In contrast to GR, this situation implies a larger size for a specific mass in the framework of $ \Ri+\beta_1{ \mathbb{T}}$.  We have demonstrated that, for a specific mass, the additional force from the gravitational theory $f(\Ri,{ \mathbb{T}})=\Ri+\beta_1{ \mathbb{T}}$ assists in partially offsetting the gravitational force, allowing for the existence of larger compact stars than would be possible otherwise. Moreover, we have shown that the Schwarzschild radius approaches the limit of the compactness  i.e.,  $C=1$ for the isotropic case, slightly surpassing the limit of $C=8/9$ for the GR case which provided by Buchdahl, as depicted in Fig. \ref{Fig:CompMRGR}\subref{fig:CompGR}. The highest allowed compactness for $f(\Ri,{ \mathbb{T}})=\Ri+\beta_1{ \mathbb{T}}$ is found to be $C_\text{max}=0.868$, which is still lower than the Buchdahl limit, i.e., $C=8/9$.

Remarkably, even without imposing any specific  EoS, the model conforms a linear EoS characterized by a bag constant. What's even more intriguing is that we have discovered $c_s$ of NS at the center to be $v^2_{r}=0.332 c^2$ (radial direction) and $v^2_{t}=0.222 c^2$ (tangential direction). Unlike  GR, these values comply while adhering to the suggested confining threshold for the sound speed, guaranteeing $c_s^2\leq c^2/3$ within the core and interior of the neutron star

When considering a surface density corresponding to the saturation nuclear energy $\rho_\text{nuc}=2\times 10^{14}$ g/cm$^3$, the model allows mass of $M=4.86 M_\odot$ with  radius  $R=18.3$ km. This results in a smaller compactness value compared to the GR prediction, as evident from Figs. \ref{Fig:CompMRGR}\subref{fig:CompGR} and \ref{Fig:CompMR}\subref{fig:Comp}. Consequently, this finding leaves the possibility open for the companion of $GW190814$ to be an anisotropic  NS with necessitating any assumptions about exotic matter sources

The current investigations demonstrate that the matter-geometry coupling in $f(\Ri,{ \mathbb{T}})=\Ri+\beta_1{ \mathbb{T}}$ theory enables the speed of sound  to match the conformal upper bound, a scenario not observed in GR. This crucial finding, along with similar studies, confirms the disparity between $f(\Ri,{ \mathbb{T}})=\Ri+\beta_1{ \mathbb{T}}$ and GR.

In summary, we have successfully derived a realistic stellar model within the framework of $f(R,T)$ theory. Now, the question arises whether the same procedure can be applied to $f({\cal T},T)$ theory \cite{ElHanafy:2014tzj}, where ${\cal T}$ represents the scalar torsion. Can we utilize a similar approach in $f({\cal T},T)$ theory, incorporating the scalar torsion ${\cal T}$, to derive a realistic stellar model?


\begin{thebibliography}{0}%
\makeatletter
\providecommand \@ifxundefined [1]{%
 \@ifx{#1\undefined}
}%
\providecommand \@ifnum [1]{%
 \ifnum #1\expandafter \@firstoftwo
 \else \expandafter \@secondoftwo
 \fi
}%
\providecommand \@ifx [1]{%
 \ifx #1\expandafter \@firstoftwo
 \else \expandafter \@secondoftwo
 \fi
}%
\providecommand \natexlab [1]{#1}%
\providecommand \enquote  [1]{``#1''}%
\providecommand \bibnamefont  [1]{#1}%
\providecommand \bibfnamefont [1]{#1}%
\providecommand \citenamefont [1]{#1}%
\providecommand \href@noop [0]{\@secondoftwo}%
\providecommand \href [0]{\begingroup \@sanitize@url \@href}%
\providecommand \@href[1]{\@@startlink{#1}\@@href}%
\providecommand \@@href[1]{\endgroup#1\@@endlink}%
\providecommand \@sanitize@url [0]{\catcode `\\12\catcode `\$12\catcode
  `\&12\catcode `\#12\catcode `\^12\catcode `\_12\catcode `\%12\relax}%
\providecommand \@@startlink[1]{}%
\providecommand \@@endlink[0]{}%
\providecommand \url  [0]{\begingroup\@sanitize@url \@url }%
\providecommand \@url [1]{\endgroup\@href {#1}{\urlprefix }}%
\providecommand \urlprefix  [0]{URL }%
\providecommand \Eprint [0]{\href }%
\providecommand \doibase [0]{http://dx.doi.org/}%
\providecommand \selectlanguage [0]{\@gobble}%
\providecommand \bibinfo  [0]{\@secondoftwo}%
\providecommand \bibfield  [0]{\@secondoftwo}%
\providecommand \translation [1]{[#1]}%
\providecommand \BibitemOpen [0]{}%
\providecommand \bibitemStop [0]{}%
\providecommand \bibitemNoStop [0]{.\EOS\space}%
\providecommand \EOS [0]{\spacefactor3000\relax}%
\providecommand \BibitemShut  [1]{\csname bibitem#1\endcsname}%
\let\auto@bib@innerbib\@empty
\end{thebibliography}%


\begin{thebibliography}{99}

\bibitem{Bogdanov:2019ixe}
S.~Bogdanov, S.~Guillot, P.~S.~Ray, M.~T.~Wolff, D.~Chakrabarty, W.~C.~G.~Ho, M.~Kerr, F.~K.~Lamb, A.~Lommen and R.~M.~Ludlam, \textit{et al.}
Astrophys. J. Lett. \textbf{887}, no.1, L25 (2019)
doi:10.3847/2041-8213/ab53eb
[arXiv:1912.05706 [astro-ph.HE]].

\bibitem{Bogdanov:2019qjb}
S.~Bogdanov, F.~K.~Lamb, S.~Mahmoodifar, M.~C.~Miller, S.~M.~Morsink, T.~E.~Riley, T.~E.~Strohmayer, A.~L.~Watts, A.~J.~Dittmann and D.~Chakrabarty, \textit{et al.}
Astrophys. J. Lett. \textbf{887}, no.1, L26 (2019)
doi:10.3847/2041-8213/ab5968
[arXiv:1912.05707 [astro-ph.HE]].

\bibitem{Miller:2019cac}
M.~C.~Miller, F.~K.~Lamb, A.~J.~Dittmann, S.~Bogdanov, Z.~Arzoumanian, K.~C.~Gendreau, S.~Guillot, A.~K.~Harding, W.~C.~G.~Ho and J.~M.~Lattimer, \textit{et al.}
Astrophys. J. Lett. \textbf{887}, no.1, L24 (2019)
doi:10.3847/2041-8213/ab50c5
[arXiv:1912.05705 [astro-ph.HE]].

\bibitem{Raaijmakers:2019qny}
G.~Raaijmakers, T.~E.~Riley, A.~L.~Watts, S.~K.~Greif, S.~M.~Morsink, K.~Hebeler, A.~Schwenk, T.~Hinderer, S.~Nissanke and S.~Guillot, \textit{et al.}
Astrophys. J. Lett. \textbf{887}, no.1, L22 (2019)
doi:10.3847/2041-8213/ab451a
[arXiv:1912.05703 [astro-ph.HE]].

\bibitem{Reardon:2015kba}
D.~J.~Reardon, G.~Hobbs, W.~Coles, Y.~Levin, M.~J.~Keith, M.~Bailes, N.~D.~R.~Bhat, S.~Burke-Spolaor, S.~Dai and M.~Kerr, \textit{et al.}
Mon. Not. Roy. Astron. Soc. \textbf{455}, no.2, 1751-1769 (2016)
doi:10.1093/mnras/stv2395
[arXiv:1510.04434 [astro-ph.HE]].

\bibitem{2014HEAD...1411607B} Balsamo, E., Gendreau, K., \& Arzoumanian, Z.\ 2014, AAS/High Energy Astrophysics Division \#14



\bibitem{Fonseca:2016tux}
E.~Fonseca, T.~T.~Pennucci, J.~A.~Ellis, I.~H.~Stairs, D.~J.~Nice, S.~M.~Ransom, P.~B.~Demorest, Z.~Arzoumanian, K.~Crowter and T.~Dolch, \textit{et al.}
Astrophys. J. \textbf{832}, no.2, 167 (2016)
doi:10.3847/0004-637X/832/2/167
[arXiv:1603.00545 [astro-ph.HE]].

\bibitem{Demorest:2010bx}
P.~Demorest, T.~Pennucci, S.~Ransom, M.~Roberts and J.~Hessels,
Nature \textbf{467}, 1081-1083 (2010)
doi:10.1038/nature09466
[arXiv:1010.5788 [astro-ph.HE]].
\bibitem{NANOGRAV:2018hou}
Z.~Arzoumanian \textit{et al.} [NANOGRAV],
Astrophys. J. \textbf{859}, no.1, 47 (2018)
doi:10.3847/1538-4357/aabd3b
[arXiv:1801.02617 [astro-ph.HE]].

\bibitem{Antoniadis:2013pzd}
J.~Antoniadis, P.~C.~C.~Freire, N.~Wex, T.~M.~Tauris, R.~S.~Lynch, M.~H.~van Kerkwijk, M.~Kramer, C.~Bassa, V.~S.~Dhillon and T.~Driebe, \textit{et al.}
Science \textbf{340}, 6131 (2013)
doi:10.1126/science.1233232
[arXiv:1304.6875 [astro-ph.HE]].

\bibitem{NANOGrav:2019jur}
H.~T.~Cromartie \textit{et al.} [NANOGrav],
Nature Astron. \textbf{4}, no.1, 72-76 (2019)
doi:10.1038/s41550-019-0880-2
[arXiv:1904.06759 [astro-ph.HE]].

\bibitem{Fonseca:2021wxt}
E.~Fonseca, H.~T.~Cromartie, T.~T.~Pennucci, P.~S.~Ray, A.~Y.~Kirichenko, S.~M.~Ransom, P.~B.~Demorest, I.~H.~Stairs, Z.~Arzoumanian and L.~Guillemot, \textit{et al.}
Astrophys. J. Lett. \textbf{915}, no.1, L12 (2021)
doi:10.3847/2041-8213/ac03b8
[arXiv:2104.00880 [astro-ph.HE]].

\bibitem{Miller:2021qha}
M.~C.~Miller, F.~K.~Lamb, A.~J.~Dittmann, S.~Bogdanov, Z.~Arzoumanian, K.~C.~Gendreau, S.~Guillot, W.~C.~G.~Ho, J.~M.~Lattimer and M.~Loewenstein, \textit{et al.}
Astrophys. J. Lett. \textbf{918}, no.2, L28 (2021)
doi:10.3847/2041-8213/ac089b
[arXiv:2105.06979 [astro-ph.HE]].

\bibitem{Riley:2021pdl}
T.~E.~Riley, A.~L.~Watts, P.~S.~Ray, S.~Bogdanov, S.~Guillot, S.~M.~Morsink, A.~V.~Bilous, Z.~Arzoumanian, D.~Choudhury and J.~S.~Deneva, \textit{et al.}
Astrophys. J. Lett. \textbf{918}, no.2, L27 (2021)
doi:10.3847/2041-8213/ac0a81
[arXiv:2105.06980 [astro-ph.HE]].

\bibitem{Legred:2021hdx}
I.~Legred, K.~Chatziioannou, R.~Essick, S.~Han and P.~Landry,
Phys. Rev. D \textbf{104}, no.6, 063003 (2021)
doi:10.1103/PhysRevD.104.063003
[arXiv:2106.05313 [astro-ph.HE]].

\bibitem{Landry:2020vaw}
P.~Landry, R.~Essick and K.~Chatziioannou,
Phys. Rev. D \textbf{101}, no.12, 123007 (2020)
doi:10.1103/PhysRevD.101.123007
[arXiv:2003.04880 [astro-ph.HE]].
\bibitem{Stephani:2003tm}
H.~Stephani, D.~Kramer, M.~A.~H.~MacCallum, C.~Hoenselaers and E.~Herlt,
Cambridge Univ. Press, 2003,
ISBN 978-0-521-46702-5, 978-0-511-05917-9
doi:10.1017/CBO9780511535185

\bibitem{Delgaty:1998uy}
M.~S.~R.~Delgaty and K.~Lake,
Comput. Phys. Commun. \textbf{115} (1998), 395-415
doi:10.1016/S0010-4655(98)00130-1
[arXiv:gr-qc/9809013 [gr-qc]].

\bibitem{1967ApJ...147..310B} Buchdahl, H.~A.\ 1967, Astrophysical Journal, 147, 310. doi:10.1086/149001

\bibitem{1999GReGr..31..945N} Nariai, H.\ 1999, General Relativity and Gravitation, 31, 945

\bibitem{Tolman:1939jz}
R.~C.~Tolman,
Phys. Rev. \textbf{55} (1939), 364-373
doi:10.1103/PhysRev.55.364


\bibitem{1975JPhA....8..508K} Krori, K.~D. \& Barua, J.\ 1975, Journal of Physics A Mathematical General, 8, 508. doi:10.1088/0305-4470/8/4/012
\bibitem{Roupas:2020mvs}
Z.~Roupas and G.~G.~L.~Nashed,
Eur. Phys. J. C \textbf{80}, no.10, 905 (2020)
doi:10.1140/epjc/s10052-020-08462-1
[arXiv:2007.09797 [gr-qc]].

\bibitem{Nashed:2023pxd}
G.~G.~L.~Nashed,
Astrophys. J. \textbf{950}, no.2, 129 (2023)
doi:10.3847/1538-4357/acd182
[arXiv:2306.10273 [gr-qc]].
\bibitem{Rahaman:2011cw}
F.~Rahaman, R.~Sharma, S.~Ray, R.~Maulick and I.~Karar,
Eur. Phys. J. C \textbf{72} (2012), 2071
doi:10.1140/epjc/s10052-012-2071-5
[arXiv:1108.6125 [gr-qc]].

\bibitem{MonowarHossein:2012ec}
S.~Monowar Hossein, F.~Rahaman, J.~Naskar, M.~Kalam and S.~Ray,
Int. J. Mod. Phys. D \textbf{21} (2012), 1250088
doi:10.1142/S0218271812500885
[arXiv:1204.3558 [gr-qc]].
\bibitem{Kalam:2012sh}
M.~Kalam, F.~Rahaman, S.~Ray, S.~M.~Hossein, I.~Karar and J.~Naskar,
Eur. Phys. J. C \textbf{72} (2012), 2248
doi:10.1140/epjc/s10052-012-2248-y
[arXiv:1201.5234 [gr-qc]].

\bibitem{Rahaman:2010mr}
F.~Rahaman, S.~Ray, A.~K.~Jafry and K.~Chakraborty,
Phys. Rev. D \textbf{82} (2010), 104055
doi:10.1103/PhysRevD.82.104055
[arXiv:1007.1889 [physics.gen-ph]].

\bibitem{Turyshev:2008dr}
S.~G.~Turyshev,
Ann. Rev. Nucl. Part. Sci. \textbf{58}, 207-248 (2008)
doi:10.1146/annurev.nucl.58.020807.111839
[arXiv:0806.1731 [gr-qc]].

\bibitem{LIGOScientific:2016aoc}
B.~P.~Abbott \textit{et al.} [LIGO Scientific and Virgo],
Phys. Rev. Lett. \textbf{116}, no.6, 061102 (2016)
doi:10.1103/PhysRevLett.116.061102
[arXiv:1602.03837 [gr-qc]].

\bibitem{EventHorizonTelescope:2019dse}
K.~Akiyama \textit{et al.} [Event Horizon Telescope],
Astrophys. J. Lett. \textbf{875}, L1 (2019)
doi:10.3847/2041-8213/ab0ec7
[arXiv:1906.11238 [astro-ph.GA]].
\bibitem{Dodelson:2003ft}
S.~Dodelson,
Academic Press, 2003,
ISBN 978-0-12-219141-1
\bibitem{Unruh:2017uaw}
W.~G.~Unruh and R.~M.~Wald,
Rept. Prog. Phys. \textbf{80}, no.9, 092002 (2017)
doi:10.1088/1361-6633/aa778e
[arXiv:1703.02140 [hep-th]].

\bibitem{Penrose:1964wq}
R.~Penrose,
Phys. Rev. Lett. \textbf{14}, 57-59 (1965)
doi:10.1103/PhysRevLett.14.57

\bibitem{Senovilla:2014gza}
J.~M.~M.~Senovilla and D.~Garfinkle,
Class. Quant. Grav. \textbf{32}, no.12, 124008 (2015)
doi:10.1088/0264-9381/32/12/124008
[arXiv:1410.5226 [gr-qc]].

\bibitem{Nojiri:2017ncd}
S.~Nojiri, S.~D.~Odintsov and V.~K.~Oikonomou,
Phys. Rept. \textbf{692}, 1-104 (2017)
doi:10.1016/j.physrep.2017.06.001
[arXiv:1705.11098 [gr-qc]].

\bibitem{Stuchlik:2020rls}
Z.~Stuchl\'\i{}k, M.~Kolo\v{s}, J.~Kov\'a\v{r}, P.~Slan\'y and A.~Tursunov,
Universe \textbf{6}, no.2, 26 (2020)
doi:10.3390/universe6020026

\bibitem{Boehmer:2004nu}
C.~G.~Boehmer,
Ukr. J. Phys. \textbf{50}, 1219-1225 (2005)
[arXiv:gr-qc/0409030 [gr-qc]].

\bibitem{Stuchlik:2016xiq}
Z.~Stuchl\'\i{}k, S.~Hled\'\i{}k and J.~Novotn\'y,
Phys. Rev. D \textbf{94}, no.10, 103513 (2016)
doi:10.1103/PhysRevD.94.103513
[arXiv:1611.05327 [gr-qc]].

\bibitem{Novotny:2021zlq}
J.~Novotn\'y, Z.~Stuchl\'\i{}k and J.~Hlad\'\i{}k,
Astron. Astrophys. \textbf{647}, A29 (2021)
doi:10.1051/0004-6361/202039338
[arXiv:2101.00891 [astro-ph.CO]].

\bibitem{Stuchlik:2011zz}
Z.~Stuchlik and J.~Schee,
JCAP \textbf{09}, 018 (2011)
doi:10.1088/1475-7516/2011/09/018

\bibitem{Carroll:2000fy}
S.~M.~Carroll,
Living Rev. Rel. \textbf{4}, 1 (2001)
doi:10.12942/lrr-2001-1
[arXiv:astro-ph/0004075 [astro-ph]].

\bibitem{Einstein:1917ce}
A.~Einstein,
Sitzungsber. Preuss. Akad. Wiss. Berlin (Math. Phys. ) \textbf{1917}, 142-152 (1917)

\bibitem{ORaifeartaigh:2017uct}
C.~O'Raifeartaigh, M.~O'Keeffe, W.~Nahm and S.~Mitton,
Eur. Phys. J. H \textbf{42}, no.3, 431-474 (2017)
doi:10.1140/epjh/e2017-80002-5
[arXiv:1701.07261 [physics.hist-ph]].

\bibitem{Buchdahl:1970ynr}
H.~A.~Buchdahl,
Mon. Not. Roy. Astron. Soc. \textbf{150}, 1 (1970)

\bibitem{Sotiriou:2008rp}
T.~P.~Sotiriou and V.~Faraoni,
Rev. Mod. Phys. \textbf{82}, 451-497 (2010)
doi:10.1103/RevModPhys.82.451
[arXiv:0805.1726 [gr-qc]].

\bibitem{Copeland:2006wr}
E.~J.~Copeland, M.~Sami and S.~Tsujikawa,
Int. J. Mod. Phys. D \textbf{15}, 1753-1936 (2006)
doi:10.1142/S021827180600942X
[arXiv:hep-th/0603057 [hep-th]].

\bibitem{DeFelice:2010aj}
A.~De Felice and S.~Tsujikawa,
Living Rev. Rel. \textbf{13}, 3 (2010)
doi:10.12942/lrr-2010-3
[arXiv:1002.4928 [gr-qc]].

\bibitem{Nojiri:2010wj}
S.~Nojiri and S.~D.~Odintsov,
Phys. Rept. \textbf{505}, 59-144 (2011)
doi:10.1016/j.physrep.2011.04.001
[arXiv:1011.0544 [gr-qc]].

\bibitem{Harko:2011kv}
T.~Harko, F.~S.~N.~Lobo, S.~Nojiri and S.~D.~Odintsov,
Phys. Rev. D \textbf{84}, 024020 (2011)
doi:10.1103/PhysRevD.84.024020
[arXiv:1104.2669 [gr-qc]].

\bibitem{Harko:2014gwa}
T.~Harko and F.~S.~N.~Lobo,
Galaxies \textbf{2}, no.3, 410-465 (2014)
doi:10.3390/galaxies2030410
[arXiv:1407.2013 [gr-qc]].

\bibitem{Goncalves:2021vci}
T.~B.~Gon\c{c}alves, J.~L.~Rosa and F.~S.~N.~Lobo,
Phys. Rev. D \textbf{105}, no.6, 064019 (2022)
doi:10.1103/PhysRevD.105.064019
[arXiv:2112.02541 [gr-qc]].

\bibitem{Goncalves:2022ggq}
T.~B.~Gon\c{c}alves, J.~L.~Rosa and F.~S.~N.~Lobo,
Eur. Phys. J. C \textbf{82}, no.5, 418 (2022)
doi:10.1140/epjc/s10052-022-10371-4
[arXiv:2203.11124 [gr-qc]].

\bibitem{Hansraj:2018jzb}
S.~Hansraj and A.~Banerjee,
Phys. Rev. D \textbf{97}, no.10, 104020 (2018)
doi:10.1103/PhysRevD.97.104020

\bibitem{Bhar:2021uqr}
P.~Bhar, P.~Rej and M.~Zubair,
Chin. J. Phys. \textbf{77}, 2201-2212 (2022)
doi:10.1016/j.cjph.2021.11.013
[arXiv:2112.07581 [gr-qc]].

\bibitem{Kumar:2021vqa}
J.~Kumar, H.~D.~Singh and A.~K.~Prasad,
Phys. Dark Univ. \textbf{34}, 100880 (2021)
doi:10.1016/j.dark.2021.100880
[arXiv:2106.12560 [gr-qc]].

\bibitem{Feng:2022bvk}
Z.~Feng,
[arXiv:2210.01574 [gr-qc]].

\bibitem{Tangphati:2022arm}
T.~Tangphati, I.~Karar, A.~Pradhan and A.~Banerjee,
Eur. Phys. J. C \textbf{82}, no.1, 57 (2022)
doi:10.1140/epjc/s10052-022-10024-6
\bibitem{Tangphati:2022mur}
T.~Tangphati, S.~Hansraj, A.~Banerjee and A.~Pradhan,
Phys. Dark Univ. \textbf{35}, 100990 (2022)
doi:10.1016/j.dark.2022.100990

\bibitem{Wu:2018idg}
J.~Wu, G.~Li, T.~Harko and S.~D.~Liang,
Eur. Phys. J. C \textbf{78}, no.5, 430 (2018)
doi:10.1140/epjc/s10052-018-5923-9
[arXiv:1805.07419 [gr-qc]].

\bibitem{Faraoni:2009rk}
V.~Faraoni,
Phys. Rev. D \textbf{80}, 124040 (2009)
doi:10.1103/PhysRevD.80.124040
[arXiv:0912.1249 [astro-ph.GA]].

\bibitem{Avelino:2018rsb}
P.~P.~Avelino and R.~P.~L.~Azevedo,
Phys. Rev. D \textbf{97}, no.6, 064018 (2018)
doi:10.1103/PhysRevD.97.064018
[arXiv:1802.04760 [gr-qc]].

\bibitem{BarrientosO:2014mys}
J.~Barrientos O. and G.~F.~Rubilar,
Phys. Rev. D \textbf{90}, no.2, 028501 (2014)
doi:10.1103/PhysRevD.90.028501

\bibitem{Pretel:2021kgl}
J.~M.~Z.~Pretel, S.~E.~Jor\'as, R.~R.~R.~Reis and J.~D.~V.~Arba\~nil,
JCAP \textbf{08}, 055 (2021)
doi:10.1088/1475-7516/2021/08/055
[arXiv:2105.07573 [gr-qc]].

\bibitem{Pretel:2020oae}
J.~M.~Z.~Pretel, S.~E.~Jor\'as, R.~R.~R.~Reis and J.~D.~V.~Arba\~nil,
JCAP \textbf{04}, 064 (2021)
doi:10.1088/1475-7516/2021/04/064
[arXiv:2012.03342 [gr-qc]].





\bibitem{Rosa:2021teg}
J.~L.~Rosa,
Phys. Rev. D \textbf{103}, no.10, 104069 (2021)
doi:10.1103/PhysRevD.103.104069
[arXiv:2103.11698 [gr-qc]].

\bibitem{Rosa:2022cen}
J.~L.~Rosa and D.~Rubiera-Garcia,
Phys. Rev. D \textbf{106}, no.6, 064007 (2022)
doi:10.1103/PhysRevD.106.064007
[arXiv:2204.12944 [gr-qc]].

\bibitem{Nashed:2004pn}
G.~G.~L.~Nashed,
Mod. Phys. Lett. A \textbf{21}, 2241-2250 (2006)
doi:10.1142/S0217732306020445
[arXiv:gr-qc/0401041 [gr-qc]].

\bibitem{Nashed:2006yw}
G.~G.~L.~Nashed,
Mod. Phys. Lett. A \textbf{22}, 1047-1056 (2007)
doi:10.1142/S021773230702141X
[arXiv:gr-qc/0609096 [gr-qc]].

\bibitem{Nashed:2020kjh}
G.~G.~L.~Nashed and S.~Capozziello,
Eur. Phys. J. C \textbf{80}, no.10, 969 (2020)
doi:10.1140/epjc/s10052-020-08551-1
[arXiv:2010.06355 [gr-qc]].

\bibitem{Nashed:2022zyi}
G.~G.~L.~Nashed and W.~El Hanafy,
Eur. Phys. J. C \textbf{82}, no.8, 679 (2022)
doi:10.1140/epjc/s10052-022-10634-0
[arXiv:2208.13814 [gr-qc]].



\bibitem{1971reas.book.....Z} Zeldovich, Y.~B. \& Novikov, I.~D.\ 1971, Chicago: University of Chicago Press, 1971
\bibitem{Buchdahl:1959zz}
H.~A.~Buchdahl,
Phys. Rev. \textbf{116}, 1027 (1959)
doi:10.1103/PhysRev.116.1027

\bibitem{Ivanov:2002xf}
B.~V.~Ivanov,
Phys. Rev. D \textbf{65}, 104011 (2002)
doi:10.1103/PhysRevD.65.104011
[arXiv:gr-qc/0201090 [gr-qc]].

\bibitem{Barraco:2003jq}
D.~E.~Barraco, V.~H.~Hamity and R.~J.~Gleiser,
Phys. Rev. D \textbf{67}, 064003 (2003)
doi:10.1103/PhysRevD.67.064003

\bibitem{Boehmer:2006ye}
C.~G.~Boehmer and T.~Harko,
Class. Quant. Grav. \textbf{23}, 6479-6491 (2006)
doi:10.1088/0264-9381/23/22/023
[arXiv:gr-qc/0609061 [gr-qc]].
\bibitem{1988CQGra...5.1329K} Kolassis, C.~A., Santos, N.~O., \& Tsoubelis, D.\ 1988, Classical and Quantum Gravity, 5, 1329. doi:10.1088/0264-9381/5/10/011

\bibitem{Ivanov:2017kyr}
B.~V.~Ivanov,
Eur. Phys. J. C \textbf{77}, no.11, 738 (2017)
doi:10.1140/epjc/s10052-017-5322-7
[arXiv:1708.07971 [gr-qc]].

\bibitem{2019EPJC...79..853D} Das, S., Rahaman, F., \& Baskey, L.\ 2019, European Physical Journal C, 79, 853. doi:10.1140/epjc/s10052-019-7367-2

\bibitem{Bedaque:2014sqa}
P.~Bedaque and A.~W.~Steiner,
Phys. Rev. Lett. \textbf{114}, no.3, 031103 (2015)
doi:10.1103/PhysRevLett.114.031103
[arXiv:1408.5116 [nucl-th]].

\bibitem{Cherman:2009tw}
A.~Cherman, T.~D.~Cohen and A.~Nellore,
Phys. Rev. D \textbf{80}, 066003 (2009)
doi:10.1103/PhysRevD.80.066003
[arXiv:0905.0903 [hep-th]].


\bibitem{Herrera:1992lwz}
L.~Herrera,
Phys. Lett. A \textbf{165}, 206-210 (1992)
doi:10.1016/0375-9601(92)90036-L

\bibitem{1993MNRAS.265..533C} Chan, R., Herrera, L., \& Santos, N.~O.\ 1993, mon note royal astron socity, 265, 533. doi:10.1093/mnras/265.3.533

\bibitem{Chandrasekhar:1964zz}
S.~Chandrasekhar,
Astrophys. J. \textbf{140}, 417-433 (1964)
[erratum: Astrophys. J. \textbf{140}, 1342 (1964)]
doi:10.1086/147938

\bibitem{1975A&A....38...51H} Heintzmann, H. \& Hillebrandt, W.\ 1975, Astronomy and Astrophysics,, 38, 51


\bibitem{Oppenheimer:1939ne}
J.~R.~Oppenheimer and G.~M.~Volkoff,
Phys. Rev. \textbf{55}, 374-381 (1939)
doi:10.1103/PhysRev.55.374



\bibitem{Abubekerov:2008inw}
M.~K.~Abubekerov, E.~A.~Antokhina, A.~M.~Cherepashchuk and V.~V.~Shimanskii,
Astron. Rep. \textbf{52}, 379-389 (2008)
doi:10.1134/S1063772908050041
[arXiv:1201.5519 [astro-ph.SR]].

\bibitem{Gangopadhyay:2013gha}
T.~Gangopadhyay, S.~Ray, X.~D.~Li, J.~Dey and M.~Dey,
Mon. Not. Roy. Astron. Soc. \textbf{431}, 3216-3221 (2013)
doi:10.1093/mnras/stt401
[arXiv:1303.1956 [astro-ph.HE]].

\bibitem{1930PhRv...35..896T} Tolman, R.~C.\ 1930, Physical Review, 35, 896. doi:10.1103/PhysRev.35.896







\bibitem{Rawls:2011jw}
M.~L.~Rawls, J.~A.~Orosz, J.~E.~McClintock, M.~A.~P.~Torres, C.~D.~Bailyn and M.~M.~Buxton,
Astrophys. J. \textbf{730}, 25 (2011)
doi:10.1088/0004-637X/730/1/25
[arXiv:1101.2465 [astro-ph.SR]].


\bibitem{Naik:2011qc}
S.~Naik, B.~Paul and Z.~Ali,
Astrophys. J. \textbf{737}, 79 (2011)
doi:10.1088/0004-637X/737/2/79
[arXiv:1106.0370 [astro-ph.SR]].


\bibitem{Ozel:2008kb}
F.~Ozel, T.~Guver and D.~Psaltis,
Astrophys. J. \textbf{693}, 1775-1779 (2009)
doi:10.1088/0004-637X/693/2/1775
[arXiv:0810.1521 [astro-ph]].

\bibitem{Webb:2007tc}
N.~A.~Webb and D.~Barret,
Astrophys. J. \textbf{671}, 727 (2007)
doi:10.1086/522877
[arXiv:0708.3816 [astro-ph]].
\bibitem{Bogdanov:2016nle}
S.~Bogdanov, C.~O.~Heinke, F.~\"Ozel and T.~G\"uver,
Astrophys. J. \textbf{831}, no.2, 184 (2016)
doi:10.3847/0004-637X/831/2/184
[arXiv:1603.01630 [astro-ph.HE]].

\bibitem{Ozel:2015fia}
F.~Ozel, D.~Psaltis, T.~Guver, G.~Baym, C.~Heinke and S.~Guillot,
Astrophys. J. \textbf{820}, no.1, 28 (2016)
doi:10.3847/0004-637X/820/1/28
[arXiv:1505.05155 [astro-ph.HE]].
\bibitem{1996IAUC.6331....1M}
 {{Marshall}, F.~E. and {Angelini}, L.},
   {IAU Circ.},
  {\textbf  6331}, 1 (1996)
\bibitem{Gonzalez-Caniulef:2019wzi}
D.~Gonzalez-Caniulef, S.~Guillot and A.~Reisenegger,
Mon. Not. Roy. Astron. Soc. \textbf{490}, no.4, 5848-5859 (2019)
doi:10.1093/mnras/stz2941
[arXiv:1904.12114 [astro-ph.HE]].

\bibitem{NANOGrav:2017wvv}
Z.~Arzoumanian \textit{et al.} [NANOGrav],
Astrophys. J. Suppl. \textbf{235}, no.2, 37 (2018)
doi:10.3847/1538-4365/aab5b0
[arXiv:1801.01837 [astro-ph.HE]].
\bibitem{LIGOScientific:2020zkf}
R.~Abbott \textit{et al.} [LIGO Scientific and Virgo],
Astrophys. J. Lett. \textbf{896}, no.2, L44 (2020)
doi:10.3847/2041-8213/ab960f
[arXiv:2006.12611 [astro-ph.HE]].
\bibitem{LIGOScientific:2018cki}
B.~P.~Abbott \textit{et al.} [LIGO Scientific and Virgo],
Phys. Rev. Lett. \textbf{121}, no.16, 161101 (2018)
doi:10.1103/PhysRevLett.121.161101
[arXiv:1805.11581 [gr-qc]].


\bibitem{Rybicki:2005id}
G.~B.~Rybicki, C.~O.~Heinke, R.~Narayan and J.~E.~Grindlay,
Astrophys. J. \textbf{644}, 1090-1103 (2006)
doi:10.1086/503701
[arXiv:astro-ph/0506563 [astro-ph]].



\bibitem{Guver:2010td}
T.~Guver, P.~Wroblewski, L.~Camarota and F.~Ozel,
Astrophys. J. \textbf{719}, 1807 (2010)
doi:10.1088/0004-637X/719/2/1807
[arXiv:1002.3825 [astro-ph.HE]].






\bibitem{Nashed:2018oaf}
G.~G.~L.~Nashed,
Eur. Phys. J. Plus \textbf{133}, no.1, 18 (2018)
doi:10.1140/epjp/i2018-11849-7

\bibitem{Nashed:2018piz}
G.~G.~L.~Nashed,
Adv. High Energy Phys. \textbf{2018}, 7323574 (2018)
doi:10.1155/2018/7323574

\bibitem{Nashed:2018efg}
G.~G.~L.~Nashed,
Int. J. Mod. Phys. D \textbf{27}, no.7, 1850074 (2018)
doi:10.1142/S0218271818500748
\bibitem{ElHanafy:2014tzj}
W.~El Hanafy and G.~G.~L.~Nashed,
Eur. Phys. J. C \textbf{75}, 279 (2015)
doi:10.1140/epjc/s10052-015-3501-y
[arXiv:1409.7199 [hep-th]].
\end{thebibliography}
\end{document}